\pdfoutput=1
\documentclass[a4paper,12pt]{article}
\usepackage{amsfonts}
\usepackage{mathrsfs}
\usepackage{amsmath}
\usepackage{amssymb}
\usepackage{framed}
\usepackage[medium]{titlesec}
\usepackage{bm}
\usepackage{cite}
\usepackage[normalem]{ulem}
\usepackage{extarrows}
\usepackage{slashed}
\usepackage{isodateo}
\usepackage{graphicx}
\usepackage{xcolor}
\usepackage[bookmarksnumbered=true,bookmarksopen=true]{hyperref}
 \hypersetup{colorlinks,%
             linkcolor=[rgb]{0,0.3,0.6}, %
             citecolor=[rgb]{0,0.3,0.6}, %
             urlcolor=[rgb]{0,0.3,0.6}}
\usepackage[hmargin=.7in,vmargin=1.1in]{geometry}
\usepackage{indentfirst}
\usepackage{booktabs}

\newcommand{\FR}[2]{\displaystyle\frac{\,{#1}\,}{#2}}
\newcommand{\fr}[2]{\mbox{$\frac{\,{#1}\,}{#2}$}}
\newcommand{\n}{\nonumber}

\graphicspath{{fig/}}

\def\bge{\begin{equation}}
\def\ede{\end{equation}}
\def\bga{\begin{aligned}}
\def\eda{\end{aligned}}
\def\bgb{\begin{bmatrix}}
\def\edb{\end{bmatrix}}
\def\bgp{\begin{pmatrix}}
\def\edp{\end{pmatrix}}
\def\bgm{\begin{matrix}}
\def\edm{\end{matrix}}
\def\bgs{\begin{subequations}}
\def\eds{\end{subequations}}
\newcommand{\order}[1]{\mathcal{O}({#1})}
\def\di{{\mathrm{d}}}

\def\mb{\mathbf}

\def\pd{\partial}
\def\ld{{\mathscr{L}}}

\def\la{\langle}\def\ra{\rangle}

\def\to{\rightarrow}

\def\ii{\mathrm{i}}

\def\al{\alpha}
\def\be{\beta}
\def\ga{\gamma}
\def\de{\delta}
\def\ep{\epsilon}

\def\lam{\lambda}

\def\si{\sigma}

\def\aa{\mathsf{a}}
\def\bb{\mathsf{b}}

\usepackage{mdframed}

\newmdenv[skipabove=0pt,%
          skipbelow=5pt,%
          leftmargin=0pt,%
          rightmargin=0pt,%
          innertopmargin=-5pt,%
          innerbottommargin=7pt,%
          innerleftmargin=2pt,%
          innerrightmargin=2pt,%
          splittopskip=0pt,%
          splitbottomskip=0pt,%
          linewidth=0pt,%
          nobreak=true]%
          {keyeqn2}

\newmdenv[backgroundcolor=gray!15,%
          skipabove=0pt,%
          skipbelow=5pt,%
          leftmargin=0pt,%
          rightmargin=0pt,%
          innertopmargin=-5pt,%
          innerbottommargin=7pt,%
          innerleftmargin=2pt,%
          innerrightmargin=2pt,%
          splittopskip=0pt,%
          splitbottomskip=0pt,%
          linewidth=0pt,%
          nobreak=true]%
          {keyeqn}

\usepackage{titlesec}          
\titleformat{\section}
{\normalfont\fontsize{15}{20}\bfseries}{\thesection}{1em}{}

\newcommand{\wt}[1]{\mkern 2mu \widetilde{\mkern -2mu #1 \mkern -2mu}\mkern 2mu}
\newcommand{\wh}[1]{\mkern 2mu \widehat{\mkern-2mu#1\mkern-2mu}\mkern 2mu}

\newcommand{\fnemail}[1]{\footnote{Email: \href{mailto:#1}{\nolinkurl{#1}}}}

\newcommand{\pdd}[2]{\frac{\partial{#1}}{\partial{#2}}}

\begin{document}


\title{\Large\textbf{Bootstrapping One-Loop Inflation Correlators with the Spectral Decomposition}\\[2mm]}

\author{Zhong-Zhi Xianyu$^{1,2\,}$\fnemail{zxianyu@tsinghua.edu.cn}
~~~~~ and ~~~~~
Hongyu Zhang$^{1,\,}$\fnemail{zhy21@mails.tsinghua.edu.cn}\\[5mm]
\normalsize{${}^{1}\,$\emph{Department of Physics, Tsinghua University, Beijing 100084, China}}\\
\normalsize{${}^{2}\,$\emph{Collaborative Innovation Center of Quantum Matter, Beijing 100084, China}}}

\date{}
\maketitle

\vspace{20mm}

\begin{abstract}
\vspace{10mm}

Phenomenological studies of cosmological collider physics in recent years have identified many 1-loop inflation correlators as leading channels for discovering heavy new particles around or above the inflation scale. However, complete analytical results for these massive 1-loop correlators are currently unavailable. In this work, we embark on a program of bootstrapping inflation correlators with massive exchanges at 1-loop order, with the input of tree-level inflation correlators and the techniques of spectral decomposition in dS. As a first step, we present for the first time the complete and analytical results for a class of 4-point and 3-point inflation correlators mediated by massive scalar fields at the 1-loop order. Using the full result, we provide simple and reliable analytical approximations for the signals and the background in the squeezed limit. We also identify configurations of the scalar trispectrum where the oscillatory signal from the loop is dominant over the background.

 \end{abstract}

\newpage
\tableofcontents

\newpage
\section{Introduction}

One of the most fascinating facts in modern cosmology is that we can access the physics of the primordial universe by measuring the correlation functions of large-scale inhomogeneities and anisotropies. Examples of such measurements include the temperature and the polarizations of the cosmological microwave background \cite{Planck:2018jri,Planck:2019kim}, the large-scale structure survey \cite{Ferraro:2022cmj}, and the more futuristic 21cm tomography \cite{Munoz:2015eqa,Liu:2022iyy}. Based on existing observational data, it is now widely believed that the large-scale inhomogeneities originated from a period of inflation in the primordial universe. During this almost exponentially fast expansion, the quantum fluctuations of fields were generated through processes similar to Schwinger pair production. They were then quickly redshifted to super-horizon scales, and sourced the large-scale fluctuations we see today \cite{Achucarro:2022qrl}. 

The central objects that bridge the observations and the theory are the $n$-point correlation functions of quantum fields produced during inflation. In this work, we collectively call them \emph{inflation correlators}. Inflation correlators are, on the one hand, calculable from a quantum field theory in an inflationary spacetime \cite{Maldacena:2002vr}, and, on the other hand, measurable through the various cosmological probes mentioned above. Examples of inflation correlators include the $n$-point functions of the inflaton fluctuations and the tensor modes of the metric fluctuation $\ga$. Depending on models, $n$-point functions of isocurvature modes, such as the dark-matter isocurvature fluctuation, may also be observable. 

From a theoretical point of view, the inflationary spacetime is close to the Poincaré patch of $3+1$ dimensional de Sitter spacetime (dS), one of the three maximally symmetric spacetimes. The inflation correlators can be thought of as correlation functions of bulk quantum fields, with all external points pinned onto the future boundary of the dS. Hence, the inflation correlators are natural dS counterparts of scattering amplitudes in Minkowski spacetime and boundary correlators in anti-de Sitter spacetime (AdS). It is thus of both theoretical and phenomenological interest to study inflation correlators. 

In recent years, it was realized that the inflation correlators can be used as probes of new heavy particles and their interactions at the inflation scale \cite{Chen:2009we,Chen:2009zp,Baumann:2011nk,Chen:2012ge,Pi:2012gf,Noumi:2012vr,Gong:2013sma,Arkani-Hamed:2015bza}, which is presumably much higher than any terrestrial collider experiments. This program has been dubbed ``cosmological collider (CC) physics'' \cite{Arkani-Hamed:2015bza}. In particular, heavy particles can leave distinct oscillatory shapes in various soft limits of $n$-point inflation correlators, known as CC signals. The rich particle phenomenology of the CC has been actively explored recently \cite{Chen:2015lza,Chen:2016nrs,Chen:2016uwp,Chen:2016hrz,Lee:2016vti,An:2017hlx,An:2017rwo,Iyer:2017qzw,Kumar:2017ecc,Chen:2017ryl,Tong:2018tqf,Chen:2018sce,Chen:2018xck,Chen:2018cgg,Chua:2018dqh,Wu:2018lmx,Saito:2018omt,Li:2019ves,Lu:2019tjj,Liu:2019fag,Hook:2019zxa,Hook:2019vcn,Kumar:2018jxz,Kumar:2019ebj,Alexander:2019vtb,Wang:2019gbi,Wang:2019gok,Wang:2020uic,Li:2020xwr,Wang:2020ioa,Fan:2020xgh,Aoki:2020zbj,Bodas:2020yho,Maru:2021ezc,Lu:2021gso,Sou:2021juh,Lu:2021wxu,Pinol:2021aun,Cui:2021iie,Tong:2022cdz,Reece:2022soh,Chen:2022vzh}. From these studies, it is now clear that many properties of heavy particles can leave distinct signatures in signals, including the mass, the spin, the sound speed, the chemical potential, and the interaction types. 

In many particle models of CC physics, the leading CC signal appears at the 1-loop level rather than the tree level. See, e.g., \cite{Chen:2016uwp,Chen:2016hrz,Chen:2018xck,Lu:2019tjj,Hook:2019zxa,Hook:2019vcn,Kumar:2018jxz,Wang:2019gbi,Wang:2020ioa,Lu:2021gso,Cui:2021iie,Tong:2022cdz}. This happens in particular when the signal-generating states have to be produced and annihilated in pairs, such as fermions and states carrying conserved charges. In such cases, the tree-level process is simply absent. There are also cases in which the 1-loop processes are more enhanced relative to the tree-level process (but higher loops remain subdominant so that the perturbation theory still works). The 1-loop-dominant CC signals cover a large range of models, including most of the Standard Model states in the symmetric phase, the chemical-potential-enhanced signals in the 3-point functions, and new physics states such as heavy neutrinos, Kaluza-Klein states, etc. 

It is thus desirable to have analytical expressions for 1-loop inflation correlators, both for a better understanding of the analytical structure of dS correlators and for phenomenological applications. However, the computation of inflation correlators is difficult, due to the lack of symmetries, the build-in time ordering in the computation of inflation correlators, and the complicated mode functions (usually Hankel functions and Whittaker functions). Recent years have witnessed considerable progress toward the analytical and numerical computation of inflation correlators \cite{Arkani-Hamed:2018kmz,Baumann:2019oyu,Baumann:2020dch,Sleight:2019mgd,Sleight:2019hfp,Sleight:2020obc,Sleight:2021iix,Sleight:2021plv,Pajer:2020wnj,Pajer:2020wxk,Cabass:2021fnw,Pimentel:2022fsc,Jazayeri:2022kjy,Qin:2022lva,Qin:2022fbv,Wang:2021qez,Goodhew:2020hob,Melville:2021lst,Goodhew:2021oqg,DiPietro:2021sjt,Tong:2021wai,Bonifacio:2021azc,Hogervorst:2021uvp,Meltzer:2021zin,Heckelbacher:2022hbq,Gomez:2021qfd,Gomez:2021ujt,Baumann:2021fxj,Baumann:2022jpr}. Analytical results for general massive exchanges at the tree level have been worked out in various ways, including the simpler dS covariant case and the more complicated boost-breaking cases. The techniques developed in recent years allow us to compute many tree-level inflation correlators with massive exchanges. 

In comparison, general 1-loop massive exchanges remain challenging, and full analytical results are very rare in the literature.\footnote{There is a relatively long history in the study of \emph{massless} loops in dS. In particular, massless loops with non-derivative couplings have been extensively studied due to their peculiar infrared properties. In general, massless loops are more tractable relative to massive loops, and we do not consider them in this work.} Known nontrivial examples include the 1-loop bubble correction to the 2-point function from scalar fields of arbitrary mass in Euclidean dS \cite{Marolf:2010zp}, the 4-point function with 1-loop exchange of conformal scalar ($m^2=2H^2$, $H$ being the inflationary Hubble scale) in the position space \cite{Heckelbacher:2022hbq}.\footnote{There are also examples such as 1-loop seagull diagram which are trivial to compute. } However, these examples contain no CC signals: The 2-point function is free from any oscillatory signal by the scale symmetry of the problem. The conformal scalar mediation does not generate any CC signals, because the conformal scalar has a real scaling dimension $\Delta=3/2-\sqrt{9/4-(m/H)^2}=1$, while the oscillatory CC signals require complex scaling dimensions, namely, $m>3H/2$.

Currently, we are unaware of any complete analytical results for 1-loop inflation correlators containing CC signals, although partial results do exist. For example, the complete analytical results for 1-loop nonlocal CC signals were worked out in \cite{Qin:2022lva} using the partial Mellin-Barnes representation. There are also full numerical results for a class of signal-carrying 1-loop 3-point functions \cite{Wang:2021qez}. It turns out that numerical computation is nontrivial as well, and fast numerical computation has not been achieved yet at the moment. 
 
The lack of full results for massive 1-loop exchange has been a problem for particle model buildings and phenomenological studies of CC physics. To assess observable parameter space, one has to resort to unjustified approximations such as a late-time expansion of loop propagators. The physical reason to take late-time expansion is that the CC signal is typically generated from a resonant process in the soft limit of the correlator. At the resonant point, the massive mode carries the soft momentum and is well outside the horizon. Therefore, the time integral is expected to receive most of its contribution from the late-time part of the massive mode. From this argument, it is clear that the late-time expansion is only useful for estimating the oscillatory signal in the squeezed limit; It cannot be used to estimate the non-oscillatory ``background'' of the correlator. Worse still, in most frequently encountered 3-point functions (Fig.\ \ref{fig_1loop3pt}), it is known that the resonant argument applies only to one of the two time integrals, and thus the late-time expansion is conceptually flawed. Indeed, if one insists on working with the late-time expansion, as in most previous studies, the resulting strength of the CC signal would have a slightly wrong parameter dependence.\footnote{More precisely, we expect that the CC signal strength scales with the intermediate mass $m$ as $\wt\nu^a e^{-b\wt\nu}$, where $\wt\nu=\sqrt{(m/H)^2-9/4}$ for scalar fields, and $(a,b)$ are parameters to be determined. It was known that a naïve late-time expansion would yield correct $b$ but wrong $a$. See the appendix of \cite{Wang:2020ioa} for a discussion of this issue.}

In this work, we start a program of bootstrapping inflation correlators with massive 1-loop exchanges via spectral decomposition. As a first step, we compute the 1-loop diagram mediated by a pair of scalar fields $\si$ of the same mass $m$. By defining and working out a \emph{loop seed integral}, we can generate complete analytical results for many 1-loop correlators with massive scalar exchanges with various types of couplings. Furthermore, by taking appropriate folded limits, we can also obtain full results for 3-point functions with massive 1-loop bubble exchange.

\begin{figure}
 \centering
  \includegraphics[width=0.95\textwidth]{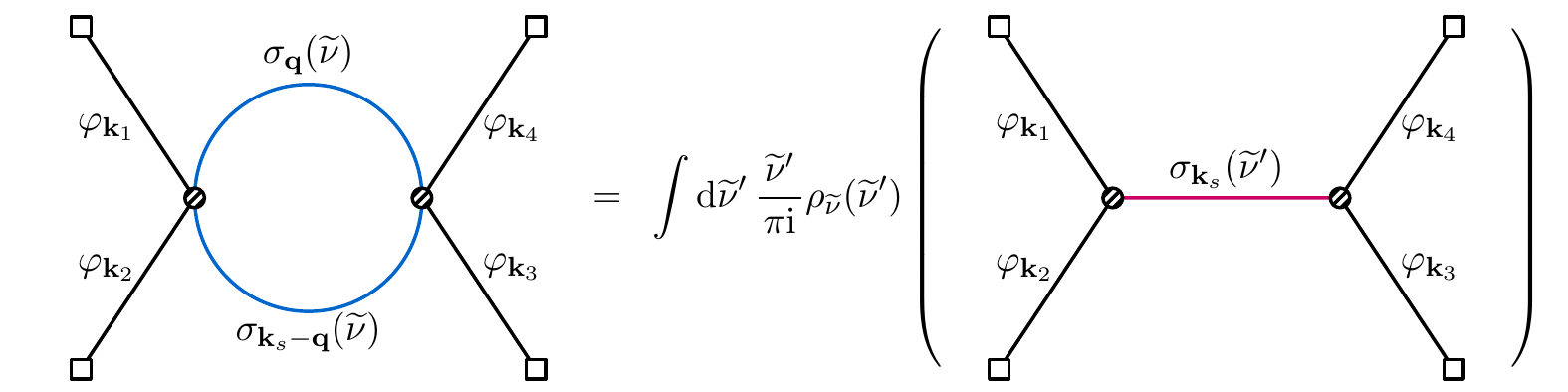} 
  \caption{Computing 1-loop inflation correlator via spectral decomposition. The diagrammatic notation follows \cite{Chen:2017ryl}.}
  \label{fig_LoopSD}
\end{figure}

We choose to work in dS$_{d+1}$ with general $d$ spatial dimensions. This makes it easier to regularize the ultraviolet (UV) divergences. Also, the massive loop correlators in general dS$_{d+1}$ might be of theoretical interest. In dS$_{d+1}$, it is more convenient to use the parameter $\wt\nu\equiv\sqrt{m^2/H^2-d^2/4}$ for scalar fields, instead of the mass $m$. We shall call $\wt\nu$ the \emph{mass parameter}.\footnote{We shall also use term like ``a scalar field of mass $\wt\nu$.'' We hope this does not confuse the readers. Also, we add a tilde for $\wt\nu$ to distinguish it from a more conventionally defined parameter $\nu\equiv\sqrt{d^2/4-m^2/H^2}$.} In this work, we only consider massive scalars that are capable of generating oscillating CC signals. Such fields have mass $m>dH/2$, or equivalently, $\wt\nu>0$. Following the terminology of the representation theory, we call them \emph{principal scalars.} (On the contrary, scalars with $0<m<dH/2$ do not generate oscillatory signals, and are called \emph{complementary scalars}.) \cite{Joung:2006gj,Joung:2007je}

As mentioned, we circumvent the difficulty of loop integral by doing spectral decomposition. This is not a new idea; Rather, it is very close to the Källen-Lehmann representation found in many ordinary quantum field theory textbooks. It has also been used to compute bubble 1-loop correction to 2-point function in dS \cite{Marolf:2010zp}. The essential idea is best explained in the position space, where the 1-loop integral is nothing but a bubble function $B(x,y)=\la\si^2(x)\si^2(y)\ra$. The spectral decomposition suggests that we rewrite the bubble function $B(x,y)$ as a superposition of free scalar propagators $D_{\wt\nu'}(x,y)$ of different masses $\wt\nu'$. Correspondingly, the loop correlator can be written as a \emph{spectral integral} of tree-level correlators mediated by scalars of masses $\wt\nu'$, weighted by a \emph{spectral function} $\rho_{\wt\nu}^\text{dS}(\wt\nu')$. Our modest new observation in this work is that, with the analytical structure of the spectral integrand known, we can finish the spectral integral by closing the contour on the complex $\wt\nu'$-plane and applying the residue theorem, and thereby get the complete analytical result for the loop correlators.  We illustrate this procedure in Fig.\ \ref{fig_LoopSD}.

As we shall see, the loop seed integral $\mathcal{J}_{\wt\nu}(r_1,r_2)$ for the 4-point correlator in Fig.\ \ref{fig_LoopSD} depends on the four external momenta $\mb k_i$ $(i=1,2,3,4)$ only through two momentum ratios $r_1\equiv k_s/(k_1+k_2)$ and $r_2\equiv k_s/(k_3+k_4)$, where $\mb k_s\equiv \mb k_1+\mb k_2$, and $k_i\equiv |\mb k_i|$ $(i=1,\cdots,4,s)$. Furthermore, it breaks into four distinct pieces according to the analytical behavior in the \emph{squeezed limit} $r_{1,2}\to 0$. Schematically, for $r_1<r_2$, we have:\footnote{The result for $r_1>r_2$ is obtained by switching $r_1$ with $r_2$.}
\begin{align}
  \mathcal{J}_{\wt\nu}(r_1,r_2)
  \sim&~ G_\text{NS}(r_1,r_2) (r_1r_2)^{\pm2\ii\wt\nu}+G_\text{LS}(r_1,r_2)\Big(\FR{r_1}{r_2}\Big)^{\pm2\ii\wt\nu} 
  + G_\text{LT}(r_1,r_2)\log r_2 +G_\text{BG}(r_1,r_2).
\end{align}
Here the four terms are called the \emph{nonlocal signal} (NS), the \emph{local signal} (LS), the \emph{logarithmic tail} (LT), and the \emph{background} (BG), respectively. We shall explain the meaning of these terms later. Here we only note that the four functions denoted by $G$ are fully analytic at $r_{1,2}=0$, and all non-analytic behaviors have been explicitly spelled out in each term. The nonlocal and local signals are of the main interest of CC physics, which already appear at the tree level. On the contrary, the logarithmic tail is a special feature of loop correlators, which does not exist for tree-level correlators. However, the logarithmic tail vanishes in $(3+1)$-dimensional dS. The background part of the loop correlators in $(3+1)$-dimensional dS is expected to possess the usual ultraviolet (UV) divergence. We use dimensional regularization to regulate the divergence, and use the familiar modified minimal subtraction ($\overline{\text{MS}}$) as our renormalization condition.  

With the analytical results for the loop seed integral, we can efficiently study the properties of loop correlators. In this work, we consider the 1-loop 4-point and 3-point correlators of the inflaton fluctuations as examples. We shall provide their full analytical results, as well as simple approximations in the squeezed limit and the large mass limit. We show that the CC signals dominate over the background in the \emph{single squeezed limit} of the 4-point correlator, namely $r_1\ll 1$ with $r_2$ fixed. On the contrary, there is no configuration of 3-point correlators where the CC signals are guaranteed to be dominant. Therefore, the single squeezed limit of the 4-point correlators can be a golden channel for discovering CC signals at the 1-loop level. 
 
\paragraph{Outline of this work.} The rest of the paper is organized as follows. In Sec.\ \ref{sec_seedint}, we discuss several examples of inflation correlators with massive 1-loop exchanges. Motivated by these examples, we define the \emph{loop seed integral} which is the central object of this work. We then introduce the idea of spectral decomposition for computing the loop seed integral. 

In Sec.\ \ref{sec_comp}, we compute the loop seed integral by carrying out the spectral integral. We first introduce the two essential ingredients for this computation, namely the spectral function and the tree seed integral in general $d$ spatial dimensions. We then carry out the spectral integral using the residue theorem. The result is summarized and briefly discussed in Sec.\ \ref{subsec_finalres}.

In Sec.\ \ref{sec_check}, we study the properties of the loop seed integral in several limits, including the $d\to 3$ limit where $d$ is the spatial dimension, the large mass limit, the squeezed limit, and the folded limit. The behavior of the loop seed integral in these limits can be either calculated by other means or inferred on physical grounds. Therefore, these limits can be used as consistency checks of our results.

In Sec.\ \ref{sec_CC}, we apply the result of the loop seed integral to compute the 4-point and 3-point inflaton correlators with massive 1-loop exchanges. These processes have direct applications in CC physics. We provide their full analytical results and discuss the squeezed limit and the large-mass limit. The conclusion and outlooks are given in Sec.\ \ref{sec_concl}.

There are six appendices following the main text. Apart from App.\ \ref{App_Formulae} which collects a couple of useful formulae, these appendices contain discussions and results that are essential for our study of 1-loop correlators. We put these materials in the appendix only because of the many technical details involved, which may become distractions had we put them in the main text. 

In App.\ \ref{App_Spectral}, we collect more discussions on the dS spectral function $\rho_{\wt\nu}^\text{dS}(\wt\nu')$. We first reproduce the derivation of the spectral function $\rho_{\wt\nu}^\text{dS}(\wt\nu')$ following the treatment of \cite{Marolf:2010zp}. Then we present the pole structure and residues of the spectral function on the complex $\wt\nu'$ plane. Next, we discuss the $d\to 3$ limit of the spectral function, with a focus on its UV divergence. Finally, we collect discussions about the $\Pi$ function, which is a variation of the spectral functions and appears at several places in the loop seed integral. 

In App.\ \ref{app_asympt}, we study the asymptotic behavior of the spectral function $\rho_{\wt\nu}^\text{dS}(\wt\nu')$ in either the large $\wt\nu$ limit or the large $\wt\nu'$ limit. The large $\wt\nu$ limit enables us to study our result in the flat-space limit, and the large $\wt\nu'$ limit is essential to apply the residue theorem when computing the spectral integral.

In App.\ \ref{App_TreeSeed}, we compute the tree seed integral, which is the basis for our bootstrapping loop correlators. We follow the method of partial Mellin-Barnes representation in \cite{Qin:2022lva,Qin:2022fbv}. In App.\ \ref{app_pMBtoBoot}, we prove the equivalence between the tree seed integral computed from the partial Mellin-Barnes method and the one from solving the bootstrap equation in \cite{Qin:2022fbv}.

Finally, in App.\ \ref{app_mink}, we bootstrap the massive 1-loop correlator in Minkowski spacetime, also using the spectral decomposition. This illustrates our method with a relatively simple setup, and the result obtained here is also useful for our consistency check of loop seed integral in dS.

\paragraph{Notations and conventions.} In this work, the spacetime metric is fixed to be $\di s^2=a^2(\tau)(-\di \tau^2+\di\mb x^2)$ where $\tau\in(-\infty,0)$ is the conformal time, $\mb x$ is the comoving spatial coordinates of $\mathbb{R}^d$ slices, $a(\tau)=-1/(H\tau)$ is the scale factor, and $H$ is the Hubble parameter and is a constant in dS. In most of this work we shall take $H=1$. 

A scalar 4-point correlator as in Fig.\ \ref{fig_LoopSD} is specified by the four external spatial momenta $\mb k_i$ $(i=1,2,3,4)$. Their magnitudes are denoted by $k_i\equiv |\mb k_i|$. The $s$-channel momentum is defined by $\mb k_s\equiv \mb k_1+\mb k_2$. Also, we use shorthand notations such as $k_{ij}=k_i+k_j$ $(i,j=1,2,3,4)$ and $k_{1234}=k_1+k_2+k_3+k_4$. We shall frequently use the momentum ratios $r_1\equiv k_s/k_{12}$ and $r_2\equiv k_s/k_{34}$. Other similar shorthands include $n_{12}=n_1+n_2$, $p_{12}=p_1+p_2$, $\bar p_{12}=p_1-p_2$, etc.

\section{Loop Seed Integral and its Spectral Decomposition}
\label{sec_seedint}

In this section, we motivate and define the \emph{loop seed integral}, which is a double-layer integral over bulk dS time variables, and is the key quantity to be computed in this work. We then introduce the spectral decomposition, which converts the loop seed (time) integral into a spectral integral over the mass parameter. 

\paragraph{Examples of 1-loop correlators.} Let us begin with the 1-loop correlator shown in the left diagram of Fig.\ \ref{fig_LoopSD}. With the field species and interaction types known, it is straightforward to build an expression for this diagram with the standard Schwinger-Keldysh (SK) formalism. See \cite{Chen:2017ryl} for a pedagogical introduction. We also follow the diagrammatic notations of \cite{Chen:2017ryl}. In particular, the four external (black) legs represent the bulk-to-boundary propagators of the nearly massless inflaton field $\varphi$. As in most previous works on this topic, we choose to Fourier-transform the $d$ spatial coordinates $\mb x$ to corresponding momenta $\mb k$, but leave the (conformal) time $\tau$ untransformed. In this representation, and specialized to $d=3$, the bulk-to-boundary propagator $G_\aa(k;\tau)$ of a massless scalar field reads:\footnote{The bulk-to-boundary massless scalar propagator in general $d$ spatial dimensions involves the Hankel function and is considerably more complicated. Fortunately, we will only need the $d=3$ result when the dimensional regularization and the counterterms are properly introduced. See the discussion and the footnote below (\ref{eq_calLinMSbar}). }
\bge
\label{eq_MLBtoB}
  G_\aa(k;\tau)=\FR{1}{2k^3}(1-\ii\aa k\tau)e^{\ii\aa k\tau}.
\ede 
Here $k\equiv|\mb k|$ is the 3-momentum carried by the propagator, $\tau$ is the time variable of the bulk point, and $\aa=\pm$ is the SK index of the bulk point.  On the other hand, the two (blue) loop lines denote the bulk propagators of a real scalar field $\si$ of mass $\wt\nu$. We only consider the case of principal scalars ($\wt\nu>0$) for $\si$ in this work. In this case, the bulk-propagator  $D_{\wt\nu,\aa\bb}(k;\tau_1,\tau_2)$ in general $d$ spatial dimensions reads:
\begin{align}
\label{eq_Dpmpm}
  D_{\wt\nu,\pm\pm}(k;\tau_1,\tau_2)
  =&~D_{\wt\nu,\gtrless}(k;\tau_1,\tau_2)\theta(\tau_1-\tau_2)+D_{\wt\nu,\lessgtr}(k;\tau_1,\tau_2)\theta(\tau_2-\tau_1),\\
\label{eq_Dpmmp}
  D_{\wt\nu,\pm\mp}(k;\tau_1,\tau_2)
  =&~D_{\wt\nu,\lessgtr}(k;\tau_1,\tau_2).
\end{align}
Here we have introduced two ``homogeneous'' propagators, which are related by $D_{\wt\nu,<}(k;\tau_1,\tau_2)=D_{\wt\nu,>}^*(k;\tau_1,\tau_2)$, and,
\begin{align}
\label{eq_Dgreater}
    D_{\wt\nu,>}(k;\tau_1,\tau_2)=\FR{\pi}{4}e^{-\pi\wt\nu}(\tau_1\tau_2)^{d/2}\text{H}_{\ii\wt\nu}^{(1)}(-k\tau_1)\text{H}_{-\ii\wt\nu}^{(2)}(-k\tau_2),
\end{align}
where $\mathrm{H}_\nu^{(1)}(z)$ and $\mathrm{H}_\nu^{(2)}(z)$ are the Hankel functions of the first and second kinds, respectively. 

It remains to specify the interaction vertices in Fig.\ \ref{fig_LoopSD}. Normally, we require the coupling to be invariant under a constant shift of the inflaton field $\varphi$, which is always true when the inflaton is coupled to other fields through its derivatives. Direct couplings are of course possible, as the shift symmetry is only approximate. Our treatment here can be applied to either case, but for definiteness, let us consider the following simple example with derivative coupling:
\bge
\label{eq_DeltaLTDC}
  \Delta\ld = -\FR{1}{4}a^{d-1}\varphi'^2\si^2.
\ede
We omit coupling constants throughout this work, which are trivial to recover. Here and below, a prime denotes conformal time derivative: $\varphi'\equiv \di\varphi/\di\tau$. The factor $a^{d-1}$ is included to ensure that the Lagrangian has the correct scaling. This operator is naturally derived from a Lorentz invariant operator $\sqrt{-g}(\pd_\mu\phi)^2\si^2$ when evaluated with the dS and inflaton background, although in this case, it is always accompanied by a spatial derivative coupling $a^2(\pd_i\varphi)^2\si^2$. It is possible to generate (\ref{eq_DeltaLTDC}) alone by integrating out a heavy degree in the underlying Lorentz invariant theory. See \cite{Qin:2022lva} for more discussions. 

With the couplings given in (\ref{eq_DeltaLTDC}) and all propagators known, it is straightforward to write down the expression for the 1-loop 4-point function in Fig.\ \ref{fig_LoopSD}, following the diagrammatic rule \cite{Chen:2017ryl}. Throughout this work, we shall only consider the $s$-channel exchange unless otherwise stated. The corresponding $t$- and $u$-channel contributions can be obtained by permuting external momenta as usual. Then, we can parameterize the $s$-channel contribution as
\bge
   \la\varphi_{\mb k_1}\varphi_{\mb k_2}\varphi_{\mb k_3}\varphi_{\mb k_4}\ra_{\varphi,s}= (2\pi)^d\de^{(d)}(\mb k_1+\mb k_2+\mb k_3+\mb k_4)\mathcal{L}_{\varphi,\wt\nu}(\mb k_1,\mb k_2,\mb k_3,\mb k_4),
\ede
where the loop amplitude $\mathcal{L}_{\varphi,\wt\nu}$ is:
\begin{align}
\label{eq_calLinf}
  \mathcal{L}_{\varphi,\wt\nu} 
  =&-\FR{1}{2}\sum_{\mathsf{a,b}=\pm}\mathsf{ab}\int_{-\infty}^{\tau_f}\FR{\di\tau_1}{(-\tau_1)^{d-1}}\FR{\di\tau_2}{(-\tau_2)^{d-1}}\n\\
   &\times \pd_{\tau_1}G_{\aa}(k_1,\tau_1)\pd_{\tau_1}G_{\aa}(k_2,\tau_1)\pd_{\tau_2}G_{\bb}(k_3,\tau_2)\pd_{\tau_2}G_{\bb}(k_4,\tau_2)\mathcal{Q}_{\wt\nu,\aa\bb}\big(k_s;\tau_1,\tau_2\big).
\end{align}
Here the pre-factor $1/2$ is a symmetric factor, and $\mathcal{Q}_{\wt\nu,\mathsf{ab}}$ is the loop momentum integral: 
\begin{keyeqn}
\begin{align}
\label{eq_calQ}
  \mathcal{Q}_{\wt\nu,\mathsf{ab}}\big(k_s;\tau_1,\tau_2\big)\equiv\int\FR{\di^d\mb q}{(2\pi)^d}D_{\wt\nu,\mathsf{ab}}\Big(q;\tau_1,\tau_2\Big)D_{\wt\nu,\mathsf{ab}}\Big(|\mb k_s-\mb q|;\tau_1,\tau_2\Big).
\end{align}
\end{keyeqn}

As our second example, let us take the four external states in Fig.\ \ref{fig_LoopSD} to be conformal scalars $\phi_c$. By a conformal scalar, we simply mean a scalar field with mass $m_c^2=(d^2-1)/4$ in dS$_{d+1}$, and we do not make any assumptions about the origin of this mass. The conformal scalar has the nice properties that its ``untilded''mass parameter $\nu=\sqrt{d^2/9- m^2}=1/2$ is independent of the spatial dimension $d$ and that its mode function is particularly simple. We denote the bulk-to-boundary propagator of a conformal scalar by $C_\aa$, and it is given by
\begin{equation}
\label{eq_confBtoB}
    C_\aa(k;\tau)=\FR{(\tau\tau_f)^{(d-1)/2}}{2k}e^{\ii\mathsf{a}k\tau},
\end{equation} 
where we have introduced a final-time cutoff $\tau_f$. To couple the conformal scalar $\phi_c$ with the massive loop field $\si$, we can simply introduce a non-derivative coupling:
\bge
  \Delta\ld =-\FR{1}{4}a^{d+1}\phi_c^2\si^2,
\ede 
where, again, we omit the coupling constant for simplicity. Then, the 4-point correlator of conformal scalars of $s$-channel loop in Fig.\ \ref{fig_LoopSD} can be written as:
\bge
   \la\phi_{\text{c},\mb k_1}\phi_{\text{c},\mb k_2}\phi_{\text{c},\mb k_3}\phi_{\text{c},\mb k_4}\ra_{s,\wt\nu}= (2\pi)^d\de^{(d)}(\mb k_1+\mb k_2+\mb k_3+\mb k_4)\mathcal{L}_{\phi_c,\wt\nu}(\mb k_1,\mb k_2,\mb k_3,\mb k_4),
\ede
where the loop amplitude is:
\begin{align}
\label{eq_LconfFR}
  \mathcal{L}_{\phi_c,\wt\nu} 
  =&-\FR{1}{2}\sum_{\mathsf{a,b}=\pm}\mathsf{ab}\int_{-\infty}^{\tau_f}\FR{\di\tau_1}{(-\tau_1)^{d+1}}\FR{\di\tau_2}{(-\tau_2)^{d+1}} \n\\
   &\times C_{\aa}(k_1,\tau_1)C_{\aa}(k_2,\tau_1)C_{\bb}(k_3,\tau_2)C_{\bb}(k_4,\tau_2)\mathcal{Q}_{\wt\nu,\aa\bb}\big(k_s;\tau_1,\tau_2\big),
\end{align}
with the loop momentum integral $\mathcal{Q}_{\wt\nu,\mathsf{ab}}$ given in (\ref{eq_calQ}).

\begin{figure}
\centering
\includegraphics[width=0.37\textwidth]{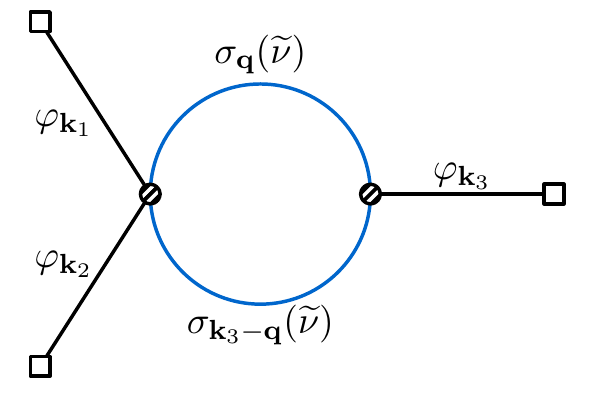}
\caption{The 3-point function of inflaton fluctuations $\varphi$ mediated by a massive scalar $\si$ at 1-loop order.}
\label{fig_1loop3pt}
\end{figure}

As the last example, let us consider a 3-point function of inflaton $\varphi$ mediated by a massive scalar loop, as shown in Fig.\ \ref{fig_1loop3pt}. 
For convenience, we parameterize the 3-point function in the following way:
\bge
   \la\varphi_{\mb k_1}\varphi_{\mb k_2}\varphi_{\mb k_3}\ra_{3,\wt\nu}= (2\pi)^d\de^{(d)}(\mb k_1+\mb k_2+\mb k_3)\mathcal{B}_{\varphi,\wt\nu}(\mb k_1,\mb k_2,\mb k_3).
\ede
Here the label $\la\cdots\ra_{3,\wt\nu}$ means that we only consider the diagram with the total loop momentum being $\mb k_3$, as in Fig.\ \ref{fig_1loop3pt}. The complete result should also include two other diagrams which can be obtained from Fig.\ \ref{fig_1loop3pt} by permutations $\mb k_{3}\leftrightarrow \mb k_1$ and $\mb k_{3}\leftrightarrow \mb k_2$, respectively. 

For the two vertices in Fig.\ \ref{fig_1loop3pt}, we choose the following couplings:
\bge
\label{eq_3ptLag}
  \Delta \ld = -\FR{1}{4}a^{d-1}\varphi'^2\si^2-\FR{1}{2}a^{d}\varphi'\si^2.
\ede
Then, the loop amplitude $\mathcal{B}_{\wt\nu}$ is given by:
\begin{align}
\label{eq_calBinf}
  \mathcal{B}_{\varphi,\wt\nu} 
  =&-\FR{1}{2}\sum_{\mathsf{a,b}=\pm}\mathsf{ab}\int_{-\infty}^{\tau_f}\FR{\di\tau_1}{(-\tau_1)^{d-1}}\FR{\di\tau_2}{(-\tau_2)^{d}}\n\\
   &\times \pd_{\tau_1}G_{\aa}(k_1,\tau_1)\pd_{\tau_1}G_{\aa}(k_2,\tau_1)\pd_{\tau_2}G_{\bb}(k_3,\tau_2) \mathcal{Q}_{\wt\nu,\aa\bb}\big(k_3;\tau_1,\tau_2\big),
\end{align}
where the loop momentum integral $\mathcal{Q}_{\wt\nu,\mathsf{ab}}$ is again given in (\ref{eq_calQ}).

\paragraph{Loop seed integral.} 
One can go on and consider more examples of 1-loop bubble diagrams with various interaction types. However, so long as the external states are massless or conformal, the resulting loop amplitudes all have similar structures. These examples thus motivate us to define a \emph{loop seed integral} $\mathcal{J}_{\wt\nu}^{p_1p_2}$, with the hope that many massive 1-loop correlators can be easily generated from $\mathcal{J}_{\wt\nu}^{p_1p_2}$. The computation of these 1-loop correlators is then reduced to the computation of the loop seed integral. We define the loop seed integral $\mathcal{J}_{\wt\nu}^{p_1p_2}$ in the following way:
\begin{keyeqn}
\begin{align}
\label{eq_LoopSeedInt}
  \mathcal{J}^{p_1p_2}_{\wt\nu}(r_1,r_2)\equiv -\FR{1}{2}\!\sum_{\aa,\bb=\pm}\!\!\aa\bb\, k_s^{d+2+p_{12}}\!\int_{-\infty}^0\!\di\tau_1\di\tau_2(-\tau_1)^{p_1}(-\tau_2)^{p_2}e^{\ii\aa k_{12}\tau_1+\ii \bb k_{34}\tau_2}\mathcal{Q}_{\wt\nu,\aa\bb}\big(k_s;\tau_1,\tau_2\big).
\end{align}
\end{keyeqn}
Several explanations are in order. First, the piece $-\frac{1}{2}\sum\aa\bb$ follows directly from the Feynman rules: The minus sign comes from the two i's for the two vertices, and the factor $1/2$ is the symmetric factor. Second, we have inserted a factor $k_s^{d+2+p_{12}}$ to make the whole integral dimensionless. As a result, the loop seed integral depends on various external momenta only through two ratios $r_1\equiv k_s/k_{12}$ and $r_2\equiv k_s/k_{34}$. Third, the powers $(-\tau_1)^{p_1}(-\tau_2)^{p_2}$ are introduced to account for various types of couplings and various choices of spatial dimensions. Here $p_1$ and $p_2$ are two arbitrary numbers that normally take integer values. Fourth, the factor $e^{\ii\aa k_{12}\tau_1+\ii\bb k_{34}\tau_2}$ comes from the bulk-to-boundary propagators of the conformal scalars; See (\ref{eq_confBtoB}). This factor also appears in massless bulk-to-boundary propagators in $d=3$; See (\ref{eq_MLBtoB}). When $d$ deviates from 3, the massless bulk-to-boundary propagators could develop new terms that contribute to a finite part of the loop amplitude. This part can be subtracted by a proper choice of counterterms, and thus we do not include this part in the definition of the loop seed integral. Finally, $\mathcal{Q}_{\wt\nu,\aa\bb}\big(k_s;\tau_1,\tau_2\big)$ is the loop momentum integral that appears in all examples we are considering, and its explicit formula is given in (\ref{eq_calQ}). 

Now, with the loop seed integral $\mathcal{J}^{p_1p_2}_{\wt\nu}(r_1,r_2)$ known, we can easily write down expressions for many 1-loop correlators. Here we show some examples from previously considered cases. First, let us consider the conformal-scalar correlator, which is the simplest one. Comparing (\ref{eq_LoopSeedInt}), (\ref{eq_LconfFR}), and (\ref{eq_confBtoB}), we have:
\begin{align}
\label{eq_Lconformal}
  \mathcal{L}_{\phi_c,\wt\nu}=\FR{(-\tau_f)^{2(d-1)}}{16k_1k_2k_3k_4k_s^{d-2}}\mathcal{J}_{\wt\nu}^{-2,-2}(r_1,r_2).~~~~~(\text{general }d)
\end{align}
The inflaton correlators can be expressed in terms of the loop seed integral only in $d=3$. By comparing (\ref{eq_LoopSeedInt}), (\ref{eq_calLinf}), and (\ref{eq_MLBtoB}), we have: 
\begin{align}
\label{eq_calLinMSbar}
  \mathcal{L}_{\varphi,\wt\nu}=\FR{1}{16k_1k_2k_3k_4k_s^{5}}\Big[\mathcal{J}_{\wt\nu}^{00}(r_1,r_2)\Big]_{\overline{\text{MS}}} .~~~~~(d=3)
\end{align}
Here the notation $[\cdots]_{\overline{\text{MS}}}$ means that we subtract the divergence of the loop seed integral at $d=3$ by the $\overline{\text{MS}}$ scheme.\footnote{When taking the $d\to 3$ limit, one might worry that the $\order{3-d}$ part of the massless bulk-to-boundary propagator would be combined with the $1/(3-d)$ part of the loop seed integral, and thus would contribute a finite piece to $\mathcal{L}_{\varphi,\wt\nu}$ that was not considered in (\ref{eq_calLinMSbar}). However, we can remove this finite part by introducing the counterterm also in $d$ spatial dimensions. In this work, we always make this ``$d$-dimensional counterterm'' prescription.} Similarly, by comparing (\ref{eq_calBinf}), (\ref{eq_LconfFR}), and (\ref{eq_confBtoB}), we see that the 3-point loop amplitude $\mathcal{B}_{\varphi,\wt\nu}$ in $d=3$ can be written as:
\begin{align}
\label{eq_calBinMSbar}
  \mathcal{B}_{\varphi,\wt\nu}=\FR{1}{8k_1k_2k_3^4}\Big[\mathcal{J}_{\wt\nu}^{0,-2}\Big(\FR{k_3}{k_{12}},1^-\Big)\Big]_{\overline{\text{MS}}},~~~~~(d=3)
\end{align}
where $1^-$ means approaching $1$ from below.

In similar ways, one can consider more general couplings. One potentially important example is the 4-point correlator in Fig.\ \ref{fig_LoopSD} with the following Lorentz covariant coupling in dS$_{3+1}$:
\bge
  \Delta\ld=-\FR{1}{4}a^{2}(\pd_\mu\varphi)^2\si^2,
\ede
in which the Lorentz indices are contracted with the Minkowski metric $\eta_{\mu\nu}$. In terms of the loop seed integral, we can write down the corresponding correlator as:
\bge
\label{eq_calLcov}
  \mathcal{L}_{\varphi,\wt\nu}=\FR{1}{16k_1^3k_2^3k_3^3k_4^3}\mathcal{O}_{12}\mathcal{O}_{34}\Big[\mathcal{J}_{\wt\nu}^{-2,-2}(r_1,r_2)\Big]_{\overline{\text{MS}}},~~~~~(d=3)
\ede
where the differential operator $\mathcal{O}_{ij}$ is defined by
\bge
  \mathcal{O}_{ij}\equiv-k_i^2k_j^2\pd_{k_{ij}}^2-\FR{1}{2}(k_s^2-k_i^2-k_j^2)(1-k_i\pd_{k_{ij}})(1-k_j\pd_{k_{ij}}).
\ede

\paragraph{Spectral decomposition.} It is the presence of the loop momentum integral $\mathcal{Q}_{\wt\nu,\mathsf{ab}}$ that makes the computation of the loop seed integral difficult. Therefore we seek for a spectral function $\rho_{\wt\nu}^\text{dS}(\wt\nu')$ which satisfies the following property:
\begin{align}
\label{eq_QIntRhoD}
  \mathcal{Q}_{\wt\nu,\mathsf{ab}}\big(k_s;\tau_1,\tau_2\big)=\int_{-\infty-\ii\ep}^{+\infty-\ii\ep}\di\wt\nu'\,\FR{\wt\nu'}{\pi\ii}\rho_{\wt\nu}^\text{dS}(\wt\nu')D_{\wt\nu',\aa\bb}\big(k_s;\tau_1,\tau_2\big).
\end{align}
The insertion of the factor $\wt\nu'/(\pi\ii)$ is conventional and can be understood as part of the definition of $\rho_{\wt\nu}^\text{dS}(\wt\nu')$. Also, as explained in App.\ \ref{App_Spectral}, the integral goes over the whole real axis on the complex $\wt\nu'$ plane. The $-\ii\ep$ term in the integral limits means that we bypass any possible poles on the real axis from below. We shall usually neglect this $-\ii\ep$ term when writing the integral.

If we can find a spectral decomposition as in (\ref{eq_QIntRhoD}), then it follows that the loop seed integral $\mathcal{J}^{p_1p_2}_{\wt\nu}(r_1,r_2)$ can be fully expressed as a superposition of tree seed integrals $\mathcal{I}^{p_1p_2}_{\wt\nu'}$ with continuously varying mass parameter $\wt\nu'$. Here by the \emph{tree seed integral}, we mean the following object:
\begin{align}
\label{eq_TreeSeed}
  \mathcal{I}^{p_1p_2}_{\wt\nu}(r_1,r_2)\equiv  -\sum_{\aa,\bb=\pm}\aa\bb k_s^{d+2+p_{12}}\int_{-\infty}^0\di\tau_1\di\tau_2(-\tau_1)^{p_1}(-\tau_2)^{p_2}e^{\ii\aa k_{12}\tau_1+\ii \bb k_{34}\tau_2}D_{\wt\nu,\aa\bb}\big(k_s;\tau_1,\tau_2\big).
\end{align}
This is a direct $d$-dimensional generalization of the scalar seed integral introduced in \cite{Qin:2022fbv}. Now, we further assume that the spectral integral over $\wt\nu'$ commutes with the time integrals. Then, we have:
\begin{keyeqn}
\begin{align}
\label{eq_JIntSpect}
  \mathcal{J}^{p_1p_2}_{\wt\nu}(r_1,r_2)=\int_{-\infty-\ii\ep}^{+\infty-\ii\ep}\di\wt\nu'\,\FR{\wt\nu'}{2\pi\ii}\rho_{\wt\nu}^\text{dS}(\wt\nu')\mathcal{I}^{p_1p_2}_{\wt\nu'}(r_1,r_2).
\end{align}
\end{keyeqn}
Therefore, if we have the explicit results for the spectral function $\rho_{\wt\nu}^\text{dS}(\wt\nu')$ and the tree seed integral $\mathcal{I}^{p_1p_2}_{\wt\nu'}(r_1,r_2)$, we can try to carry out the spectral integral (\ref{eq_JIntSpect}) directly. This will be the topic of the next section.

\section{Computation of the Loop Seed Integral}
\label{sec_comp}

 In this section, we compute the loop seed integral, defined in (\ref{eq_LoopSeedInt}), by carrying out the spectral integral (\ref{eq_JIntSpect}). This is the most technical section of this work. Readers uninterested in technical details can directly go to Sec.\ \ref{subsec_finalres} for the final result.
 
Our strategy is that we insert the explicit expressions for the spectral function $\rho_{\wt\nu}^\text{dS}(\wt\nu')$ and the tree seed integral $\mathcal{I}_{\wt\nu'}^{p_1p_2}$ in to (\ref{eq_JIntSpect}), properly close the integral contour on the complex $\wt\nu'$ plane, and then evaluate the integral with the residue theorem.

\subsection{Ingredients}

To successfully bootstrap the loop seed integral from the tree seed integral, we need explicit analytical expressions for the spectral function and the tree seed integrals. Fortunately, both of them have been worked out in previous works. Here we present the full results as the starting point of our calculation. 

\paragraph{Spectral Function.} First, we need an analytical expression for the spectral function $\rho_{\wt\nu}^\text{dS}(\wt\nu')$ that satisfies the relation (\ref{eq_QIntRhoD}) in dS$_{d+1}$. Such a spectral function can be extracted from the 1-loop bubble diagram $D^2(x,y)$ formed by two massive scalar lines. The 1-loop bubble diagram can be evaluated in either the Euclidean dS$_{d+1}$ or in AdS$_{d+1}$. The spectral function in dS$_{d+1}$ can then be obtained by proper analytical continuations of these results. 

Both the Euclidean dS approach and the AdS approach have been investigated in the literature. In \cite{Marolf:2010zp}, the spectral function was computed in Euclidean dS (EdS). The Euclidean dS$_{d+1}$ is simply the $(d+1)$-dimensional sphere $S^{d+1}$. The spectral function can thus be computed by exploiting various relations of the $(d+1)$-dimensional spherical harmonics. On the other hand, as shown in \cite{DiPietro:2021sjt} and \cite{Sleight:2021plv}, the dS spectral function can be obtained by the analytical continuation of the AdS 1-loop bubble function \cite{Penedones:2010ue,Fitzpatrick:2011hu}. Both approaches lead to spectral functions expressed in terms of a generalized hypergeometric function ${}_7\text{F}_6$ of argument unity, but with a slightly different appearance in their parameters. Owing to the existence of a large number of connection formulae that transform the generalized hypergeometric function of argument unity, we suspect that both results are mathematically equivalent for the parameter domain where both results are well defined. The equivalence of the two results can also be checked by simplified expressions in certain dimensions such as $d=2$, while the equivalence can be conveniently verified numerically for other dimensions. 

It turns out that the result from the EdS approach \cite{Marolf:2010zp} is easier to implement in our computation, in part because the UV divergence of the spectral function in $d=3$ is made explicit by the Euler $\Gamma$ factor. Here we quote the final result and present the details of the derivation in App.\ \ref{App_Spectral}.

\begin{align}
\label{eq_rhodS}
    \rho_{\wt{\nu}}^\text{dS}(\wt{\nu}') 
=&~\frac1{(4\pi)^{(d+1)/2}}\frac{\cos[\pi(\frac d2-\ii\wt{\nu})]}{\sin(-\pi \ii \wt{\nu})}      \Gamma\left[\begin{matrix}
        \frac{3-d}{2}, \frac{d}{2}-\ii\wt\nu \\
         \frac{2-d}{2}-\ii\wt\nu
    \end{matrix}\right]\n \\
    &~\times {}_7\mathcal{F}_6\left[\begin{matrix}
        \fr{2-d}2+\ii \wt{\nu}'-\ii \wt{\nu}, \tfrac{3-d/2+\ii \wt{\nu}'-\ii \wt{\nu}}2,  \fr{2-d}2, \fr{2-d}2-\ii \wt{\nu}, \fr{2-d}2+\ii \wt{\nu}', \tfrac{\ii \wt{\nu}'-2\ii \wt{\nu}+d/2}2, \tfrac{\ii \wt{\nu}'+d/2}2 \\
        \tfrac{1-d/2+\ii \wt{\nu}'-\ii \wt{\nu}}2, 1+\ii \wt{\nu}'-\ii \wt{\nu}, 1+\ii \wt{\nu}', 1-\ii \wt{\nu}, \tfrac{4+\ii \wt{\nu}'-3d/2}2, \tfrac{4+\ii \wt{\nu}'-2\ii \wt{\nu}-3d/2}2
    \end{matrix}\middle|1\right]\n\\
    &~+(\wt\nu\to-\wt\nu).
\end{align}
Here ${}_7\mathcal{F}_{6}$ is a dressed (generalized) hypergeometric function, defined in (\ref{eq_DressedF}) in App.\ \ref{App_Formulae}. When evaluating the spectral integral (\ref{eq_JIntSpect}) using the residue theorem, we shall need the pole structure of the spectral function, which we review in App.\ \ref{App_Spectral} as well.

\paragraph{Tree seed integral.} The tree seed integral (\ref{eq_TreeSeed}) for arbitrary $(p_1,p_2,d)$ is not directly available, although several special cases have been worked out in the literature using various methods. In \cite{Arkani-Hamed:2018kmz}, the integral with $p_1=p_2=-2$ and $d=3$ was computed by solving the bootstrap equations. In \cite{Sleight:2019mgd} a result with $p_1=p_2=-2$ and arbitrary $d$ was obtained by working in the Mellin space. In \cite{Qin:2022fbv}, a result with arbitrary $(p_1,p_2)$ and $d=3$ was computed using the partial Mellin-Barnes representation. We compute the most general case with arbitrary $(p_1,p_2,d)$ in App.\ \ref{App_TreeSeed} using the method of partial Mellin-Barnes representation introduced in \cite{Qin:2022lva}, and here we quote the final result.

In general, the tree seed integral $\mathcal{I}_{\wt\nu}^{p_1p_2}$ defined in (\ref{eq_TreeSeed}) is a function of two independent momentum ratios $r_1=k_s/k_{12}$ and $r_2=k_s/k_{34}$. It is often convenient to break the tree seed integral into three distinct pieces according to their analytic properties at $r_{1,2}=0$:
\begin{align}
\label{eq_Itotal}
     \mathcal{I}^{p_1p_2}_{\wt\nu}(r_1,r_2)=\mathcal{I}_{\text{NL},\wt\nu}^{p_1p_2}(r_1,r_2)+\mathcal{I}_{\text{L},\wt\nu}^{p_1p_2}(r_1,r_2)+\mathcal{I}_{\text{BG},\wt\nu}^{p_1p_2}(r_1,r_2).
\end{align}  
The three terms on the right-hand side correspond to the nonlocal-signal piece (NL), the local-signal piece (L), and the background piece (BG), respectively. When $r_1<r_2$, the explicit expressions for the three pieces are given below. The result with $r_1>r_2$ can be obtained by switching $r_1,p_1\leftrightarrow r_2,p_2$ in the following expressions. 
\begin{align}
    \label{eq_INL}
    &\mathcal{I}_{\text{NL},\wt\nu}^{p_1p_2}(r_1,r_2)
    = \mathcal{C}_{\ii\wt\nu,d}^{p_1p_2}\mb{F}_{\ii\wt\nu,d}^{p_1}(r_1)\mb{F}_{\ii\wt\nu,d}^{p_2}(r_2)(r_1r_2)^{+\ii \wt{\nu}}+\text{c.c.},\\
    \label{eq_IL}
    &\mathcal{I}_{\text{L},\wt\nu}^{p_1p_2}(r_1,r_2)
    = -\mathcal{C}_{\ii\wt\nu,d}^{p_1p_2}\mb{F}_{\ii\wt\nu,d}^{p_1}(r_1)\mb{F}_{-\ii\wt\nu,d}^{p_2}(r_2)\Big(\FR{r_1}{r_2}\Big)^{+\ii \wt{\nu}}+\text{c.c.},\\
    \label{eq_IBG}
    &\mathcal{I}_{\text{BG},\wt\nu}^{p_1p_2} (r_1,r_2)
    =\sum_{\ell,m=0}^\infty\frac{(-1)^{\ell+1}\sin[\frac{\pi}{2}(p_{12}+d)](\ell+1)_{2m+d+p_{12}+1}}{2^{2m+1}\big(\tfrac{\ell-\ii \wt{\nu}+p_2+1}2+\tfrac d4\big)_{m+1}\big(\tfrac{\ell+\ii \wt{\nu}+p_2+1}2+\tfrac d4\big)_{m+1}}r_1^{2m+d+p_{12}+2}\Big(\frac{r_1}{r_2}\Big)^{\ell},
\end{align}
where the coefficient $\mathcal{C}_{\ii\wt\nu,d}^{p_1p_2}$ and the function $\mb{F}_{\ii\wt\nu,d}^{p}(r)$ are defined by\footnote{Note that our definitions of $\mathcal{C}_{\ii\wt\nu,d}^{p_1p_2}$ and $\mb{F}_{\ii\wt\nu,d}^{p}(r)$ are slightly different from the ones given in \cite{Qin:2022fbv}.}:
\begin{align}
\label{eq_calC}
  \mathcal{C}_{\ii\wt\nu,d}^{p_1p_2}
  \equiv &~\FR{1}{8}\csc^2(\pi\ii\wt\nu)\Big\{\cos \FR{\pi \bar p_{12}}2 +\cos\Big[\pi\Big(\ii \wt{\nu}+\FR{p_{12}+d}2\Big)\Big]\Big\},\\
  \label{eq_boldF}
  \mb{F}_{\ii\wt\nu,d}^{p}(r)
  \equiv &~(2r)^{p+d/2+1}\times{}_2\mathcal{F}_1\left[ \begin{matrix}
        \tfrac d4+\tfrac12+\tfrac{p}2+\tfrac{\ii \wt{\nu}}2, \tfrac d4+1+\tfrac{p}2+\tfrac{\ii \wt{\nu}}2 \\
        1+\ii \wt{\nu}
    \end{matrix}\middle|r^2\right],
\end{align}
and we have introduced the shorthands $p_{12}\equiv p_1+p_2$ and $\bar p_{12}\equiv p_1-p_2$. Again, the function $_{2}\mathcal{F}_1$ is a dressed hypergeometric function as defined in (\ref{eq_DressedF}). Therefore, we see that, apart from the unimportant factor $(2r)^{p+d/2+1}$ in $\mb{F}_{\ii\wt\nu,d}^p$, the nonlocal piece behaves like $\mathcal{I}_\text{NL} \sim (r_1r_2)^{\pm\ii\wt\nu}$ in the squeezed limit $r_{1,2}\to 0$, and the local piece behaves like $(r_1/r_2)^{\pm\ii\wt\nu}$. That is, the nonlocal and local signals are nonanalytic in $r_1r_2$ and $r_1/r_2$, respectively. Given that $\wt\nu>0$ for principal scalars ($m>dH/2$), we see that the nonlocal and local pieces give rise to terms of the form $\cos[\wt\nu\log(r_1r_2)+\vartheta_\text{NL}]$ and $\cos[\wt\nu\log(r_1/r_2)+\vartheta_\text{L}]$, and this oscillatory behavior is what we call the signal. On the other hand, the background piece is written as a Taylor series in $r_1$ and $r_1/r_2$ (when $r_1<r_2$), or in $r_2$ and $r_2/r_1$ (when $r_1>r_2$). Therefore we see that the background piece is analytic in both $r_{1}$ and $r_2$ when $r_{1,2}\to 0$.

As shown in App.\ \ref{App_TreeSeed}, the background piece $\mathcal{I}_{\text{BG},\wt\nu}^{p_1p_2}$ obtained by the partial Mellin-Barnes representation has a form in which the $r_1/r_2$ series is resummed into a hypergeometric function. See (\ref{eq_IBGpMB}). However, for our analysis of ultraviolet divergence in the 1-loop process, it turns out useful to have an expression with $r_1/r_2$ series fully expanded, as shown in (\ref{eq_IBG}). Such a series can be more directly obtained by solving the inhomogeneous bootstrap equation for general $(p_1,p_2,d)$, as was done for special cases in \cite{Arkani-Hamed:2018kmz} and \cite{Qin:2022fbv}. However, it is possible to derive (\ref{eq_IBG}) by directly expanding the hypergeometric function in the partial Mellin-Barnes result (\ref{eq_IBGpMB}). This would be a direct proof of the equivalence between the bootstrapped series (\ref{eq_IBG}) and the partial Mellin-Barnes series (\ref{eq_IBGpMB}). We give the details of this proof in App.\ \ref{app_pMBtoBoot}.

\paragraph{Strategy.} With the explicit expressions for the spectral function and the tree seed integral at hand, we are now ready to perform the spectral integral (\ref{eq_JIntSpect}). Since the tree seed integral is broken into three pieces, we will compute the spectral integral for the three pieces separately. Thus we define the following three integrals:
\begin{align}
\label{eq_Jint1}
  \mathcal{J}_{\text{(1)}} 
  \equiv&\int_{-\infty}^{+\infty}\di\wt\nu'\,\FR{\wt\nu'}{2\pi\ii}\rho_{\wt\nu}^\text{dS}(\wt\nu')\mathcal{I}^{p_1p_2}_{\text{NL},\wt\nu'}(r_1,r_2), \\
\label{eq_Jint2}
  \mathcal{J}_{\text{(2)}} 
  \equiv&\int_{-\infty}^{+\infty}\di\wt\nu'\,\FR{\wt\nu'}{2\pi\ii}\rho_{\wt\nu}^\text{dS}(\wt\nu')\mathcal{I}^{p_1p_2}_{\text{L},\wt\nu'}(r_1,r_2), \\
\label{eq_Jint3}
\mathcal{J}_{\text{(3)}} 
  \equiv&\int_{-\infty}^{+\infty}\di\wt\nu'\,\FR{\wt\nu'}{2\pi\ii}\rho_{\wt\nu}^\text{dS}(\wt\nu')\mathcal{I}^{p_1p_2}_{\text{BG},\wt\nu'}(r_1,r_2).
\end{align}
Here we have suppressed all indies for $\mathcal{J}$ integrals to avoid unnecessary complications of notations. Also, we use ($\mathcal{J}_{(1)},\mathcal{J}_{(2)},\mathcal{J}_{(3)}$) instead of the more obvious choice ($\mathcal{J}_\text{NL},\mathcal{J}_\text{L},\mathcal{J}_\text{BG}$), because the analytic properties of these integrals are unclear for the moment. We devote the next subsection to the computations of these three integrals. 

\subsection{Contributions from the nonlocal tree integral}

In this subsection, we compute the integral $\mathcal{J}_{\text{(1)}}$ in (\ref{eq_Jint1}). Note that all the momentum dependence comes from the nonlocal tree seed integral, which is the sum of two terms as shown in (\ref{eq_INL}). The term proportional to $(r_1r_2)^{+\ii\wt\nu'}$ is explicitly spelled out in (\ref{eq_INL}), while the term proportional to $(r_1r_2)^{-\ii\wt\nu'}$ is contained in the complex conjugate, namely ``c.c.'' in (\ref{eq_INL}). For physical configurations, we have $0<r_1,r_2<1$ and thus $0<r_1r_2<1$. Thus, for the term proportional to  $(r_1r_2)^{+\ii\wt\nu'}$, we should close the integral contour with a large semi-circle in the lower-half $\wt\nu'$-plane. On the contrary, for the term proportional to $(r_1r_2)^{-\ii\wt\nu'}$, we should close the contour with a large semi-circle in the upper-half $\wt\nu'$-plane. It is thus necessary to treat $(r_1r_2)^{+\ii\wt\nu'}$ term and $(r_1r_2)^{-\ii\wt\nu'}$ term separately. Below, we list all the poles on the lower-half $\wt\nu'$-plane in the integrand of (\ref{eq_Jint1}) involving the $(r_1r_2)^{+\ii\wt\nu'}$ term, and all the poles on the upper-half $\wt\nu'$-plane in the integrand of (\ref{eq_Jint1}) involving the $(r_1r_2)^{-\ii\wt\nu'}$ term. These poles can be conveniently classified into the following three sets: 
\begin{align}
  &\text{Poles}   
  &&(r_1r_2)^{+\ii\wt\nu'}~\text{term} 
  &&(r_1r_2)^{-\ii\wt\nu'}~\text{term} \n\\
  \label{eq_Set1A}
  &\text{Set 1A:} &&\text{-}  &&\wt{\nu}'=\ii d/2\pm2\wt{\nu}+2\ii n, \\
  \label{eq_Set1B}
  &\text{Set 1B:} &&\text{-}  &&\wt{\nu}'=\ii d/2+2\ii n,\\
  \label{eq_Set1C}
  &\text{Set 1C:} &&\wt{\nu}'=-\ii n,~~(n\neq0)  &&\wt{\nu}'=+\ii n.
\end{align}
In all these expressions $n$ goes over all nonnegative integers, except for the Set 1C poles $\wt\nu'=-\ii n$ for the $(r_1r_2)^{+\ii\wt\nu'}$ term, in which case the pole at $n=0$ is outside the integral contour, due to the ``$-\ii\ep$'' prescription of the integral contour in (\ref{eq_QIntRhoD}). Poles in Set 1A and Set 1B are from the spectral density $\rho_{\wt\nu}^\text{dS}(\wt\nu')$, and poles in Set 1C are from the factor $\csc^2(\pi\ii\wt\nu')$ in the non-local part of the tree seed integral $\mathcal{I}_{\text{NL},\wt\nu'}^{p_1p_2}$. See (\ref{eq_INL}) and (\ref{eq_calC}). There are also poles from the dressed hypergeometric functions in (\ref{eq_INL}), as shown in (\ref{eq_boldF}), but these poles are not inside the integral contour, and thus do not contribute to the integral.

\paragraph{Set 1A: nonlocal Signal.} The poles in Set 1A are all simple poles, coming from an Euler $\Gamma$ factor in the spectral function. At these poles, the residue of the spectral function is given in (\ref{eq_rhoPoleA}). Only the $(r_1r_2)^{-\ii\wt\nu}$ term in (\ref{eq_INL}) makes nonzero contributions since the poles are in the upper-half $\wt\nu'$ plane. The summation of residues at these poles is straightforward, and the result is:
\begin{align}
\label{eq_J1A}
  \mathcal{J}_{\text{(1A)}} 
  =&-\frac{(r_1r_2)^{d/2\pm2\ii\wt{\nu}}\sin[\pi(\fr{d}{2}+2\ii \wt{\nu})]}{8\pi^{d/2}\Gamma\big(\tfrac d2\big)\sin^2(\pi\ii\wt{\nu})}\sum_{n=0}^\infty\FR{(1+n)_{\frac d2-1} \big[(1+\ii\wt\nu+n)_{\frac d2-1}\big]^2(1+2\ii\wt\nu+n)_{\frac d2-1}}{(1+2\ii\wt\nu+2n)_{d-1}}\n\\
  &~\times(\fr{d}{2}+2\ii\wt{\nu}+2n)\mathcal{C}_{2\ii\wt{\nu}+d/2+2n,d}^{p_1p_2}\mb{F}_{2\ii\wt{\nu}+d/2+2n,d}^{p_1}(r_1)\mb{F}_{2\ii\wt{\nu}+d/2+2n,d}^{p_2}(r_2)(r_1r_2)^{2n}+\text{c.c.}.
\end{align}
The poles of Set 1A give rise to terms proportional to $(r_1r_2)^{\pm 2\ii\wt\nu}$. In the terminology of CC physics, they correspond to the nonlocal signal of the 1-loop process.

\paragraph{Set 1B: background.} Similarly, the poles in Set 1B are all simple poles, coming from an Euler $\Gamma$ factor in the spectral function. At these poles, the residue of the spectral function is given in (\ref{eq_rhoPoleB}). Again, only the $(r_1r_2)^{-\ii\wt\nu}$ term in (\ref{eq_INL}) makes nonzero contributions, since the poles are in the upper-half $\wt\nu'$ plane. The result of summing the residues of these poles is: 
\begin{align}
\label{eq_J1B}
    \mathcal{J}_{\text{(1B)}}=&~\frac{(r_1r_2)^{d/2}\sin\fr{\pi d}{2}}{4\pi^{d/2}\Gamma\big(\tfrac d2\big)\sin^2(\pi\ii\wt{\nu})}\sum_{n=0}^\infty\FR{\big[(1+n)_{\frac d2-1}\big]^2 (1+\ii\wt\nu+n)_{\frac d2-1} (1-\ii\wt\nu+n)_{\frac d2-1}}{(1+2n)_{d-1}}\n\\
    &\times(\fr{d}{2}+2n)\mathcal{C}_{d/2+2n,d}^{p_1p_2}\mb{F}_{d/2+2n,d}^{p_1}(r_1)\mb{F}_{d/2+2n,d}^{p_2}(r_2)(r_1r_2)^{2n}. 
\end{align}
We see that this result is analytic in $r_1$ and $r_2$ as $r_{1,2}\to 0$, apart from the unimportant prefactor $(r_1r_2)^{d/2}$ and similar factors in $\mb{F}$ functions. Therefore, this result belongs to the background piece of the loop seed integral. 
 
\paragraph{Set 1C: background and nonlocal logarithmic tail.} 
The poles in Set 1C are from the factor $\csc^2(\pi\ii\wt\nu')$ in $\mathcal{I}_{\text{NL},\wt\nu}^{p_1p_2}$, as is clear from (\ref{eq_INL}) and (\ref{eq_calC}). The pole at $\wt{\nu}'=0$ is present only for the $(r_1r_2)^{-\ii\wt\nu'}$ term, since our prescription of the integral contour requires that we bypass any poles on the real axis from the negative imaginary direction. Therefore, the $\wt\nu'=0$ pole is outside the integral contour for the $(r_1r_2)^{+\ii\wt\nu'}$ term, which is entirely in the lower-half plane. Also, the $\wt\nu'=0$ pole is a simple pole, since there is a factor of $\wt\nu'$ in the integrand of (\ref{eq_Jint1}). On the other hand, the poles at $\wt\nu'=\pm\ii n$ with $n\neq 0$ are of second order and are present for both of $(r_1r_2)^{\pm\ii\wt\nu'}$ terms. So, we treat $n=0$ and $n\neq 0$ cases separately. The contribution of $\wt{\nu}'=0$ is
\begin{align}
\label{eq_J1C0}
    \mathcal{J}_{(\text{1C0})}=&-\frac{\rho_{\wt{\nu}}^\text{dS}(0)}{8\pi^2}\Big\{\cos\frac{\pi\bar{p}_{12}}2+\cos\Big[\pi\Big(\frac{p_{12}+d}2\Big)\Big]\Big\}\mb{F}_{0,d}^{p_1}(r_1)\mb{F}_{0,d}^{p_2}(r_2).
\end{align}
As we shall see, this result will be canceled by a similar term from the integration of the local tree seed integral. 

Then we consider the second-order poles at $\wt\nu'=\pm\ii n$ with $n=1,2,\cdots$ for the $(r_1r_2)^{\mp\ii\wt\nu'}$ term. The residues at these poles are a little more complicated, as they necessarily involve the derivatives of the integrand (with the pole-generating function $\csc^2$ removed) with respect to $\wt\nu'$. Combining contributions from both of $(r_1r_2)^{\mp\ii\wt\nu'}$ terms, the result is:
\begin{align}
\label{eq_J1C}
    \mathcal{J}_{(\text{1C})}=&~ \frac{1}{8\pi^2}\sum_{n=1}^{\infty}\mathcal{Z}_{n,d}^{p_1p_2}\Big\{\Pi_{n,d}(\wt\nu)
    \Big[\log(r_1r_2) +\mb{G}_{n,d}^{p_1}(r_1)+ \mb{G}_{n,d}^{p_2}(r_2)-\mathcal{Y}_{n,d}^{p_1p_2}+\fr{1}{n}\Big]- \Xi_{n,d}(\wt\nu)\Big\} \n\\
     &\times \mb{F}_{n,d}^{p_1}(r_1)\mb{F}_{n,d}^{p_2}(r_2)(r_1r_2)^n .
\end{align}
Here we have defined the following functions:
\begin{align}
\label{eq_boldG}
  \mb{G}_{\mu,d}^p(r)\equiv&~\FR{\di}{\di\mu}\log\mb F_{\mu,d}^p(r), \\
\label{eq_Pi}
  \Pi_{n,d}(\wt\nu)\equiv&~\rho_{\wt\nu}^\text{dS}(-\ii n)-\rho_{\wt\nu}^\text{dS}(+\ii n),\\
\label{eq_Xi}
  \Xi_{n,d}(\wt\nu)\equiv&~ \ii \bigg[\FR{\di \rho_{\wt\nu}^\text{dS}(\wt\nu')}{\di\wt\nu'}\bigg|_{\wt\nu'=-\ii n}+\FR{\di \rho_{\wt\nu}^\text{dS}(\wt\nu')}{\di\wt\nu'}\bigg|_{\wt\nu'=+\ii n}\bigg],
\end{align}
as well as the following coefficients: 
\begin{align}
\label{eq_calZ}
   \mathcal{Z}_{n,d}^{p_1p_2}\equiv&~ n\bigg[\cos\frac{\pi\bar{p}_{12}}2+(-1)^n\cos\FR{\pi(p_{12}+d)}2\bigg], \\ 
\label{eq_calY}
   \mathcal{Y}_{n,d}^{p_1p_2}\equiv &~ \FR{n(-1)^n \pi\sin[\fr{\pi}{2}(p_{12}+d)]}{\mathcal{Z}_{n,d}^{p_1p_2}}.
\end{align}
As we show in App.\ \ref{app_Pi}, the function $\Pi_{n,d}(\wt\nu)$ is free from the UV divergence for any $d$. In several even spatial dimensions such as $d=2$ and $d=4$, $\Pi_{n,d}(\wt\nu)$ is an exponentially small function which scales as $e^{-2\pi\wt\nu}$. In  $d=3$ which is our main interest, the function $\Pi_{n,d}(\wt\nu)$ vanishes when $n$ is an integer. On the other hand, the function $\Xi_{n,d}(\wt\nu)$ in (\ref{eq_Xi}) does not vanish in $d=3$ for integer $n$, but it is still an exponentially small quantity, and drops out in the final result for the loop seed integral. The explicit expression for $\Xi_{n,3}(\wt\nu)$ with integer $n$ is given in (\ref{eq_Xin3}).

In $\mathcal{J}_{(\text{1C})}$, both $\mb{F}(r)$ and $\mb{G}(r)$ are regular as $r\to 0$. Therefore, no ``signals'' are present in $\mathcal{J}_{(\text{1C})}$. However, there is a notable logarithmic tail $\propto \log(r_1r_2)$ that is present is $\mathcal{J}_{(\text{1C})}$, which has no counterpart in the tree seed integral. This logarithmic tail is absent in $d=3$ due to the vanishing of $\Pi_{n,3}$ for $n\in\mathbb{Z}$.

\subsection{Contributions from the local tree integral}

Next, we consider the spectral integral $\mathcal{J}_{(2)}$ in (\ref{eq_Jint2}) with the local tree seed integral, which is quite similar to the computation of $\mathcal{J}_{(1)}$ in the last subsection. The local tree seed integral $\mathcal{I}_{\text{L},\wt\nu}^{p_1p_2}$ also consists of two terms, one proportional to $(r_1/r_2)^{+\ii\wt\nu'}$ and the other proportional to $(r_1/r_2)^{-\ii\wt\nu'}$. Here we concentrate on the case with $0<r_1<r_2<1$. So, we close the contour from the lower half $\wt{\nu}'$-plane for the $(r_1/r_2)^{+\ii\wt\nu'}$ term which is explicitly displayed in (\ref{eq_IL}), and we should close the contour from the upper-half $\wt\nu'$-plane for the $(r_1/r_2)^{-\ii\wt\nu'}$ which is contained in the ``c.c.'' term in (\ref{eq_IL}). The relevant poles of the integrand of (\ref{eq_Jint2}) can be classified into the following four sets:
\begin{align}
  &\text{Poles}   
  &&(r_1/r_2)^{+\ii\wt\nu'}~\text{term} 
  &&(r_1/r_2)^{-\ii\wt\nu'}~\text{term} \n\\
  \label{eq_Set2A}
  &\text{Set 2A:} &&\text{-}  &&\wt{\nu}'=\ii d/2\pm2\wt{\nu}+2\ii n, \\
  \label{eq_Set2B}
  &\text{Set 2B:} &&\text{-}  &&\wt{\nu}'=\ii d/2+2\ii n,\\
  \label{eq_Set2C}
  &\text{Set 2C:} &&\wt{\nu}'=-\ii n,~~(n\neq0)  &&\wt{\nu}'=+\ii n,\\
  \label{eq_Set2D}
  &\text{Set 2D:} &&\wt\nu'=-\ii(d/2+p_2+1+n), &&\wt\nu'=\ii(d/2+p_2+1+n).
\end{align}
Again, in all these expressions $n$ goes over all nonnegative integers, except in Set 2C for $(r_1/r_2)^{\ii\wt\nu'}$, where $n=0$ should be excluded. Similar to the previous subsection, the poles in Set 2A and Set 2B are from the spectral function, and the poles in Set 2C are from the $\csc^2(\pi\ii\wt\nu')$ function in the local tree seed integral through the $\mathcal{C}_{\ii\wt\nu',d}^{p_1p_2}$ factor. However, a new set of poles emerge, marked as Set 2D, from the dressed hypergeometric function in the $\mb{F}$ function in (\ref{eq_IL}), which is absent in $\mathcal{J}_{(1)}$. The computation of $\mathcal{J}_{(2)}$ is thus very similar to that of $\mathcal{J}_{(1)}$ in the previous subsection, and we present the result below.

\paragraph{Set 2A: local Signal.}  The residues at poles in Set 2A give rise to the 1-loop local signal. Explicitly, 
\begin{align}
\label{eq_J2A}
  \mathcal{J}_{\text{(2A)}} 
  =&~\frac{(r_1/r_2)^{d/2+2\ii\wt{\nu}}\sin[\pi(\fr{d}{2}+2\ii \wt{\nu})]}{8\pi^{d/2}\Gamma\big(\tfrac d2\big)\sin^2(\pi\ii\wt{\nu})}\sum_{n=0}^\infty\FR{(1+n)_{\frac d2-1} \big[(1+\ii\wt\nu+n)_{\frac d2-1}\big]^2(1+2\ii\wt\nu+n)_{\frac d2-1}}{(1+2\ii\wt\nu+2n)_{d-1}}\n\\
  &\times(\fr{d}{2}+2\ii\wt{\nu}+2n)\mathcal{C}_{2\ii\wt{\nu}+d/2+2n,d}^{p_1p_2}\mb{F}_{2\ii\wt{\nu}+d/2+2n,d}^{p_1}(r_1)\mb{F}_{-2\ii\wt{\nu}-d/2-2n,d}^{p_2}(r_2)(r_1/r_2)^{2n}+\text{c.c.}.
\end{align}

\paragraph{Set 2B: background.} The contribution of the second group of poles is
\begin{align}
\label{eq_J2B}
  \mathcal{J}_\text{(2B)}=&-\frac{(r_1/r_2)^{d/2}\sin\fr{\pi d}{2}}{4\pi^{d/2}\Gamma\big(\tfrac d2\big)\sin^2(\pi\ii\wt{\nu})}\sum_{n=0}^\infty\FR{\big[(1+n)_{\frac d2-1}\big]^{2} (1+\ii\wt\nu+n)_{\frac d2-1} (1-\ii\wt\nu+n)_{\frac d2-1}}{(1+2n)_{d-1}}\n\\
  &\times(\fr{d}{2}+2n)\mathcal{C}_{d/2+2n,d}^{p_1p_2}\mb{F}_{d/2+2n,d}^{p_1}(r_1)\mb{F}_{-d/2-2n,d}^{p_2}(r_2)(r_1/r_2)^{2n} .
\end{align}

\paragraph{Set 2C: background and local logarithmic tail.}  The poles of Set 2C are from the $\csc^2(\pi\ii\wt\nu')$ factor in $\mathcal{I}_{\text{L},\wt\nu}^{p_1p_2}$. Similar to the poles of Set 1C in the last subsection, the $\wt\nu'=0$ pole is a simple pole, together with the $(r_1/r_2)^{-\ii\wt\nu'}$ term, it contributes the following result:
\begin{align}
  \mathcal{J}_{(\text{2C}0)}=&~\frac{\rho_{\wt\nu}^\text{dS}(0)}{8\pi^2}\Big\{\cos\frac{\pi\bar{p}_{12}}2+\cos\Big[\pi\Big(\frac{p_{12}+d}2\Big)\Big]\Big\}\mb{F}_{0,d}^{p_1}(r_1)\mb{F}_{0,d}^{p_2}(r_2).
\end{align}
As has been mentioned in the last subsection, $\mathcal{J}_\text{(2C0)}$ is canceled by $\mathcal{J}_\text{(1C0)}$ in (\ref{eq_J1C0}), so these two terms do not appear in the final result. On the other hand, the poles at $\wt\nu'=\pm\ii n$ with $n\neq0$ are second-order poles. Their contributions to the integral can be found to be: 
\begin{align}
\label{eq_J2C}
    \mathcal{J}_{(\text{2C})}=&-\frac{1}{8\pi^2}\sum_{n=1}^{\infty}\mathcal{Z}_{n,d}^{p_1p_2}\Big\{\Pi_{n,d}(\wt\nu)
    \Big[\log \FR{r_1}{r_2} +\mb{G}_{n,d}^{p_1}(r_1)-\mb{G}_{-n,d}^{p_2}(r_2) -\mathcal{Y}_{n,d}^{p_1p_2}+\fr{1}{n}\Big]-\Xi_{n,d}(\wt\nu)\Big\}\n\\
    &\times \mb{F}_{n,d}^{p_1}(r_1)\mb{F}_{-n,d}^{p_2}(r_2)\Big(\FR{r_1}{r_2}\Big)^n .
\end{align} 
The various quantities in this expression have been defined in (\ref{eq_boldG})-(\ref{eq_calY}). Similar to the result of Sec 1C poles, we find a term $\propto \log(r_1/r_2)$ which is nonanalytic in $r_1/r_2$ when $r_1/r_2\to 0$. We call it the local logarithmic tail, as it is from the spectral integral of the local scalar seed integral. Again, all terms in $\mathcal{J}_{(\text{2C})}$ except the one proportional to $\Xi_{n,d}(\wt\nu)$ vanish when $d=3$.

\paragraph{Set 2D: background.}  Finally, there are poles arising from the Euler $\Gamma$ factors in the dressed hypergeometric function in $\mb{F}_{\mp\ii\wt{\nu}',d}^{p_2}(r_2)$. See (\ref{eq_IL}). These are simple poles that have no counterparts in the integrand of $\mathcal{J}_{(1)}$. The contribution from these poles is:
\begin{align}
\label{eq_J2D}
  \mathcal{J}_\text{(2D)}=&~ (2r_1)^{d/2+1+p_2}\sum_{n=0}^\infty\frac{(-1)^n2^{1+n}\sqrt{\pi}\big(\frac d2+p_2+1+n\big)}{n!}\Pi_{d/2+p_2+1+n,d}(\wt\nu)\n\\
  &\times\mathcal{C}_{d/2+p_2+1+n,d}^{p_1p_2}\mathbf{F}_{d/2+p_2+1+n,d}^{p_1}(r_1){\,}_2\wt{\mathrm{F}}_1\left[\begin{matrix}
    -\tfrac n2, \tfrac{1-n}2 \\
    -\tfrac d2-p_2-n
  \end{matrix}\middle|r_2^2\right]\Big(\frac{r_1}{r_2}\Big)^n.
\end{align}

\subsection{Contributions from the background tree integral}

Finally, we consider the integral (\ref{eq_Jint3}). Unlike all the previous integrals, in (\ref{eq_Jint3}), the powers of $r_1$ and $r_2$ are independent of the integral variable $\wt\nu'$, and thus we cannot use the power of $r_{1,2}$ to decide on which side of the complex $\wt\nu'$-plane to close the integral contour. On the other hand, as we show in App.\ \ref{app_largenup}, the spectral function $\rho_{\wt\nu}^\text{dS}(\wt\nu')$ behaves for large $\wt\nu'$ like $|\wt\nu'|^{d-3}$, while the background tree seed integral $\mathcal{I}_{\text{BG},\wt\nu'}^{p_1p_2}$ behaves like $ |\wt\nu'|^{-2}$. So, the integrand of (\ref{eq_Jint3}) dies away for large $\wt\nu'$ like $|\wt\nu'|^{d-4}$, and we can close the contour from either the upper-half or lower-half plane, so long as $d<3$. 

It turns out simpler to close the contour from the lower-half $\wt\nu'$ plane, where the poles are entirely from the background tree seed integral (\ref{eq_IBG}) through the Pochhammer symbol $(\tfrac{\ell-\ii \wt{\nu}'+p_2+1}2+\tfrac d4)_{m+1}$ in the denominator. We assume that the spectral integral commutes with the summations over $m$ and $\ell$. Then, for each fixed $m$ and $\ell$, the poles of integrand of $\mathcal{J}_{(3)}$ are:
\begin{align}
  \label{eq_Set3}
  &\text{Set 3:} &&\wt{\nu}'=-2\ii n-\ii\big(\fr{1}{2}d+p_2+1+\ell\big), ~~~(n=0,1,\cdots, m)
\end{align} 
Summing the residues at these poles, we get the following result: 
\begin{align}
\label{eq_J3}
    \mathcal{J}_{\text{(3)}} 
    =& \sum_{\ell,m=0}^\infty\sum_{n=0}^{m}\frac{(-1)^{\ell+n}\sin[\frac{\pi}{2}(p_{12}+d)](\ell+1)_{2m+d+p_1+p_2+1}}{2^{2m}n!(m-n)!\big(p_2+1+\fr{d}{2}+\ell+n\big)_{m+1}}\big(\fr{1}{2}d+p_2+1+\ell+2n\big)\n\\
     &\times \rho_{\wt\nu}^\text{dS}(-\fr{\ii d}{2}-\ii p_2-\ii-\ii\ell-2\ii n) r_1^{2m+d+p_{12}+2}\Big(\frac{r_1}{r_2}\Big)^{\ell}.
\end{align}

\subsection{Final result}
\label{subsec_finalres}

At this point we have finished the computation of the loop seed integral by working out all three contributions in (\ref{eq_Jint1}), (\ref{eq_Jint2}), and (\ref{eq_Jint3}). The final result for the loop seed integral $\mathcal{J}_{\wt\nu}^{p_1p_2}(r_1,r_2)$ in (\ref{eq_LoopSeedInt}) can thus be found by summing up all contributions obtained from the previous three subsections, including (\ref{eq_J1A}), (\ref{eq_J1B}), (\ref{eq_J1C}), (\ref{eq_J2A}), (\ref{eq_J2B}), (\ref{eq_J2C}), (\ref{eq_J2D}), and (\ref{eq_J3}). We find it helpful to regroup the many terms in the final result in terms of their analytic properties in the squeezed limit $r_{1,2}\to 0$. From the explicit results in the previous three subsections, we can identify four types of terms with distinct analytic properties in the squeezed limit, including a \emph{nonlocal signal} piece $\mathcal{J}_\text{NS}^{p_1p_2}(r_1,r_2)$, which is proportional to $(r_1r_2)^{\pm2\ii\wt\nu}$ in the squeezed limit; a \emph{local signal} piece $\mathcal{J}_\text{LS}^{p_1p_2}(r_1,r_2)$, proportional to $(r_1/r_2)^{\pm2\ii\wt\nu}$; a \emph{logarithmic tail} piece $\mathcal{J}_\text{LT}^{p_1p_2}(r_1,r_2)$, proportional to $\log r_2$; and, finally, a \emph{background} piece $\mathcal{J}_\text{BG}^{p_1p_2}(r_1,r_2)$, which is analytic in the squeezed limit. In summary, we have:
\begin{keyeqn}
\begin{align}
\label{eq_Jresult}
  \mathcal{J}_{\wt\nu}^{p_1p_2}(r_1,r_2)=\mathcal{J}_\text{NS}^{p_1p_2}(r_1,r_2)+\mathcal{J}_\text{LS}^{p_1p_2}(r_1,r_2)+\mathcal{J}_\text{LT}^{p_1p_2}(r_1,r_2)+\mathcal{J}_\text{BG}^{p_1p_2}(r_1,r_2).
\end{align}\vspace{-6.5mm}
\end{keyeqn}
Below we give the explicit expressions for the four pieces defined here. Similar to the expressions for the tree seed integral, the expressions below apply to the case of $r_1<r_2$. The result for $r_1>r_2$ can be found from the following expressions by switching the variables $r_1,p_1\leftrightarrow r_2,p_2$.

\paragraph{Nonlocal signal.} The nonlocal signal $\mathcal{J}_\text{NS}^{p_1p_2}(r_1,r_2)$ comes totally from the spectral integral (\ref{eq_Jint1}) over the nonlocal tree seed integral through the Set-1A poles. See (\ref{eq_Set1A}) and (\ref{eq_J1A}). The nonlocal signal features a pair of terms proportional to $(r_1r_2)^{\pm 2\ii\wt\nu}$ in the squeezed limit $r_{1,2}\to 0$. The full expression is:
\begin{align}
\label{eq_JNS}
  \mathcal{J}_{\text{NS}}^{p_1p_2}
  =&-\frac{(r_1r_2)^{d/2+2\ii\wt{\nu}}\sin[\pi(\fr{d}{2}+2\ii \wt{\nu})]}{8\pi^{d/2}\Gamma\big(\tfrac d2\big)\sin^2(\pi\ii\wt{\nu})}\sum_{n=0}^\infty\FR{(1+n)_{\frac d2-1} \big[(1+\ii\wt\nu+n)_{\frac d2-1}\big]^2(1+2\ii\wt\nu+n)_{\frac d2-1}}{(1+2\ii\wt\nu+2n)_{d-1}}\n\\
  &~\times(\fr{d}{2}+2\ii\wt{\nu}+2n)\mathcal{C}_{2\ii\wt{\nu}+d/2+2n,d}^{p_1p_2}\mb{F}_{2\ii\wt{\nu}+d/2+2n,d}^{p_1}(r_1)\mb{F}_{2\ii\wt{\nu}+d/2+2n,d}^{p_2}(r_2)(r_1r_2)^{2n}+\text{c.c.}.
\end{align}
Here the coefficient $\mathcal{C}$ is defined in (\ref{eq_calC}) and the function $\mb{F}$ is defined in (\ref{eq_boldF}). For intermediate scalars with $m>dH/2$, we have $\wt\nu>0$. So, in the squeezed limit, the nonlocal signal piece gives rise to oscillatory functions of $r_1r_2$ in the form of $\mathcal{J}_\text{NS}^{p_1p_2}\propto (r_1r_2)^{2+3d/2+p_{12}}\cos[2\wt\nu\log(r_1r_2)+\vartheta_\text{NS}]$. Here we see that the frequency of the oscillation is precisely $2\wt\nu$ at the 1-loop level, which is twice the frequency for a tree-level mediation by a single scalar line of mass parameter $\wt\nu$. This confirms the observation originally made in \cite{Arkani-Hamed:2015bza}. Here we have worked out the size and also the phase $\vartheta_\text{NS}$ associated with this signal. Also, we have found all the subleading corrections to the squeezed-limit result, in the form of power series in $r_1r_2$ and also in the two $\mb F$ functions. We note that this part has also been found analytically in \cite{Qin:2022lva}. The result in \cite{Qin:2022lva} was expressed as a four-layer summation, while our result here has only one layer of summation. An analytical proof of the equivalence of the two results would be nontrivial. We have checked that the two results agree perfectly numerically. Also, we have checked that the two results agree analytically at several leading terms in the powers of $r_1$ and $r_2$.

\paragraph{Local signal.} The local signal is completely from the spectral integral (\ref{eq_Jint2}) over the local tree seed integral through the Set-2A poles. See (\ref{eq_Set2A}) and (\ref{eq_J2A}). The local signal $\mathcal{J}_\text{LS}^{p_1p_2}(r_1,r_2)$ contains a factor $(r_1/r_2)^{\pm2\ii\wt\nu}$ in the squeezed limit, and thus is nonanalytic in either $r_1$ and $r_2$, but is analytic in the intermediate momentum $k_s$ (since $k_s$ is canceled in the ratio $r_1/r_2$). The full result is:
\begin{align}
\label{eq_JLS}
  \mathcal{J}_{\text{LS}}^{p_1p_2}
  =&~\frac{(r_1/r_2)^{d/2+2\ii\wt{\nu}}\sin[\pi(\fr{d}{2}+2\ii \wt{\nu})]}{8\pi^{d/2}\Gamma\big(\tfrac d2\big)\sin^2(\pi\ii\wt{\nu})}\sum_{n=0}^\infty\FR{(1+n)_{\frac d2-1} \big[(1+\ii\wt\nu+n)_{\frac d2-1}\big]^2(1+2\ii\wt\nu+n)_{\frac d2-1}}{(1+2\ii\wt\nu+2n)_{d-1}}\n\\
  &\times(\fr{d}{2}+2\ii\wt{\nu}+2n)\mathcal{C}_{2\ii\wt{\nu}+d/2+2n,d}^{p_1p_2}\mb{F}_{2\ii\wt{\nu}+d/2+2n,d}^{p_1}(r_1)\mb{F}_{-2\ii\wt{\nu}-d/2-2n,d}^{p_2}(r_2)(r_1/r_2)^{2n}+\text{c.c.}.
\end{align}
Again, the local signal generates an oscillatory behavior $\propto (r_1/r_2)^{2+3d/2+p_{12}}\cos[2\wt\nu\log(r_1/r_2)+\vartheta_\text{LS}]$. To our best knowledge, the local CC signals in 1-loop processes have not been worked out in the literature, and we believe that our result for $\mathcal{J}_{\text{LS}}$ is new.

\paragraph{Logarithmic tail.} The logarithmic tail is from the spectral integrals over the nonlocal and local tree seed integral, namely (\ref{eq_Jint1}) and (\ref{eq_Jint2}), through the poles in Set 1C and Set 2C. The ``nonlocal'' tail $\propto \log r_1r_2$ and the ``local'' tail $\propto \log(r_1/r_2)$ can be combined into a single term, which we call the logarithmic tail:
\begin{align}
    \mathcal{J}_\text{LT}^{p_1p_2}=&~ \frac{1}{4\pi^2}\sum_{n=1}^{\infty}\mathcal{Z}_{n,d}^{p_1p_2}\Pi_{n,d}(\wt\nu)  \mb{F}_{n,d}^{p_1}(r_1) \mb{F}_{n,d}^{p_2}(r_2)(r_1r_2)^n\log r_2 .
\end{align}
Recall that $r_2=k_s/k_{34}$. Thus the logarithmic tail is nonanalytic in $k_s$ as $k_s\to 0$ but does not exhibit any oscillatory behavior in momentum ratios. The existence of the logarithmic tail is a special feature of the 1-loop correlator, which has no counterparts in tree-level correlators. However, as we show in App.\ \ref{app_Pi}, in the case of $d=3$, the function $\Pi_{n,3}(\wt\nu)=0$ when $n\in \mathbb{Z}$. Therefore, the logarithmic tail vanishes identically in $(3+1)$-dimensional dS, although it does not vanish in general $d$. For example, the logarithmic tail is nonzero in $d=2$. 

\paragraph{Background.} Finally, the background piece $\mathcal{J}_\text{BG}^{p_1p_2}$ is fully analytic in $r_1$ and $r_2$ around $r_{1,2}= 0$. It receives contributions from all three integrals in (\ref{eq_Jint1})-(\ref{eq_Jint3}). We can put it in the following way:
\bge
\label{eq_JBG}
  \mathcal{J}_\text{BG}^{p_1p_2}=\mathcal{J}_{\text{(3)}}+\mathcal{J}_{\text{(B)}}+\mathcal{J}_{(\text{C})}+\mathcal{J}_\text{(2D)}.
\ede
Here $\mathcal{J}_{\text{(3)}}$ is from the integral (\ref{eq_Jint3}), namely, it is from the background tree integral (\ref{eq_IBG}).
\begin{align}
\label{eq_J3BG}
    \mathcal{J}_{\text{(3)}} 
    =& \sum_{\ell,m=0}^\infty\sum_{n=0}^{m}\frac{(-1)^{\ell+n}\sin[\frac{\pi}{2}(p_{12}+d)](\ell+1)_{2m+d+p_1+p_2+1}}{2^{2m}n!(m-n)!\big(p_2+1+\fr{d}{2}+\ell+n\big)_{m+1}}\big(\fr{1}{2}d+p_2+1+\ell+2n\big)\n\\
     &\times \rho_{\wt\nu}^\text{dS}(-\fr{\ii d}{2}-\ii p_2-\ii-\ii\ell-2\ii n) r_1^{2m+d+p_{12}+2}\Big(\frac{r_1}{r_2}\Big)^{\ell}.
\end{align}
As we shall see, $\mathcal{J}_{\text{(3)}}$ is the only term in the loop seed integral that is divergent when $d\to 3$. 

Next, $\mathcal{J}_{\text{(B)}}$ comes from the Set 1B and Set 2B poles of $\mathcal{J}_{\text{(1)}}$ and $\mathcal{J}_{\text{(2)}}$, respectively:
\begin{align}
    \mathcal{J}_{\text{(B)}}=&~\frac{\sin\fr{\pi d}{2}\csc^2(\pi\ii\wt{\nu})}{4\pi^{d/2}\Gamma\big(\tfrac d2\big)}\sum_{n=0}^\infty\FR{\big[(1+n)_{\frac d2-1}\big]^2 (1+\ii\wt\nu+n)_{\frac d2-1} (1-\ii\wt\nu+n)_{\frac d2-1}}{(1+2n)_{d-1}}(\fr{d}{2}+2n)\mathcal{C}_{d/2+2n,d}^{p_1p_2}\n\\
    &\times\mb{F}_{d/2+2n,d}^{p_1}(r_1)r_1^{2n+d/2}\Big[\mb{F}_{d/2+2n,d}^{p_2}(r_2)r_2^{2n+d/2}+\mb{F}_{-d/2-2n,d}^{p_2}(r_2)r_2^{-2n-d/2}\Big] . 
\end{align} 

The third piece $\mathcal{J}_{(\text{C})}$ comes from the Set 1C and Set 2C poles of $\mathcal{J}_{\text{(1)}}$ and $\mathcal{J}_{\text{(2)}}$, respectively. Most terms in (\ref{eq_J1C}) and (\ref{eq_J2C}) combine to zero, and we end up with the following result:
\begin{align}
     \mathcal{J}_{(\text{C})} 
   =&~ \frac{1}{8\pi^2}\sum_{n=1}^{\infty}\mathcal{Z}_{n,d}^{p_1p_2}
    \Pi_{n,d}(\wt\nu)\mb{F}_{n,d}^{p_1}(r_1)\n\\
    &\times\Big\{
    \mb{G}_{n,d}^{p_2}(r_2) \mb{F}_{n,d}^{p_2}(r_2)(r_1r_2)^n    +\mb{G}_{-n,d}^{p_2}(r_2)\mb{F}_{-n,d}^{p_2}(r_2)\Big(\FR{r_1}{r_2}\Big)^n\Big\}.
\end{align}
When $d\to 3$, this result also vanishes since $\Pi_{n,d}(\wt\nu)=0$ when $d=3$ and $n\in\mathbb{Z}$.
  
Finally, the piece $\mathcal{J}_\text{(2D)}$ comes from the Set 2D poles of the integral $\mathcal{J}_{(2)}$:
\begin{align}
\label{eq_J2Dresult}
  \mathcal{J}_\text{(2D)}=&~ (2r_1)^{d/2+1+p_2}\sum_{n=0}^\infty\frac{(-1)^n2^{1+n}\sqrt{\pi}\big(\frac d2+p_2+1+n\big)}{n!}\Pi_{d/2+p_2+1+n,d}(\wt\nu)\n\\
  &\times\mathcal{C}_{d/2+p_2+1+n,d}^{p_1p_2}\mathbf{F}_{d/2+p_2+1+n,d}^{p_1}(r_1){\,}_2\wt{\mathrm{F}}_1\left[\begin{matrix}
    -\tfrac n2, \tfrac{1-n}2 \\
    -\tfrac d2-p_2-n
  \end{matrix}\middle|r_2^2\right]\Big(\frac{r_1}{r_2}\Big)^n.
\end{align}   
In $d=3$ and integer $p_2$, this piece involves the $\Pi_{n,d}$ function with $n+1/2\in\mathbb{Z}$. As we can see from the explicit expression of $\Pi_{n,3}$ in (\ref{eq_Pin3}), $\Pi_{1/2,3}$ is finite when $n=1/2$; $\Pi_{n,3}=0$ when $n=2k+1/2$ with $k=1,2,\cdots$, and is divergent as a simple pole when $n=2k+3/2$ with $k=0,1,2,\cdots$. However, these poles happen to be zeros of the second degree of the factor $\mathcal{C}_{d/2+p_2+1+n,d}^{p_1p_2}$ when $p_{1,2}$ are even integers, which is always true for the special examples considered in this work. Therefore, when $d=3$ and $p_1, p_2$ being even integers, $\mathcal{J}_\text{(2D)}$ vanishes unless $p_2$ taking a value equal to or smaller than $-2$. When $p_2=-2$, there will be a single term with $n=0$ in (\ref{eq_J2Dresult}) that contributes to the background. This happens in several examples considered in Sec.\ \ref{sec_seedint}; See (\ref{eq_Lconformal}), (\ref{eq_calBinMSbar}), and (\ref{eq_calLcov}). On the other hand, we shall never encounter $p_2<-2$ in this work.

At this point, we have finished the computation of the loop seed integral. The result is rather complicated in general $d$ dimensions, as is summarized in (\ref{eq_Jresult}). We shall examine its property and make several consistency checks in the next section, at least to make sure that our result reduces to known results in certain limits. Then, in Sec.\ \ref{sec_CC}, we shall use the loop seed integral to compute several realistic 1-loop inflaton correlators in $d=3$. There we shall get significantly simplified expressions. 

\section{Properties and Consistency Checks}
\label{sec_check}

In this section, we discuss the properties of the loop seed integrals in several limits, including the $d\to 3$ limit, the large mass ($\wt\nu\gg 1$) limit, the squeezed limit where $r_{1},r_{2}\to 0$, the folded limit where $r_{1}$ or $r_2\to 1$, and the $r_1=r_2$ limit. In these limits, either the loop seed integral can be (at least partially) computed using other methods, or it should have expected properties on physical grounds. Therefore, examining these limits can serve as consistency checks of our results. 

\subsection{Divergence in 3+1 dimensions}

While the seed integral is computed in arbitrary $d$ spatial dimensions, we are ultimately interested in the case of $d=3$. In dS$_{3+1}$, the UV divergence is expected to appear in the 1-loop correlator. In the loop seed integral $\mathcal{J}_{\wt\nu}^{p_1p_2}$ in (\ref{eq_Jresult}), the nonlocal signal $\mathcal{J}_\text{NS}^{p_1p_2}$ and the local signal $\mathcal{J}_\text{LS}^{p_1p_2}$ are manifestly finite when $d\to 3$, as one can directly see from (\ref{eq_JNS}) and (\ref{eq_JLS}). The logarithmic tail $\mathcal{J}_\text{LT}^{p_1p_2}$ vanishes when $d\to 3$, due to the vanishing of the $\Pi$ function: $\Pi_{n,d}(\wt\nu)=0$ at $d=3$ and integer $n$. So any possible UV divergence must come from the background piece $\mathcal{J}_\text{BG}^{p_1p_2}$, as we should expect. 

For the background piece $\mathcal{J}_\text{BG}^{p_1p_2}$ in (\ref{eq_JBG}), we can see that the last three terms, including $\mathcal{J}_{\text{(B)}}$, $\mathcal{J}_{(\text{C})}$, and $\mathcal{J}_\text{(2D)}$, all remain finite or vanish at $d\to 3$. Therefore, the only potential UV divergence comes from the first term $\mathcal{J}_{(3)}$, which is explicitly given in (\ref{eq_J3BG}). Closer examination shows that the divergence of $\mathcal{J}_{(3)}$ comes from the spectral density $\rho_{\wt\nu}^\text{dS}(\wt\nu')$. As we show in App.\ \ref{app_rho3plus1}, when $d\to 3$, the spectral density behaves like:
\bge
\label{eq_rhodSd3limit}
  \lim_{d\to 3}\rho_{\wt\nu}^\text{dS}(\wt\nu')\sim -\FR{1}{(4\pi)^2} \FR{2}{3-d} +\text{finite terms}.
\ede
As expected, the divergent part is independent of either $\wt\nu$ or $\wt\nu'$, as UV physics should be insensitive to physics at finite scales. Now, we insert (\ref{eq_rhodSd3limit}) back into $\mathcal{J}_{\text{(3)}}$ in (\ref{eq_J3BG}), and find:
\begin{align}
\label{eq_J3div}
    \mathcal{J}_{\text{(3)}} 
    =&~  \FR{1}{(4\pi)^2}\FR{2}{3-d}\sum_{\ell,m=0}^\infty\sum_{n=0}^{m}\frac{(-1)^{\ell+n}\cos\fr{\pi p_{12}}{2}(\ell+1)_{2m+p_{12}+4}}{2^{2m}n!(m-n)!\big(p_2+\fr{5}{2}+\ell+n\big)_{m+1}}\n\\
     &\times\big(p_2+\fr{5}{2}+\ell+2n\big) r_1^{2m+p_{12}+5}\Big(\frac{r_1}{r_2}\Big)^{\ell}+\mathcal{O}\Big((3-d)^0\Big).
\end{align}
At this point, the summation in (\ref{eq_J3div}) can be completed, because the $n$-sum is zero except when $m=0$. To see this, consider the following slightly more general summation: 
\begin{align}
 & \sum_{n=0}^{M}\frac{(-1)^{n} (p_2+\fr{5}{2}+\ell+2n )  }{n!(m-n)!\big(p_2+\fr{5}{2}+\ell+n\big)_{m+1}}=\FR{(-1)^M}{m}\Gamma\bgb p_2+\fr{7}{2}+\ell+M \\ 1+M, m-M, p_2+\fr{7}{2}+\ell+M+m\edb  .
\end{align}
The $n$-sum in (\ref{eq_J3div}) is obtained by setting $M=m$, which gives zero when $m\neq 0$, due to an Euler $\Gamma$ factor $\Gamma(m-M)$ in the denominator. However, when $m=0$, the two factors, $m$ and $\Gamma(m-M)$, are combined to give a finite result. Therefore, we see that we only need to retain the $m=0$ term in the summation in (\ref{eq_J3div}), and the result is:
\begin{equation}
    \mathcal{J}_{\text{(3)}} 
    = \FR{1}{(4\pi)^2}\FR{2}{3-d}\cos\big(\fr{\pi p_{12}}{2}\big) \Gamma(5+p_{12}) \Big(\FR{r_1r_2}{r_1+r_2}\Big)^{5+p_{12}}+\mathcal{O}\Big((3-d)^0\Big).
\end{equation}
Note that the combination $r_1r_2/(r_1+r_2)=k_s/k_{1234}$ is nothing but the factor we would expect to see from the 4-point correlator generated by a contact interaction. This shows that the UV divergence of the 1-loop correlator is completely local and can be subtracted by a local counterterm.

To be more specific, we can take an example with four conformal scalars directly coupled to the massive scalars. As shown in (\ref{eq_Lconformal}), this corresponds to $p_1=p_2=-2$. So, the above expression gives
\begin{equation}
     \mathcal{L}_{\phi_c,\wt\nu}\Big|_{s,\text{div.}}=\FR{\tau_f^{4}}{16k_1k_2k_3k_4k_s}\mathcal{J}_{\text{(3)}}^{-2,-2}
    = \FR{1}{(4\pi)^2}\FR{2}{3-d}\FR{\tau_f^{4}}{16k_1k_2k_3k_4k_{1234}} +\mathcal{O}\Big((3-d)^0\Big).
\end{equation}
There are two identical divergent contributions from the $t$-channel and $u$-channel exchanges. So the total result for the UV divergence is the above expression multiplied by three. On the other hand, we can directly compute the 4-point correlator of conformal scalars with direct quartic interaction $\Delta\ld=-\fr{1}{24}\de_\lam\phi_c^4$, with $\de_\lam$ understood as the coefficient of the counterterm. It is straightforward to find the result from a bulk calculation:
\begin{align}
  \la\phi_{c,\mb k_1}\phi_{c,\mb k_2}\phi_{c,\mb k_3}\phi_{c,\mb k_4}\ra'
  =&-\ii \de_\lam \sum_{\aa=\pm}\aa\int_{-\infty}^{\tau_f}\FR{\di\tau}{\tau^4} C_{\aa}(k_1,\tau)C_{\aa}(k_2,\tau)C_{\aa}(k_3,\tau)C_{\aa}(k_4,\tau)\n\\
  =&~\FR{-\de_\lam \tau_f^4}{8k_1k_2k_3k_4k_{1234}}.
\end{align} 
Therefore, we see that the UV divergence can be subtracted by a local counterterm $-\fr{1}{24}\de_\lam\phi_c^4$ with $\de_\lam=3/[(4\pi)^2(3-d)]$, just as what we would expect from a flat-space computation. The reason for this agreement is clear: For a massive loop, the divergence at $d=3$ is totally from the physics in the ultraviolet region of the momentum space, where the finite curvature of the spacetime should be negligible.

\subsection{Large mass limit}

As shown in the previous subsection, the flat-space intuition applies to the UV divergent part of the 1-loop correlator. Moreover, when the mass $m$ running in the loop is much greater than the Hubble scale $H$, we should also expect that the \emph{finite} part of the dS 1-loop correlator approaches the corresponding result in the flat space. The easiest way to see this is that there are only two scales involved in our problem, namely the mass $m$ and the Hubble scale $H$. Therefore, taking the large mass limit $m/H\gg 1$ is equivalent to sending $H\to 0$, which is just the flat-space limit.  

Even in the Minkowski space, the 1-loop in-in correlator is not a very familiar result. It turns out that a direct computation via the standard diagrammatic rule in the SK formalism is not trivial in flat space. In App.\ \ref{app_mink}, we use the spectral decomposition method to compute the same 1-loop correlator in Fig.\ \ref{fig_LoopSD} with the direct coupling $-\frac{1}{4}\varphi^2\si^2$ but in Minkowski spacetime. Here we quote the result in the $d\to 3$ limit:
\begin{align}
  \mathcal{L}_{m}^\text{Mink} 
  =&~\FR{1}{256\pi^2E_1E_2E_3E_4E_{1234}}\bigg[\FR{2}{3-d}-\ga_E+\log4\pi+2\n\\
  &~+\FR{2}{E_{12}-E_{34}}\int_0^1\di\xi\,\bigg(E_{34}\log\FR{E_{12}+E_\text{min}}{\mu_R}-E_{12}\log\FR{E_{34}+E_\text{min}}{\mu_R}\bigg) \bigg].
\end{align}
Here $E_i\equiv\sqrt{\mb k_i^2+m_i^2}$ with $i=1,2,3,4$, (It does not bring any pain to add nonzero masses to external fields in this case.) and $E_\text{min}\equiv \sqrt{\mb k_s^2+m^2/[\xi(1-\xi)]}$. The mass scale $\mu_R$ is a renormalization scale, and can be thought of as coming from the mass dimension of the coupling constant in $d$ spatial dimensions. In the large mass limit $m\to\infty$,
\begin{align}
\label{eq_LMinkLM}
  \Big[\mathcal{L}_{m}^\text{Mink}\Big]_{\overline{\text{MS}}}
  =&~\FR{1}{256\pi^2E_1E_2E_3E_4E_{1234}}\bigg[\log\FR{\mu_R^2}{m^2}-\FR{E_{12}E_{34}+\mb k_s^2}{6m^2}+\mathcal{O}\Big(\FR{1}{m^{3}}\Big)\bigg].
\end{align}

Now let us return to the dS case. Let us define a renormalized spectral function under the modified minimal subtraction scheme ($\overline{\text{MS}}$):
\begin{align}
\label{eq_rhoMSbar}
  \wh{\rho}_{\wt\nu}^\text{dS}(\wt\nu')\equiv \lim_{d\to 3}\bigg[\rho_{\wt\nu}^\text{dS}(\wt\nu')+\FR{1}{(4\pi)^2}\Big(\FR{2}{3-d}-\ga_E+\log4\pi\Big)\bigg].
\end{align}
In App.\ \ref{app_asympt}, we show that the large mass limit ($\wt\nu\gg 1$) of the renormalized spectral function $\wh{\rho}_{\wt\nu}^\text{dS}(\wt\nu')$ with $\wt\nu'$ held fixed is given by (\ref{eq_rhoLargeNu}). Specialized to $d=3$, the result is:
\bge 
\label{eq_rhoLargeNu3}
  \lim_{\wt\nu\to \infty}\wh{\rho}_{\wt\nu}^\text{dS}(-\ii\nu')\sim \FR{1}{(4\pi)^2}\bigg[\log(\wt\nu^2)+\order{\wt\nu^{-2}}\bigg],
\ede
where the logarithmic term is obtained from explicit analytical derivations, while the $1/\wt\nu^2$ term can be obtained from a numerical fit. To get a loop correlator that can be directly compared with the flat-space result (\ref{eq_LMinkLM}), we again take the example of the non-derivatively coupled conformal scalar in (\ref{eq_Lconformal}). We replace the spectral function $\rho_{\wt\nu}^\text{dS}$ in $\mathcal{J}_\text{(3)}$ by (\ref{eq_rhoLargeNu3}), and take $p_1=p_2=-2$. Then, we get:
\begin{equation}
\label{eq_J3LargeNu}
    \lim_{\wt\nu\to\infty}\FR{\tau_f^4}{16k_1k_2k_3k_4k_s}\mathcal{J}_{\text{(3)}}^{-2,-2} 
    \sim -\FR{1}{16k_1k_2k_3k_4k_{1234}}\FR{1}{(4\pi)^2}\Big[\log\FR{m^2}{\mu_R^2}+\order{\wt\nu^{-2}}\Big].
\end{equation}
Here we have set $\wt\nu\simeq m/H$ in the final expression. Also, we have introduced a renormalization scale $\mu_R$, which comes from the dimension of the coupling constant in general $d$ dimensions. It is clear that the coefficient of $\log (m^2/\mu_R^2)$ in (\ref{eq_J3LargeNu}) matches exactly the coefficient of $\log (m^2/\mu_R^2)$ in (\ref{eq_LMinkLM}), as expected.

\subsection{Squeezed limit}

Now we consider a particular kinematic configuration, called the squeezed limit, where $r_{1,2}\ll 1$, namely, the momentum $k_s$ mediated by the loop is much smaller than any of the external momenta $k_i$. This is the limit of most interest to CC physics, where we expect to see oscillatory signals. From a bulk perspective, the signal is contributed dominantly by the superhorizon modes of the loop particles, and therefore it is possible to compute at least the nonlocal signal by taking the late-time expansion of the loop propagator. Below we shall perform this late-time computation directly in the bulk, and compare the resulting nonlocal signal with the leading term in $\mathcal{J}_\text{NS}^{p_1p_2}$ in (\ref{eq_JNS}).

It turns out that the late-time computation is most easily done in position space. Therefore, we Fourier-transform the loop momentum integral (\ref{eq_calQ}) back to position space:\footnote{
After the Fourier transformation, the SK indices $\aa\bb$ of the momentum-space propagator $D_{\wt\nu,\aa\bb}$ are translated to various types of $\ii\ep$-prescriptions for the coordinates. The detail is irrelevant here, since the required nonlocal part of the propagator is real and is independent of SK indices. See (\ref{eq_D2exp}).
}
\begin{align}
\label{eq_calQinPS}
  \mathcal{Q}_{\wt\nu,\mathsf{ab}}\big(k_s;\tau_1,\tau_2\big)
  =&\int\FR{\di^d\mb q}{(2\pi)^d}D_{\wt\nu,\mathsf{ab}}\Big(q;\tau_1,\tau_2\Big)D_{\wt\nu,\mathsf{ab}}\Big(|\mb k_s-\mb q|;\tau_1,\tau_2\Big)\n\\
  =&\int\di^d\mb x\,e^{\ii\mb k_s\cdot\mb x}\Big[D_{\wt\nu}(\tau_1,\mb x;\tau_2,\mb 0)\Big]^2,
\end{align}
where $D_{\wt\nu}(\tau_1,\mb x_1;\tau_2,\mb x_2)$ is the position-space massive scalar propagator in $d$ spatial dimensions:
\begin{align}
  D_{\wt\nu}(\tau_1,\mb x_1;\tau_2,\mb x_2)=\FR{1}{(4\pi)^{(d+1)/2}}
  {}_2\mathcal{F}_1\left[\bgm \fr{d}{2}+\ii\wt\nu,\fr{d}{2}-\ii\wt\nu \\ \fr{d+1}{2}\edm \middle|\FR{(\tau_1+\tau_2)^2-|\mb x_1-\mb x_2|^2}{4\tau_1\tau_2}\right].
\end{align}
The good thing is that the loop momentum integral reduces to the square of a single propagator. Therefore, we can simply make the late-time expansion of the propagator at $\tau_{1,2}\to 0$, from which we get:
\begin{align}
\label{eq_D2exp}
  \Big[D_{\wt\nu}(\tau_1,\mb x;\tau_2,\mb 0)\Big]_\text{NL}^2\sim \FR{1}{16\pi^{d+2}}\Gamma^2\big[\fr{d}{2}+\ii\wt\nu,-\ii\wt\nu\big]\Big(\FR{\tau_1\tau_2}{\mb x^2}\Big)^{d+2\ii\wt\nu}+\text{c.c.}.
\end{align}
Here the notation $[\cdots]_\text{NL}$ means that we only retain nonlocal terms, i.e., terms of noninteger powers of $\mb x^2$, with the anticipation that such terms are the only source of the nonlocal signal.

Then the nonlocal part of the loop momentum integral can be directly got by substituting (\ref{eq_D2exp}) back into (\ref{eq_calQinPS}), and the result is:
\begin{align}
  \big[\mathcal{Q}_{\wt\nu}\big]_\text{NL}\sim \FR{(k_s^2\tau_1\tau_2)^{d+2\ii\wt\nu}}{8\pi^{(d+5)/2}k_s^d}\cos\big[\fr{\pi}{2}(d+4\ii\wt\nu)\big]\Gamma^2\big[\fr{d}{2}+\ii\wt\nu,-\ii\wt\nu\big]\Gamma\bgb\fr{1+d}{2}+2\ii\wt\nu,-d-4\ii\wt\nu \\ d+2\ii\wt\nu \edb+\text{c.c.}.
\end{align}
Note in particular that the nonlocal part of the loop momentum integral is independent of the SK indices, a feature we should expect. Then, the bulk time integral in (\ref{eq_LoopSeedInt}) can be directly finished, and we get:
\begin{align}
 &\big[\mathcal{J}_{\wt\nu}^{p_1p_2}\big]_\text{NL}
  \sim \FR{r_1^{p_1}r_2^{p_2}(r_1r_2)^{1+d+2\ii\wt\nu}}{2^{1+d+2\ii\wt\nu}\pi^{2+d/2}}\cos\big[\fr{\pi}{2}(d+p_1+2\ii\wt\nu)\big]\cos\big[\fr{\pi}{2}(d+p_2+2\ii\wt\nu)\big]\cos\big[\fr{\pi}{2}(d+4\ii\wt\nu)\big]\n\\
  &\times\Gamma^2(-\ii\wt\nu)\Gamma\bgb 1+d+p_1+2\ii\wt\nu,1+d+p_2+2\ii\wt\nu,-d-4\ii\wt\nu,\fr{1+d}{2}+2\ii\wt\nu,\fr{d}{2}+\ii\wt\nu \\ \fr{1+d}{2}+\ii\wt\nu \edb +\text{c.c.}.
\end{align}
This agrees exactly with the leading ($n=0$) term of the nonlocal signal in (\ref{eq_JNS}).

We note that the computation of the local signal and the background from the late-time expansion would be nontrivial, in part because these pieces are contributed by the ``local'' terms of the propagator $D(\tau_1,\mb x,\tau_2,\mb 0)$, namely, the terms analytic in $\mb x^2$. When Fourier-transforming such terms back to the momentum space, one encounters a divergence, which is essentially the UV divergence of the original loop momentum integral. In comparison, the nonlocal signal is automatically free from such divergences and thus is more tractable in the late-time calculation.

\begin{figure}[t]
 \centering
  \includegraphics[width=0.49\textwidth]{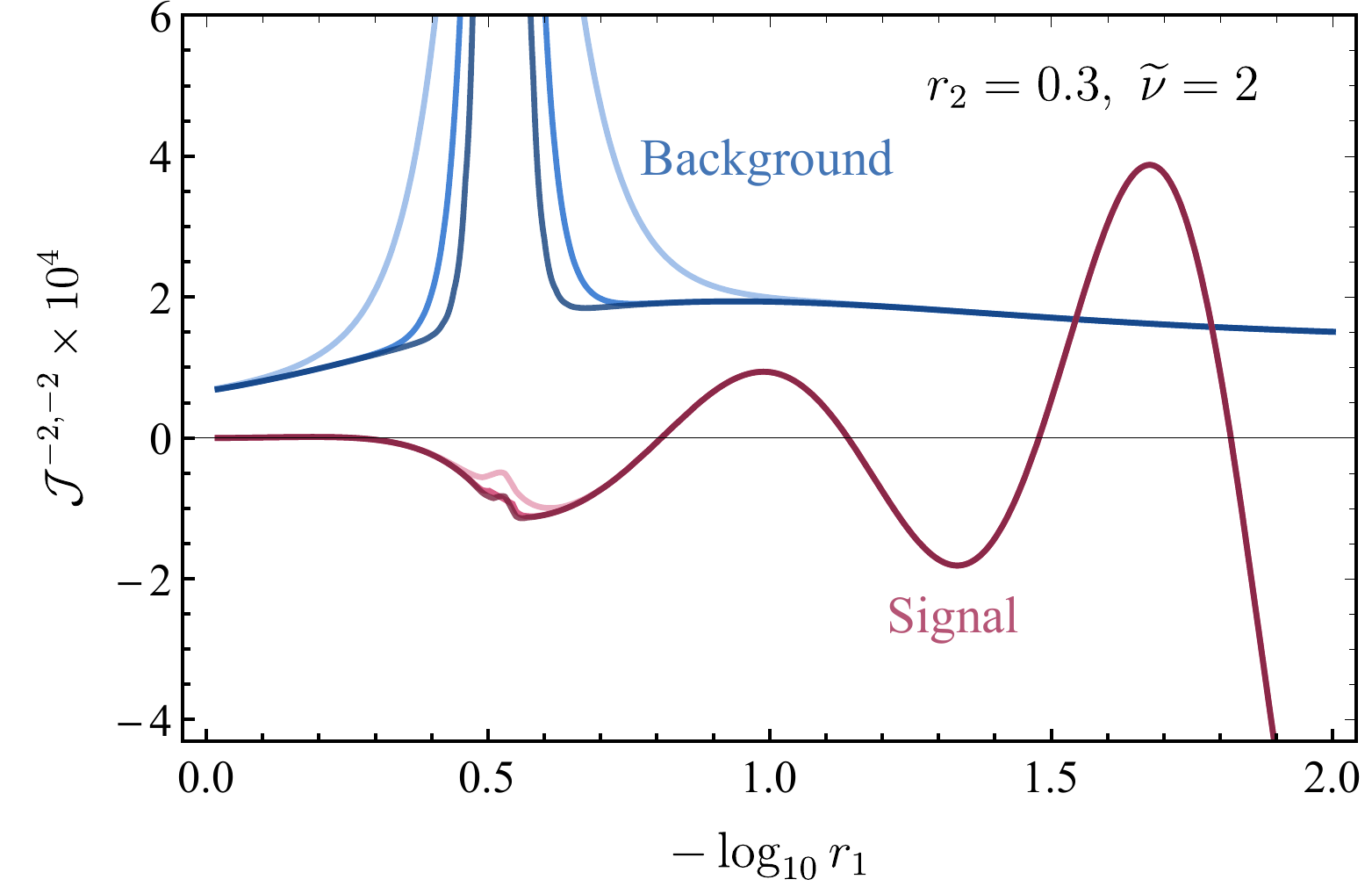} 
  \includegraphics[width=0.49\textwidth]{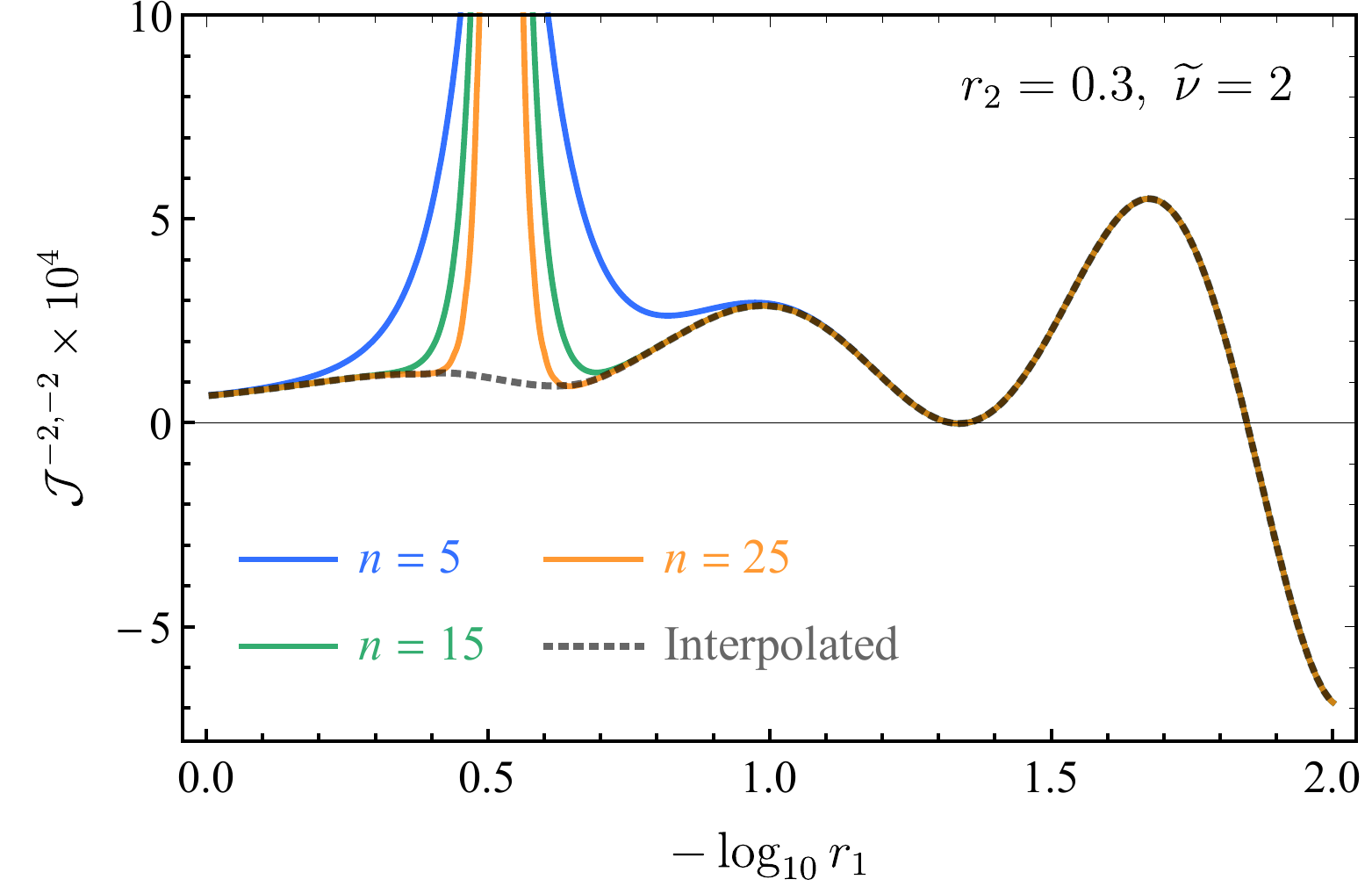} 

\caption{The convergence of the loop seed integral. In the left panel, we show the convergence of the signals $\mathcal{J}_\text{NS}^{-2,-2}+\mathcal{J}_\text{LS}^{-2,-2}$ (magenta) and the background $\mathcal{J}_\text{BG}^{-2,-2}$ (blue) separately. In each of the two classes of curves, the three curves from lighter to darker colors show the sums of the series up to $n$'th power of $r_1/r_2$, with $n=5,15,25$, respectively. In the right panel, we show the sum of the signals and the background, again, with three choices of $n$. The grey dotted curve shows the interpolated function of the whole seed integral. In this plot we fix $d=3$, $p_1=p_2=-2$, $r_2=0.3$, and $\wt\nu=2$. }
\label{fig_conv} 
\end{figure}

\subsection[Folded limit and $r_1=r_2$ limit]{Folded limit and $\bm{r_1=r_2}$ limit}

Now we briefly comment on the two limits where our expressions exhibit superficial divergences. One is the folded limit, the other is the $r_1=r_2$ limit.

The folded limit of the loop seed integral means that one or both of $r_1$ and $r_2$ go to 1 from below. (Note that the momentum conservation at both vertices of the 1-loop diagram requires that $0\leq r_{1,2}\leq 1$.) All terms in our final result for the loop seed integral (\ref{eq_Jresult}) are expressed as series or partially resummed power series in $r_1$ and $r_2$, and we would expect many of these series to be divergent as $r_{1,2}\to 1$. On the other hand, we expect that the 1-loop correlator is free from any divergences in the folded limit, as a consequence of choosing Bunch-Davies initial state for all the modes under consideration. Therefore, it would be useful to show that all superficial divergences in the folded limit cancel in (\ref{eq_Jresult}). Our procedure of spectral decomposition suggests that we can check the cancelation of folded divergences directly at the tree level, which was done in previous works \cite{Arkani-Hamed:2018kmz,Qin:2022fbv}. Then, so long as taking the folded limit commutes with the spectral integral which we shall assume, it would follow automatically that the loop correlator is also free from folded divergence. We also note that sending the single-side folded limit ($r_2\to 1^-$ while keeping $r_1$ fixed) is simple enough. We shall do this explicitly in the next section when computing a 3-point inflaton correlator.  

Finally, let us look at the $r_1=r_2$ limit. Physically, we expect our result of the loop seed integral to be smooth at this point. Also, the tree seed integral is smooth at $r_1=r_2$, although some pieces (the local signal and the background) in the tree seed integral may exhibit non-smooth behavior at $r_1=r_2$. This non-smooth behavior is purely an artifact of our definition of the local signal and the background. See the comments below (\ref{eq_Itotal}).  On the other hand, the various terms of the loop seed integral in (\ref{eq_Jresult}) are expressed in power series of $r_1/r_2$ whose convergences at $r_1/r_2=1$ are far from clear. This superficial discontinuity may be resolved by resumming the series, as we can do for the tree correlators with the Partial Mellin-Barnes representation. (See App.\ \ref{App_TreeSeed}.) We leave this for a future work. For the moment, we only note that this superficial discontinuity does not bring any obstacle to numerical implementation. This is shown in Fig.\ \ref{fig_conv}, where we show the convergence of the various series in (\ref{eq_Jresult}), and in particular, the discontinuity as $r_1$ goes across $r_2$. From this figure, we see that the signal series in (\ref{eq_JNS}) and (\ref{eq_JLS}) converge very quickly. On the other hand, the background series $\mathcal{J}_\text{BG}^{-2,-2}$ in (\ref{eq_J3BG}) converges rather slow around $r_1=r_2$. However, it is easy to get an interpolated function that smoothly joins the two sides of $r_1=r_2$ where the series converges quickly. This shows that our result can be easily implemented in numerical computations.

\section{Applications to Cosmological Collider Physics}
\label{sec_CC}
 
 With the full result for the loop seed integral at hand, we can directly compute many 1-loop inflaton correlators mediated by massive scalar fields with various types of couplings. In this section, we provide full results for two examples that are most relevant to the study of CC physics. One is the 1-loop trispectrum $\mathcal{L}_{\varphi,\wt\nu}$ in (\ref{eq_calLinMSbar}), and the other is the 1-loop bispectrum $\mathcal{B}_{\varphi,\wt\nu}$ in (\ref{eq_calBinMSbar}).

\subsection{1-loop trispectrum}

Here we present the result for the 1-loop inflaton correlator in Fig.\ \ref{fig_LoopSD}, with the coupling given in (\ref{eq_DeltaLTDC}). As suggested by the expression (\ref{eq_calLinMSbar}), the result can be found from (\ref{eq_Jresult}) by setting $p_1=p_2=0$, and then taking the $d\to 3$ limit. Many terms in (\ref{eq_Jresult}) drop out in this limit, and the result is much simplified compared to the full seed integral. In particular, the logarithmic tail vanishes identically for this process. 

It turns out that the result can be more conveniently written in the following way:
\begin{keyeqn}
\begin{align}
\label{eq_Lresult}
  \mathcal{L}_{\varphi,\wt\nu}=\FR{1}{16k_1k_2k_3k_4(k_{12}k_{34})^{5/2}}\Big[\wh{\mathcal{J}}\!_\text{NS}(r_1,r_2)+ \wh{\mathcal{J}}\!_\text{LS}(r_1,r_2)+\wh{\mathcal{J}}\!_\text{BG}(r_1,r_2)\Big].
\end{align}
\end{keyeqn}
Here $\wh{\mathcal{J}}\!_{\text{NS}}\equiv (r_1r_2)^{-5/2}\mathcal{J}_\text{NS}^{00}$ with the unhatted $\mathcal{J}_\text{NS}^{p_1p_2}$ given by (\ref{eq_JNS}). This part represents the nonlocal signal:
\begin{align}
\label{eq_JNS00}
  \wh{\mathcal{J}}\!_{\text{NS}}
  =&~\frac{2(r_1r_2)^{3/2+2\ii\wt{\nu}}}{\pi^{2}\cos(2\pi\ii\wt\nu)}\sum_{n=0}^\infty\FR{(1+n)_{\frac{1}{2}} \big[(1+\ii\wt\nu+n)_{\frac{1}{2}}\big]^2(1+2\ii\wt\nu+n)_{\frac{1}{2}}}{(1+2\ii\wt\nu+2n)_{2}}(\fr{3}{2}+2\ii\wt{\nu}+2n)\n\\
  &~\times{}_2\mathcal{F}_1\left[\bgm 2+\ii\wt\nu+n,\fr{5}{2}+\ii\wt\nu+n \\ \fr{5}{2}+2\ii\wt\nu+2n
  \edm\middle|r_1^2\right]
  {}_2\mathcal{F}_1\left[\bgm 2+\ii\wt\nu+n,\fr{5}{2}+\ii\wt\nu+n \\ \fr{5}{2}+2\ii\wt\nu+2n
  \edm\middle|r_2^2\right]
  (r_1r_2)^{2n}+\text{c.c.}.
\end{align} 
Similarly, $\wh{\mathcal{J}}\!_{\text{LS}}\equiv (r_1r_2)^{-5/2}\mathcal{J}_\text{LS}^{00}$ with the unhatted $\mathcal{J}_\text{LS}^{p_1p_2}$ given by (\ref{eq_JLS}). This part represents the local signal:
\begin{align}
\label{eq_JLS00}
  \wh{\mathcal{J}}\!_{\text{LS}}
  =&-\frac{2(r_1/r_2)^{3/2+2\ii\wt{\nu}}}{\pi^{2}\cos(2\pi\ii\wt\nu)} \sum_{n=0}^\infty\FR{(1+n)_{\frac{1}{2}} \big[(1+\ii\wt\nu+n)_{\frac{1}{2}}\big]^2(1+2\ii\wt\nu+n)_{\frac{1}{2}}}{(1+2\ii\wt\nu+2n)_{2}}(\fr{3}{2}+2\ii\wt{\nu}+2n)\n\\
  &~\times{}_2\mathcal{F}_1\left[\bgm 2+\ii\wt\nu+n,\fr{5}{2}+\ii\wt\nu+n \\ \fr{5}{2}+2\ii\wt\nu+2n
  \edm\middle|r_1^2\right]
  {}_2\mathcal{F}_1\left[\bgm \fr{1}{2}-\ii\wt\nu-n,1-\ii\wt\nu-n \\ -\fr{1}{2}-2\ii\wt\nu-2n
  \edm\middle|r_2^2\right]
  \Big(\FR{r_1}{r_2}\Big)^{2n}+\text{c.c.}.
\end{align}  
Finally, $\wh{\mathcal{J}}\!_{\text{BG}}\equiv (r_1r_2)^{-5/2}\mathcal{J}_\text{BG}^{00}$ with the unhatted $\mathcal{J}_\text{BG}^{p_1p_2}$ given by (\ref{eq_JBG}). It turns out that all terms in (\ref{eq_JBG}), except $\mathcal{J}_{(3)}$, vanish in $d\to 3$ limit with $p_1=p_2=0$. On the contrary, the term in $\mathcal{J}_{(3)}$ possesses the usual UV divergence as $d\to 3$. Therefore we use the $\overline{\text{MS}}$ spectral function $\wh{\rho}_{\wt\nu}^\text{dS}(\wt\nu')$ in (\ref{eq_rhoMSbar}) in place of the original spectral function, and get the following expression for the background $\wh{\mathcal{J}}\!_{\text{BG}}$ under the $\overline{\text{MS}}$ scheme:
\begin{align}
\label{eq_JBG00}
  \wh{\mathcal{J}}\!_{\text{BG}}
    =&\sum_{\ell,m=0}^\infty\sum_{n=0}^{m}\frac{(-1)^{\ell+n+1}(\ell+1)_{2m+4}(\fr{5}{2}+\ell+2n ) }{2^{2m}n!(m-n)!\big(\fr{5}{2}+\ell+n\big)_{m+1}}\n\\
     &\times\bigg[ 
      \wh\rho_{\wt\nu}^\text{dS}(-\fr{\ii 5}{2}-\ii\ell-2\ii n)-\FR{1}{(4\pi)^2}\log \mu_R^2\bigg] r_1^{2m}\Big(\frac{r_1}{r_2}\Big)^{5/2+\ell}.
\end{align}
An explicit expression for the $\overline{\text{MS}}$ spectral function $\wh{\rho}_{\wt\nu}^\text{dS}(\wt\nu')$ is given in (\ref{eq_rhodSMSbarExplicit}), which we find useful for numerical implementation. Here we also restore the renormalization scale $\mu_R$, which comes from the change of the mass dimension of the coupling constant when the spatial dimension deviates from $d=3$.\footnote{More explicitly, in $d$ dimensions, the mass dimensions of scalar fields are given by $[\varphi]=[\si]=(d-1)/2$, and the Lagrangian has mass dimension $[\ld]=d+1$. Thus, the coupling term in the Lagrangian should be written as $\mu_R^{3-d}\varphi'^2\si^2/(4\Lambda^2)$ where $\Lambda$ is a dim-1 cutoff scale, while the counterterm should be written as $\mu_R^{3-d}\varphi'^4/(24\Lambda_c^4)$ where $\Lambda_c^4$ is another dim-1 cutoff scale. When $d\to 3$, the two couplings in the 1-loop diagram produce a finite piece $-[2/(4\pi)^2]\log\mu_R^2$ when combined with the divergent term $-2/[(4\pi)^2(3-d)]$ of the spectral function. See (\ref{eq_rhodSd3limit}). The counterterm gives another finite piece $+[1/(4\pi)^2]\log\mu_R^2$. Combining the two pieces, we get the total dependence on $\mu_R$ as $-[1/(4\pi)^2]\log\mu_R^2$.  }

\paragraph{Squeezed limit.} The squeezed limit is of most interest for CC applications, and therefore it would be useful to have an explicit result for the trispectrum in the squeezed limit. Here we take a ``hierarchical'' squeezed limit $r_1\ll r_2\ll 1$ in (\ref{eq_Lresult}). Then we can keep the leading terms in both $r_1$ and $r_2$, and also in $r_1/r_2$ for all three terms in (\ref{eq_Lresult}). The result is:
\begin{align}
\label{eq_JNSSq}
  \lim_{r_1\ll r_2\ll 1}\wh{\mathcal{J}}\!_{\text{NS}}  
  =&~\FR{4^{1+2\ii\wt\nu}\sec(2\pi\ii\wt\nu)}{\pi^2}\FR{(1+\ii\wt\nu)\Gamma^2[\fr{3}{2}+\ii\wt\nu,\fr{5}{2}+\ii\wt\nu]}{\Gamma(4+4\ii\wt\nu)}(r_1r_2)^{3/2+2\ii\wt\nu}+\text{c.c.}, \\
  \label{eq_JLSSq}
  \lim_{r_1\ll r_2\ll 1}\wh{\mathcal{J}}\!_{\text{LS}} 
  =&-\FR{\sec(2\pi\ii\wt\nu)}{4\sqrt\pi}\Gamma\bgb\fr{3}{2}+\ii\wt\nu,\fr{5}{2}+\ii\wt\nu,1-2\ii\wt\nu \\ 1+\ii\wt\nu,1+\ii\wt\nu,-\fr{1}{2}-2\ii\wt\nu\edb\Big(\FR{r_1}{r_2}\Big)^{3/2+2\ii\wt\nu}+\text{c.c.},\\
  \lim_{r_1\ll r_2\ll 1}\label{eq_JBGSq}
  \wh{\mathcal{J}}\!_{\text{BG}} 
  =&-24\bigg[\wh{\rho}_{\wt\nu}^\text{dS}(-\fr{5\ii}{2})-\FR{1}{(4\pi)^2}\log \mu_R^2\bigg]\Big(\FR{r_1}{r_2}\Big)^{5/2} .
\end{align}
An interesting observation here is that, if we take the single squeezed limit, i.e., we keep $r_2$ fixed, and take the $r_1\ll r_2$ limit, then the overall sizes of both the nonlocal and local signals decay as $r_1^{3/2}$, while the background decays as $r_1^{5/2}$. Therefore, the signals dominate over the background in the single squeezed limit even at the 1-loop level. Although we made this observation from the squeezed-limit results (\ref{eq_JNSSq})-(\ref{eq_JBGSq}), this conclusion holds for any fixed $r_2$ which is not necessarily small.

\begin{figure}[t]
 \centering
  \includegraphics[width=0.49\textwidth]{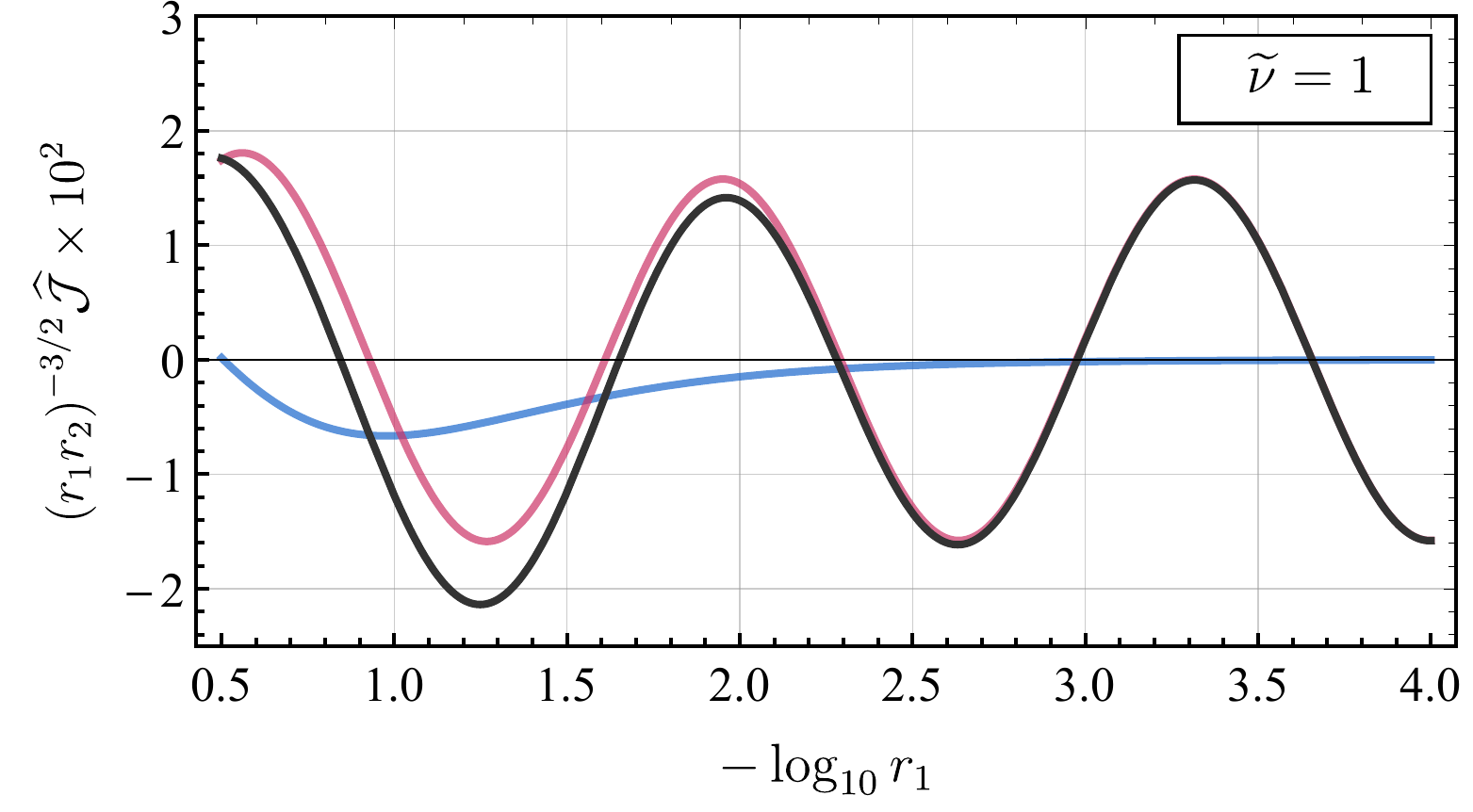} 
  \includegraphics[width=0.49\textwidth]{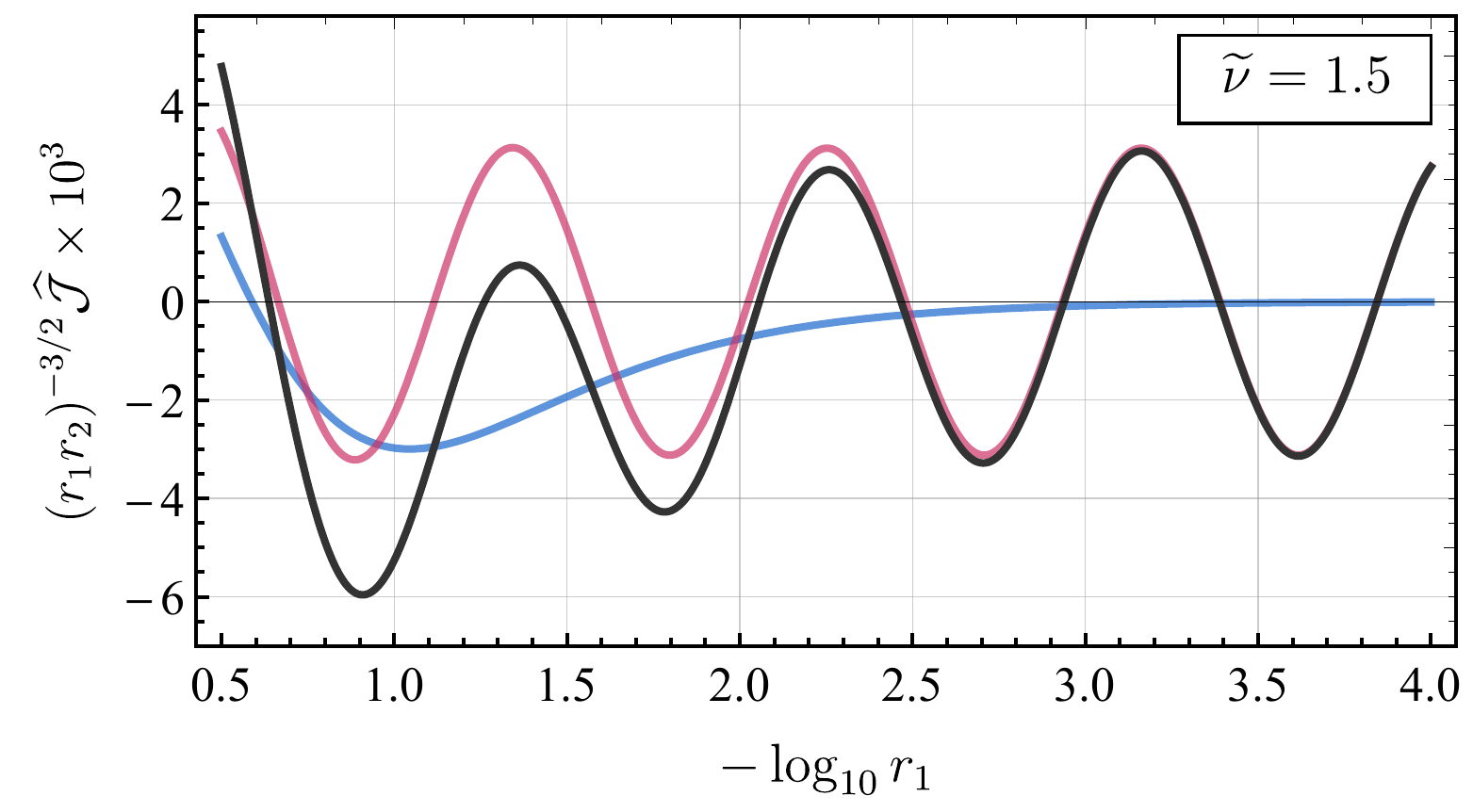} \\
  \includegraphics[width=0.49\textwidth]{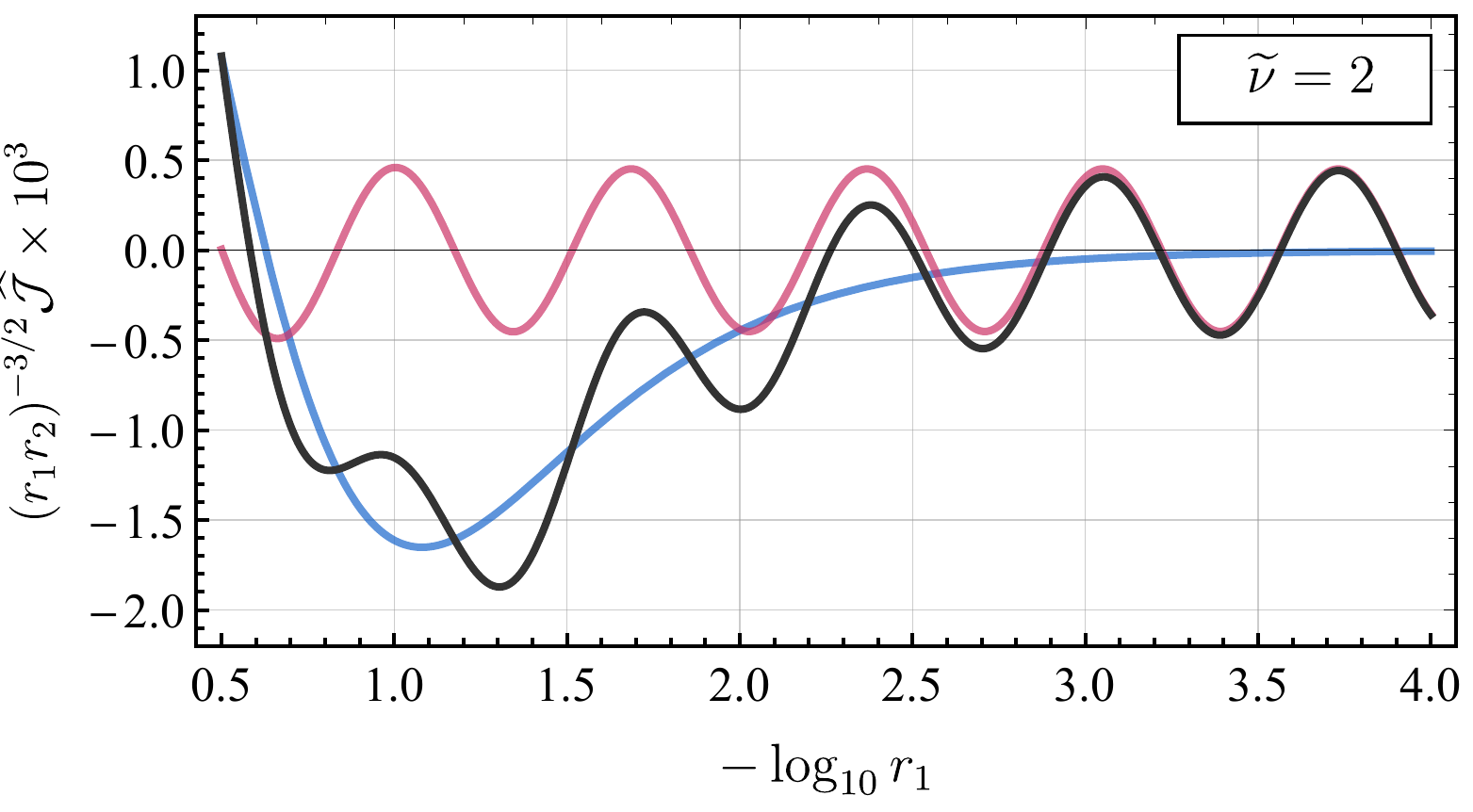} 
  \includegraphics[width=0.49\textwidth]{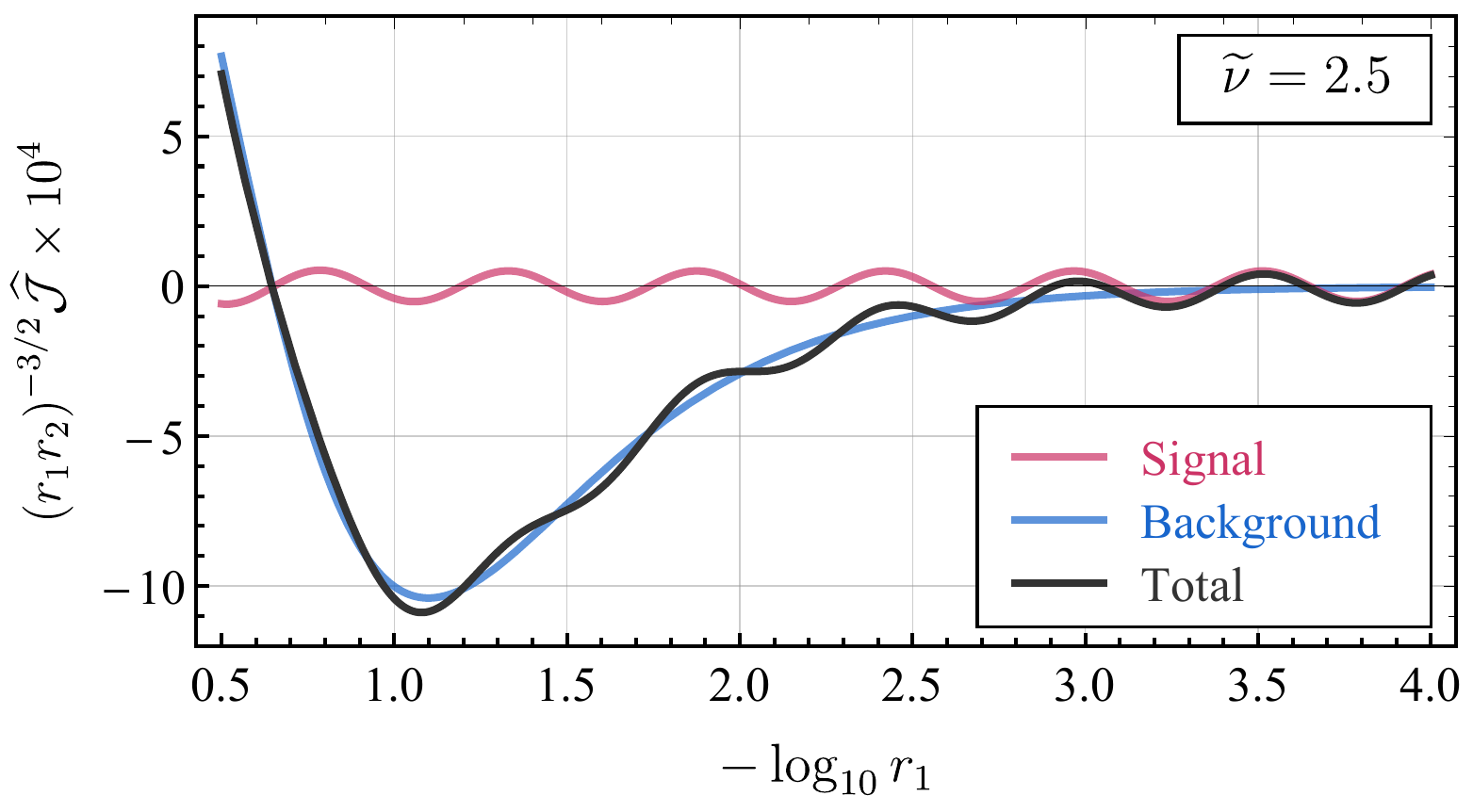} 

\caption{The 1-loop inflaton trispectra of 1-loop scalar exchange with mass parameter $\wt\nu$. The magenta, blue, and black curves show the signal $\wh{\mathcal{J}}\!_{\text{NS}}+\wh{\mathcal{J}}\!_{\text{LS}}$, the background $\wh{\mathcal{J}}\!_{\text{BG}}$, and the total result $\wh{\mathcal{J}}\!_{\text{NS}}+\wh{\mathcal{J}}\!_{\text{LS}}+\wh{\mathcal{J}}\!_{\text{BG}}$, respectively. In this figure we fix $r_2=0.9$ and vary $r_1\in(10^{-4},10^{-1/2})$. We use the dimensional regularization and the $\overline{\text{MS}}$ scheme, with the renormalization scale chosen at $\mu_R=\wt\nu H$.}
\label{fig_trispectrum} 
\end{figure}

\paragraph{Squeezed and large mass limit.} It is often useful to further take the large mass limit $\wt\nu\gg 1$ on top of the squeezed limit. In this case, we get
\begin{align}
  \lim_{\wt\nu\gg 1}\,\lim_{r_1\ll r_2\ll 1}\wh{\mathcal{J}}\!_{\text{NS}} 
  =&~\sqrt{\FR{2}{\pi}} e^{-\ii\pi/4} \wt\nu^{7/2}e^{-2\pi\wt\nu}\Big(\FR{r_1r_2}{4}\Big)^{3/2+2\ii\wt\nu}+\text{c.c.},\\
  \lim_{\wt\nu\gg 1}\,\lim_{r_1\ll r_2\ll 1}\wh{\mathcal{J}}\!_{\text{LS}} 
  =&-\sqrt{\FR{2}{\pi}}e^{+\ii\pi/4} \wt\nu^{7/2}e^{-2\pi\wt\nu}\Big(\FR{r_1}{r_2}\Big)^{3/2+2\ii\wt\nu}+\text{c.c.},\\
  \lim_{\wt\nu\gg 1}\,\lim_{r_1\ll r_2\ll 1}\wh{\mathcal{J}}\!_{\text{BG}} 
  =&-\FR{3}{2\pi^2}\bigg(\log\FR{\wt\nu^2}{\mu_R^2}+\FR{1}{\wt\nu^2}\bigg)\Big(\FR{r_1}{r_2}\Big)^{5/2}.
\end{align}
This result is simple enough and can be used for quick estimates of the signal size and observability in phenomenological CC studies. 

With the above analytical expressions, we can nicely reproduce several previously known properties of 1-loop correlators: The signals are suppressed by a factor of $e^{-2\pi\wt\nu}$ due to the doubled Boltzmann suppression of the ``on-shell'' loop mode in the large mass limit. The background piece, on the other hand, is not exponentially suppressed. Apart from a renormalization-dependent logarithmic term, the background piece scales as $1/\wt\nu^2$. At the same time, our result provides new information that cannot be revealed by a bulk computation with late-time expansion. For example, besides the exponential factor $e^{-2\pi\wt\nu}$, we have also fixed the power dependence $\wt\nu^{7/2}$ in the signals for both the nonlocal and the local signal. In particular, our result explicitly shows that the local signal, although being analytic in $k_s$, is still free from any UV divergence, a fact we should expect but very hard to see directly from the late-time expansion. Also, if we take a late-time expansion of the loop propagator, one would naively suspect that there would be ``singly oscillated'' signals proportional to $(r_1r_2)^{\pm\ii\wt\nu}$ or $(r_1/r_2)^{\pm\ii\wt\nu}$. Our results show that no such oscillatory terms appear in the final result. Therefore, if such terms are generated in any intermediate step of the late-time calculation, they must be canceled out in the final answer.

In Fig.\ \ref{fig_trispectrum}, we plot the 1-loop trispectrum in (\ref{eq_Lresult}) with the full expressions (\ref{eq_JNS00})-(\ref{eq_JBG00}) for several choices of the mass parameter $\wt\nu$. In this figure, we fix $r_2=0.9$ and vary $r_1$ on a logarithmic scale from $10^{-4}$ to $10^{-1/2}$. Here the signal represents the sum of the nonlocal signal and the local signal $\wh{\mathcal{J}}\!_{\text{NS}}+\wh{\mathcal{J}}\!_{\text{LS}}$ while the background is given by $\wh{\mathcal{J}}\!_{\text{BG}}$. The dominance of the signal in the squeezed limit (the right side of each panel) is evident in all cases.

\subsection{1-loop bispectrum}

As the second example of this section, we now look at the 3-point correlator in Fig.\ \ref{fig_1loop3pt}. With the interactions given in (\ref{eq_3ptLag}), the corresponding bispectrum $\mathcal{B}_{\varphi,\wt\nu}$ can be written as (\ref{eq_calBinf}). Then, with the explicit result for the loop seed integral (\ref{eq_Jresult}), we can compute the bispectrum $\mathcal{B}_{\varphi,\wt\nu}$ according to (\ref{eq_calBinMSbar}). 

The evaluation of (\ref{eq_calBinMSbar}) involves the folded limit $r_2\to 1^-$ in the second argument of the loop seed integral. Similar to what we would find from a tree-level process, the folded limit of the local and nonlocal signals diverges, but the divergent pieces cancel themselves. On the other hand, the background part remains finite in the $r_2\to 1$ limit. The cancelation of divergence can be checked term by term in the series expressions for $\mathcal{J}_\text{NS}$ and $\mathcal{J}_\text{LS}$. Alternatively, one can also cancel the divergence directly at the tree level, and then perform the spectral integral over the tree-level 3-point function.\footnote{See Eq.\ (256) in \cite{Qin:2022fbv} for an explicit result of tree-level 3-point function mediated by a massive scalar field.} Either way, we find the result to be:
\begin{keyeqn}
\begin{align}
  \mathcal{B}_{\varphi,\wt\nu} =\FR{1}{8k_1k_2k_3^4}\Big[\mathcal{S}_\text{S}(r)+\mathcal{S}_\text{BG}(r)\Big].
\end{align}
\end{keyeqn}
Here and below, the momentum ratio $r\equiv k_3/k_{12}$. The signal $\mathcal{S}_\text{S}(r)$ is given by the sum $\mathcal{S}_\text{S}(r)= \mathcal{J}_\text{NS}^{0,-2}(r,1^-)+\mathcal{J}_\text{LS}^{0,-2}(r,1^-)$, and the explicit result is:
\begin{align}
   \mathcal{S}_\text{S}(r)
  =&~\FR{ r^{4+2\ii\wt\nu}}{8\pi\sin(-2\pi\ii\wt\nu)}
  \sum_{n=0}^\infty (3+4\ii\wt\nu+4n)
  \FR{(1+n)_{\frac{1}{2}}(1+2\ii\wt\nu+n)_{\frac{1}{2}}}{(\fr{1}{2}+\ii\wt\nu+n)_{\frac{1}{2}}(\fr{3}{2}+\ii\wt\nu+n)_{\frac{1}{2}}}\n\\
  &~\times {}_2\mathcal{F}_1\left[\bgm
        2+\ii\wt{\nu}+n, \tfrac52+\ii\wt{\nu}+n \\
        \tfrac52+2\ii\wt{\nu}+2n
    \edm\middle| r^2\right] r^{2n}+\text{c.c.}
\end{align}
The background part is given by $\mathcal{S}_\text{BG}(r)=\mathcal{J}_{\text{BG}}^{0,-2}(r,1^-)$. This can be obtained from (\ref{eq_JBG}) by setting $p_1=0$, $p_2=-2$, $r_2=1$ and taking the limit $d\to 3$. Once again, most terms vanish in this limit, except  $\mathcal{J}_\text{(3)}$ in (\ref{eq_J3BG}) and the $n=0$ term of $\mathcal{J}_\text{(2D)}$ in (\ref{eq_J2Dresult}). For the UV divergent term, we again use the renormalized spectral function $\wh\rho_{\wt\nu}^\text{dS}(\wt\nu')$ in the $\overline{\text{MS}}$ scheme. Also, when taking $d\to 3$ limit, it is essential to know the $d=3$ result for the $\Pi_{n,d}$ function for integer and half-integer $n$, which we work out in App.\ \ref{app_Pi}. After a bit of calculation, we find the result of the background to be: 
 \begin{align}
    \mathcal{S}_\text{BG}(r) 
  =&-\frac{r^3\wt{\nu}\;\text{csch}(2\pi\wt{\nu})}{2\pi(1-r^2)^2} 
    + \sum_{\ell,m=0}^\infty\sum_{n=0}^m\frac{\left(-1\right)^{\ell+n}(\ell+1)_{2m+2}(\fr12+\ell+2n)}{2^{2m}n!(m-n)!(\fr{1}{2}+\ell+n)_{m+1}}\n\\
   &\times
    \bigg[\wh{\rho}_{\wt{\nu}}^\text{dS}\big(-\fr{\ii}2-\ii \ell-2\ii n\big)-\FR{1}{(4\pi)^2}\log\mu_R^2\bigg]r^{3+2m+\ell}.
\end{align}
Here, we also restore the renormalization scale $\mu_R$ as in the previous example. As expected, the nonlocal and local signal of the 4-point function become degenerate in the 3-point limit, and both contribute the signal piece $\mathcal{S}_\text{S}$. Therefore, we see that it is not quite right to use the late-time expansion method to compute the signal in the 3-point correlator, since this late-time calculation can only capture the signal in the $r_{1,2}\to 0$ limit. However, the actual signal in the 3-point function comes from both the nonlocal and local signals with $r_1\to 0$ but $r_2\to 1$. 

It is again useful to look at the squeezed limit $r\to 0$, where the bispectrum simplifies to:
\begin{align}
  \lim_{r\ll 1}\mathcal{S}_\text{S}(r)
  =&~ 2\csc(-2\pi\ii\wt\nu)(1+\ii\wt\nu)_{\frac{1}{2}}(1+\ii\wt\nu)_{\frac{3}{2}}\Big(\FR{r}{2}\Big)^{4+2\ii\wt\nu}+\text{c.c.},\n\\
  \lim_{r\ll 1}\mathcal{S}_\text{BG}(r)
  =&~ 2\bigg[\wh\rho_{\wt\nu}^\text{dS}(-\fr{\ii}{2})-\FR{1}{(4\pi)^2}\log\mu_R^2-\FR{\wt\nu\text{csch}(2\pi\wt\nu)}{2\pi}\bigg]r^3.
\end{align}
Here we see that, in the squeezed limit $r\to 0$, the loop signal decays as $r^4$ while the background decays as $r^3$. Therefore, the signal decays faster than the background. This shows that, if we want to look for signals from 1-loop 3-point correlators, we should look at the moderately large momentum ratio $k_{12}/k_3$ instead of the deep squeezed limit. In other words, there is no parameter space where the signals are dominant. Comparing this behavior with the single squeezed limit of the trispectrum as discussed above, we see that it may be more advantageous to look at the 4-point function if we want to discover CC signals from 1-loop processes. 

Finally, we take the large mass limit on top of the squeezed limit. The result is:
\begin{align}
  \lim_{\wt\nu\gg 1}\,\lim_{r\ll 1}\mathcal{S}_\text{S}(r)
  =&-4\ii\wt\nu^2 e^{-2\pi\wt\nu}\Big(\FR{r}{2}\Big)^{4+2\ii\wt\nu}+\text{c.c.},\\
  \lim_{\wt\nu\gg 1}\,\lim_{r\ll 1}\mathcal{S}_\text{BG}(r)
  =&~\FR{1}{8\pi^2}\bigg(\log\FR{\wt\nu^2}{\mu_R^2}+\FR{1}{25\wt\nu^2}\bigg)r^3.
\end{align}
Again, the signal is suppressed by $e^{-2\pi\wt\nu}$ and the background is suppressed by $1/\wt\nu^2$, as expected. Furthermore, we can pin down the power dependence $\wt\nu^2$ in the signal besides the exponential factor. 

\begin{figure}[t]
 \centering
  \includegraphics[width=0.49\textwidth]{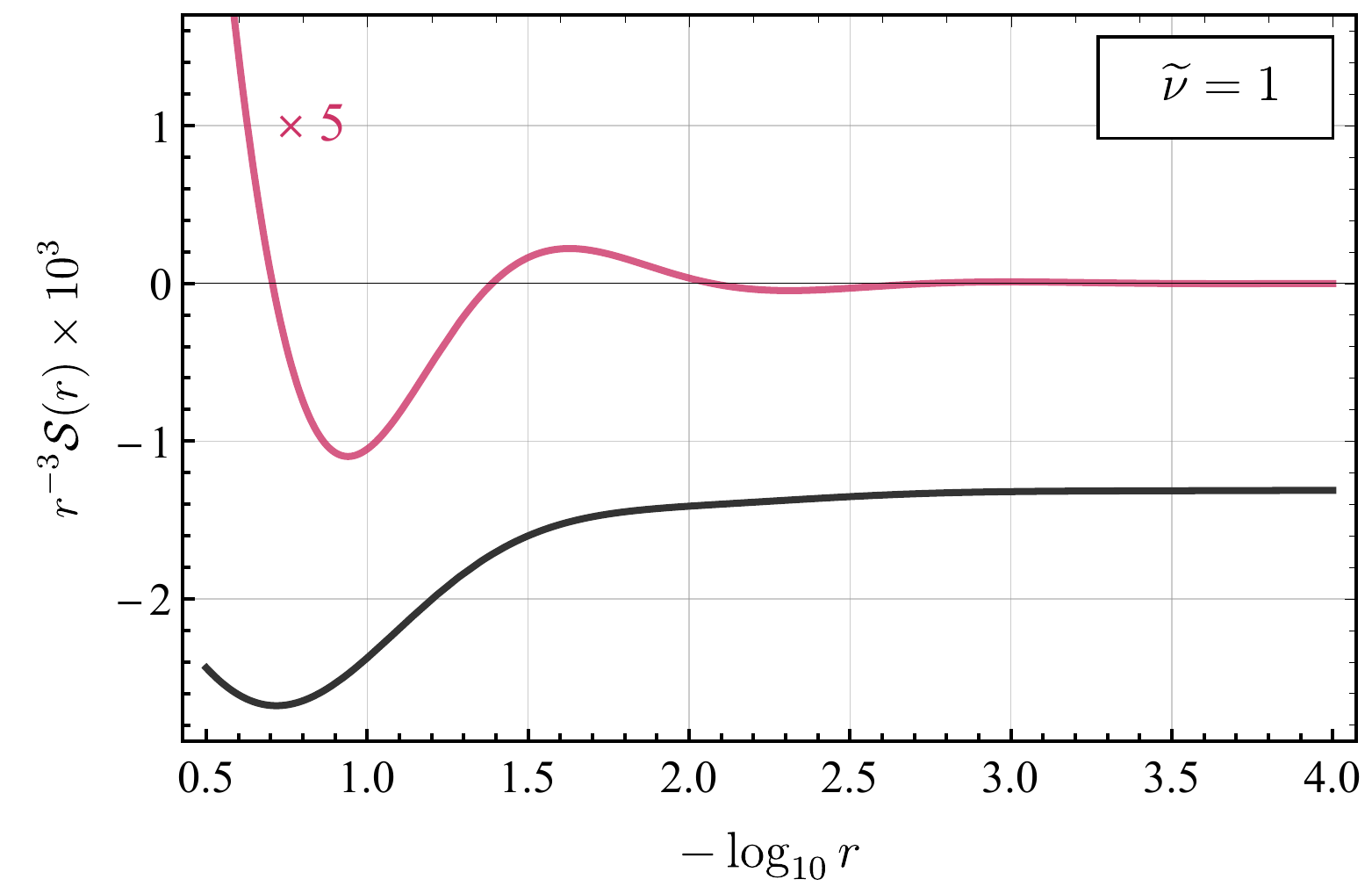} 
  \includegraphics[width=0.49\textwidth]{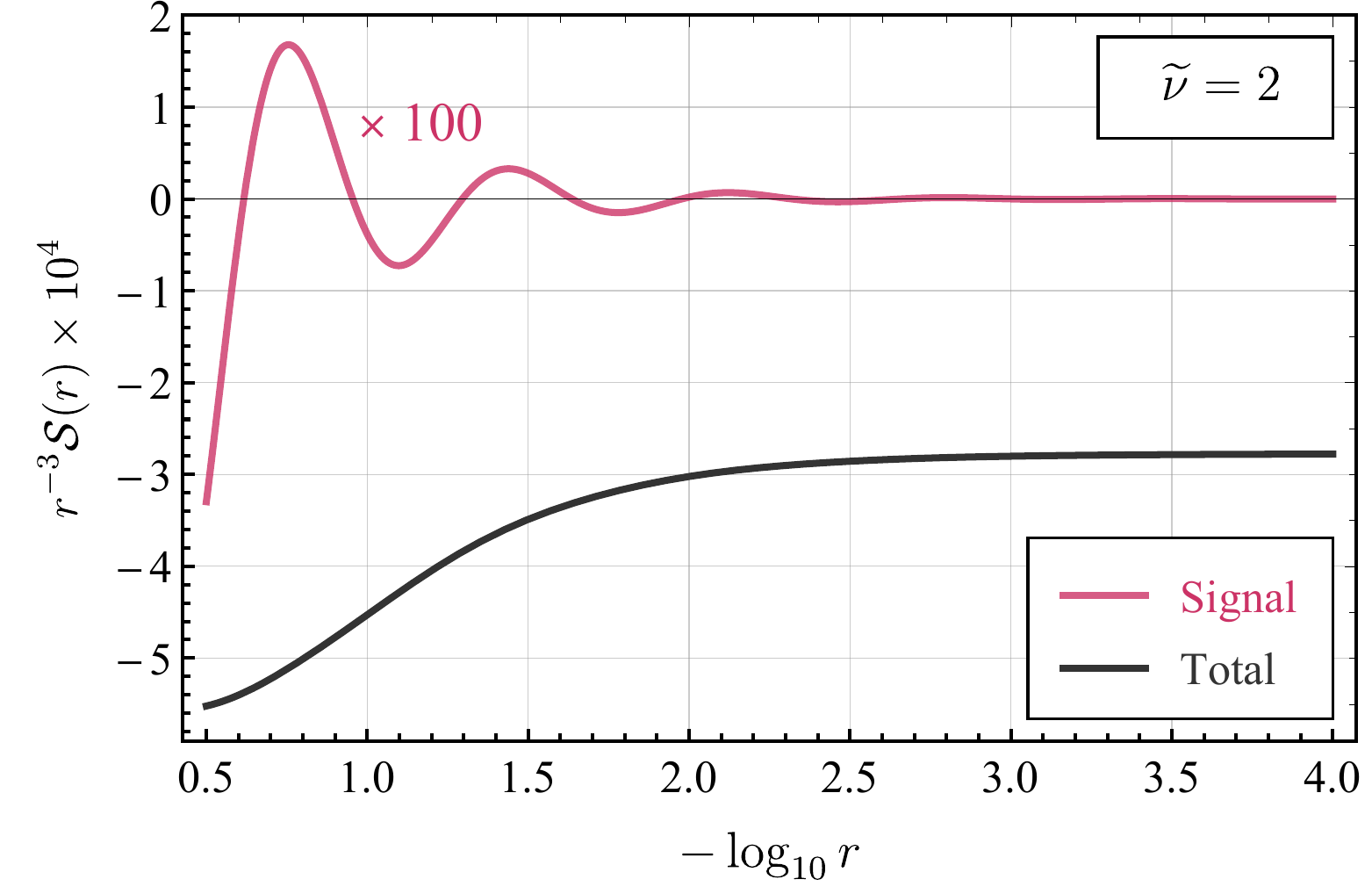} 

\caption{The 1-loop inflaton bispectra of 1-loop scalar exchange with mass parameter $\wt\nu$. The magenta and black curves show the signal and the total result, respectively. We multiply the signal by a factor of $5$ for $\wt\nu=1$ and a factor of $100$ for $\wt\nu=2$ to make it visible. We use the dimensional regularization and the $\overline{\text{MS}}$ scheme, with the renormalization scale chosen at $\mu_R=\wt\nu H$.}
\label{fig_bispectrum} 
\end{figure}

In Fig.\ \ref{fig_bispectrum}, we show two examples of the bispectrum from 1-loop scalar exchange with the mass parameter $\wt\nu=1$ and 2, respectively. Contrary to the case of the trispectrum, the signals in the bispectrum are always too small to be directly visible.  This is consistent with what was found from a full numerical approach in \cite{Wang:2021qez}. In Fig.\ \ref{fig_bispectrum}, we multiply the signals by a factor indicated in each panel to make them visible. In practice, when there is a small signal buried in the background, we can use appropriate filtering techniques to dig it out, as demonstrated in \cite{Wang:2021qez}.

\section{Conclusions and Outlooks}
\label{sec_concl}

The inflation correlators at the 1-loop order belong to a largely unexplored area. On the other hand, complete analytical results for these 1-loop correlators can help us to better understand the analytical structures of loop correlators in dS. At the same time, they are also useful for the study of cosmological collider physics. In this work, we have obtained the analytical results for a class of inflation correlators mediated by massive scalar fields at 1-loop order. We use the spectral decomposition to bootstrap the loop correlators directly from tree-level correlators, and thus bypass the difficulty of computing the loop momentum integral. By defining and computing a loop seed integral, we can obtain complete analytical expressions for many correlators with 1-loop massive exchange. We have also identified the single squeezed limit of the 4-point function as a preferred configuration for detecting CC signals from 1-loop processes, where the signals dominate over the background. 

There are several directions to be pursued along the line of this work, listed below. We shall address some of them in the next publication, and leave more open questions for future works.

\begin{itemize}

 \item In this work, we only considered intermediate states of principal series ($m>dH/2$), which is most relevant to CC physics. It is in principle straightforward to apply our method to fields of complementary series, namely $0<m<dH/2$ for scalar fields. We do not expect to see oscillatory signals in this case, and the analytic structure of the spectral integrand is also very different from the principal fields. However, we do expect to see some enhancements of the result for small mass. Therefore they could be of interest to phenomenological studies of inflation physics. See \cite{Lu:2021wxu} for an example. 
 
 \item We have assumed in this work that the masses of the two loop propagators are equal. Although this is the most encountered case, the two masses in the loop can certainly be different, and it will be interesting to generalize our method to that case as well. An interesting new feature of the 1-loop with masses $\nu_1\neq \nu_2$ is that we can have two classes of oscillatory signals, one with frequency $\wt\nu_1+\wt\nu_2$, and the other with frequency $\wt\nu_1-\wt\nu_2$. The phenomenology of unequal-mass loop processes has also been explored previously; See, e.g., \cite{Cui:2021iie}.
 
 \item It should be straightforward to generalize our method to 1-loop diagrams with nonzero spin exchanges. These are the most relevant 1-loop processes in CC physics. It is also interesting to include boost-breaking effects, such as the helicity-dependent chemical potential, the non-unit sound speed, and the slow-roll corrections. The spectral decomposition technique used in this work relies heavily on the full dS isometry. Thus it would be very interesting to see whether this method can be generalized to boost breaking scenarios.
 
 \item The spectral decomposition used in this work can in principle be applied recursively. With the 1-loop bubble correlator obtained in this work, it is in principle possible to bootstrap multi-loop diagrams, such as the sunset diagram at the 2-loop level. 
 
 \item The results presented in this work are expressed as series, or partially resummed series, in several momentum ratios $r_1=k_s/k_{12}$, $r_2=k_s/k_{12}$, and $r_1/r_2=k_{34}/k_{12}$. Although the convergence of the series is guaranteed by construction when these ratios approach 1, many individual terms in the result are superficially divergent, which is inconvenient for numerical implementation. Therefore it would be helpful if we can resum at least part of these power series, as was done in \cite{Qin:2022fbv} using the partial Mellin-Barnes representation. We leave these questions for future work. 

\end{itemize}

\paragraph{Acknowledgments.} We thank Qianshu Lu, Zhehan Qin, Matthew Reece, Lian-Tao Wang, and Yi-Ming Zhong for useful discussions. We also thank Xingang Chen and Zhehan Qin for helpful comments on a draft version of this paper. This work is supported by the National Key R\&D Program of China (2021YFC2203100), NSFC under Grant No.\ 12275146, an Open Research Fund of the Key Laboratory of Particle Astrophysics and Cosmology, Ministry of Education of China, and a Tsinghua University Initiative Scientific Research Program. 

\newpage
\begin{appendix}

\section*{Appendix}
\section{Useful Formulae}
\label{App_Formulae}

In this appendix, we collect definitions and formulae frequently used in the main text and the rest of the appendix. First, we use the following shorthand notations for the products of Euler $\Gamma$ functions:
\begin{align}
  \Gamma\left[ z_1,\cdots,z_m \right]
  \equiv&~ \Gamma(z_1)\cdots \Gamma(z_m) ,\\
  \Gamma\left[\bgm z_1,\cdots,z_m \\w_1,\cdots, w_n\edm\right]
  \equiv&~\FR{\Gamma(z_1)\cdots \Gamma(z_m)}{\Gamma(w_1)\cdots \Gamma(w_n)}.
\end{align}
The expression $\Gamma [\cdots,\{z\}^{(n)},\cdots ]$ means that the entry $z$ is repeated $n$ times in the $\Gamma$ products, namely, it represents a factor of $\Gamma^n(z)$.

The Pochhammer symbol $(z)_n$ is frequently used, which is defined by
\begin{align}
\label{eq_pochhammer}
  (z)_n\equiv\Gamma\left[\bgm z+n \\ z\edm\right].
\end{align}
In this work, we use various types of (generalized) hypergeometric functions. The original generalized hypergeometric function of $(p,q)$-type is defined by
\begin{align}
\label{eq_HGF}
  {}_p\mathrm{F}_q\left[\bgm a_1,\cdots,a_p \\ b_1,\cdots,b_q \edm  \middle| z \right]=\sum_{n=0}^\infty\FR{(a_1)_n\cdots (a_p)_n}{(b_1)_n\cdots (b_q)_n}\FR{z^n}{n!}.
\end{align}
We shall also use the regularized hypergeometric function ${}_p\wt{\mathrm{F}}_q$:
\begin{equation}
\label{eq_RegF}
    {}_p\wt{\mathrm{F}}_q\left[\begin{matrix}
        a_1, \cdots, a_p \\
        b_1, \cdots, b_q
    \end{matrix}\middle|z\right]=\frac1{\Gamma\left[b_1, \cdots, b_q\right]}{}_p\mathrm{F}_q\left[\begin{matrix}
        a_1, \cdots, a_p \\
        b_1, \cdots, b_q
    \end{matrix}\middle|z\right].
\end{equation}
The regularized hypergeometric function has the nice property that it is an entire function of all its parameters $(a_1,\cdots,a_p,b_1,\cdots,b_q)$ when $z$ is away from the singular points such as $z=1$ and $z=\infty$. 

We also frequently use the ``dressed" hypergeometric function to simplify expressions, which is defined below.
\begin{equation}
\label{eq_DressedF}
    {}_p\mathcal{F}_q\left[\begin{matrix}
        a_1, \cdots, a_p \\
        b_1, \cdots, b_q
    \end{matrix}\middle|z\right]=\Gamma\left[\begin{matrix}
        a_1, \cdots, a_p \\
        b_1, \cdots, b_q
    \end{matrix}\right]{}_p\mathrm{F}_q\left[\begin{matrix}
        a_1, \cdots, a_p \\
        b_1, \cdots, b_q
    \end{matrix}\middle|z\right].
\end{equation}
 
We occasionally need to take derivatives of the hypergeometric functions with respect to their parameters. The first derivative of the hypergeometric function ${}_p\mathrm{F}_q (a_1,\cdots,a_p; b_1,\cdots,b_q; z )$ with respect to $a_i$ and $b_i$ has been worked out in the literature. See, for instance, \cite{Ancarani:2010}:
\begin{align}
\label{eq_dFda}
    \pdd{}{a_1}{}_p\mathrm{F}_q\left[\begin{matrix}
        a_1, \cdots, a_p \\
        b_1, \cdots, b_q
    \end{matrix}\middle|z\right]=&~\FR{z}{a_1}\FR{a_1\cdots a_p}{b_1\cdots b_q}{}_p\Theta_q^{(1)}
    \left[\bgm 1,1 \\ a_1+1 \edm\middle|
    \begin{matrix}
        a_1, a_1+1, \cdots, a_p+1 \\
        2, b_1+1, \cdots, b_q+1
    \end{matrix}\middle|z, z\right], \\
  \label{eq_dFdb}
    \pdd{}{b_1}{}_p\mathrm{F}_q\left[\begin{matrix}
        a_1, \cdots, a_p \\
        b_1, \cdots, b_q
    \end{matrix}\middle|z\right]=&-\FR{z}{b_1}\FR{a_1\cdots a_p}{b_1\cdots b_q}{}_p\Theta_q^{(1)}\left[
    \bgm 1,1 \\ b_1+1 \edm \middle|
    \bgm b_1, a_1+1, \cdots, a_p+1 \\
         2, b_1+1, \cdots, b_q+1 \edm
    \middle|z, z\right], 
\end{align}
where $_p\Theta_q^{(1)}$ is a type of hypergeometric function of two variables defined as below.
\begin{align}
\label{eq_ThetaFunc}
    &{}_p^{}\Theta_q^{(1)}\left[
    \bgm \alpha_1, \alpha_2 \\
         \gamma_1 \edm \middle|
    \bgm    \beta_1, \beta_2, \cdots, \beta_{p+1} \\
         \delta_1, \delta_2, \cdots, \delta_{q+1} \edm\middle|
         z_1, z_2\right]\n \\
    \equiv&\sum_{m_1,m_2=0}^\infty
     \FR{(\al_1)_{m_1}(\al_2)_{m_2}(\be_1)_{m_1}}{(\ga_1)_{m_1}} \FR{(\beta_2)_{m_1+m_2}\cdots(\beta_{p+1})_{m_1+m_2}}{(\de_1)_{m_1}\cdots(\de_{q+1})_{m_1+m_2}} 
     \FR{z_1^{m_1}z_2^{m_2}}{m_1!m_2!}.
\end{align}

Many relations and theorems regarding hypergeometric functions are employed in this work, and we introduce them when they are needed.

\section{More on the Spectral Function in dS}
\label{App_Spectral}

The spectral function $\rho_{\wt\nu}^\text{dS}(\wt\nu')$ is of central importance in our bootstrap program for 1-loop diagrams. In this appendix, we provide more discussions about the spectral function in dS. First, for completeness, we reproduce the derivation of the bubble function $B^\text{EdS}$ in Euclidean dS, namely the $d$-dimensional sphere $S^d$. following the treatment of \cite{Marolf:2010zp}. Then, we make an analytic continuation from EdS to dS to obtain the spectral function $\rho_{\wt\nu}^\text{dS}(\wt\nu')$ in dS. Then, we present the pole structure of the spectral function on the complex $\wt\nu'$ plane, which is useful for our evaluation of the spectral integral. We then discuss the divergence of the spectral function in 3 spatial dimensions. Finally, we derive some useful properties of the $\Pi$ function defined in (\ref{eq_Pi}).

\subsection{Derivation of the spectral function}

The derivations in this subsection closely follow \cite{Marolf:2010zp}.

\paragraph{Bubble function on a sphere.}
On a $(d+1)$-dimensional sphere $S^{d+1}$, the 2-point function $D(x,y)$ of the massive scalar field can be decomposed in terms of spherical harmonics $Y_{\vec L}(x)$. Here $\vec L = (L,L_{d},L_{d-1},\cdots, L_1)$ are a vector of $d+1$ integer entries satisfying $L\geq L_d\geq\cdots\geq L_2\geq |L_1|$.
\begin{align}
  D_{\wt\nu}(x,y)=\sum_{\vec{L}}\FR{1}{\lam_{L\wt\nu}}Y_{\vec L}(x) Y_{\vec L}^*(y)=\sum_{L=0}^\infty\FR{\Gamma(d/2)(L+d/2)}{2\pi^{d/2+1}\lam_{L\wt\nu}} \mathrm{C}_L^{d/2}(Z),
\end{align}
where $\lam_{L\wt\nu}=(L+d/2)^2+\wt\nu^2$, $Z$ is the imbedding distance between $x$ and $y$, $\mathrm{C}_m^{\al}(z)$ is the Gegenbauer polynomial, and we have used the following summation formula:
\begin{align}
  \sum_{L_1,\cdots,L_{d}}Y_{\vec L}(x)Y_{\vec L}^*(y)=\FR{\Gamma(d/2)(2L+d)}{4\pi^{d/2+1}}\mathrm{C}_L^{d/2}(Z_{xy}),
\end{align}

On the other hand, the products of two propagators $D^2(x,y)$ is again a rotational invariant 2-point function. Suppose we can decompose it in the following way
\begin{align}
\label{eq_D2xySum}
  D_{\wt\nu}^2(x,y)=\sum_{\vec L}\rho_{\wt\nu}^\text{EdS}(L)Y_{\vec L}(x) Y_{\vec L}^*(y)=&\FR{\Gamma(d/2)}{2\pi^{d/2+1}}\sum_{L=0}^\infty(L+d/2)B_{\wt\nu}^\text{EdS}(L)\mathrm{C}_L^{d/2}(Z_{xy})\n\\
  =&\FR{\Gamma(d/2)}{2\pi^{d/2+1}}\sum_{L=0}^\infty(-1)^L(L+d/2)B_{\wt\nu}^\text{EdS}(L)\mathrm{C}_L^{d/2}(-Z_{xy}),
\end{align}
where $B_{\wt\nu}^\text{EdS}$ is the bubble function in the $L$ space. Using the orthogonality of the Gegenbauer polynomials: 
\begin{align}
\label{eq_GegenOrtho}
  \int_{-1}^1\di x\,(1-x^2)^{\al-1/2}\mathrm{C}_m^{\al}(x)\mathrm{C}_n^{\al}(x)=\FR{2^{1-2\al}\pi}{n+\al}\Gamma\bgb n+2\al\\n+1,\al,\al\edb\de_{mn},
\end{align}
we can find
\begin{align}
    B_{\wt\nu}^\text{EdS}(L)
  =&~(4\pi)^{d/2}\Gamma\left[\bgm L+1,d/2\\ L+d \edm\right]\int_{-1}^1\di Z\,(1-Z^2)^{(d-1)/2}\mathrm{C}_{L}^{d/2}(Z)D_{\wt\nu}^2(Z)\n\\
  =&~(4\pi)^{d/2}\Gamma\left[\bgm L+1,d/2\\ L+d \edm\right]\int_{-1}^1\di Z\,(1-Z^2)^{(d-1)/2} \n\\
  &~\times\sum_{M,N=0}^\infty\FR{\Gamma^2(d/2)(M+d/2)(N+d/2)}{4\pi^{d+2}\lam_{M\wt\nu}\lam_{N\wt\nu}}\mathrm{C}_{L}^{d/2}(Z)\mathrm{C}_{M}^{d/2}(Z)\mathrm{C}_{N}^{d/2}(Z)\n\\
  =&~\FR{4^{d/2-1}}{\pi^{d/2+2}}\Gamma\left[\bgm L+1,\{d/2\}^{(3)}\\ L+d \edm\right]\sum_{M,N=0}^\infty\FR{(M+d/2)(N+d/2)}{\lam_{M\wt\nu}\lam_{N\wt\nu}}\n\\
  &~\times\int_{-1}^1\di Z\,(1-Z^2)^{(d-1)/2} \mathrm{C}_{L}^{d/2}(Z)\mathrm{C}_{M}^{d/2}(Z)\mathrm{C}_{N}^{d/2}(Z).
\end{align}
Here we need to quote a result for the integral of three Gegenbauer polynomials of the same degree:
\begin{align}
\label{eq_IntTripGegen}
   & \int_{-1}^1\di Z\,(1-Z^2)^{\al-1/2} \mathrm{C}_{L}^{\al}(Z)\mathrm{C}_{M}^{\al}(Z)\mathrm{C}_{N}^{\al}(Z)\n\\
  =&~\left\{
  \begin{array}{l}
  2^{1-2\al}\pi\Gamma\left[\bgm K+2\al,K-L+\al,K-M+\al,K-N+\al \\ \{\al\}^{(4)},K+\al+1,K-L+1,K-M+1,K-N+1\edm\right];\\
  \text{(when $L,M,N\in \mathbb{N}_0$ satisfy the triangle inequalities and $K\equiv \fr{1}{2}(L+M+N)\in \mathbb{N}_0$)}\\
  0.~~~\text{(otherwise)}
  \end{array}\right. 
\end{align}
where $K\equiv (L+M+N)/2$.  
Using this formula, we get
\begin{align}
\label{eq_BEdSsumMid}
    B_{\wt\nu}^\text{EdS}(L)
  =&~\FR{1}{8\pi^{d/2+1}}\Gamma\left[\bgm L+1\\ L+d, \fr{d}{2} \edm\right]\sum_{(M,N)}  \Lambda_{M\wt\nu}\Lambda_{N\wt\nu}\n\\
   &~\times \Gamma\left[\bgm K+d,K-L+\fr{d}{2},K-M+\fr{d}{2},K-N+\fr{d}{2} \\K+\fr{d}{2}+1,K-L+1,K-M+1,K-N+1\edm\right],
\end{align}
where $\Lambda_{M\wt\nu}\equiv(2M+d)/\lambda_{M\wt\nu}$, and the notation $(M,N)$ means that the summation goes over all values satisfying the triangle inequality. 
The next step is to remove the constraint on the variables $(M,N)$. Following the method in \cite{Marolf:2010zp} , we define new variables:
\begin{align}
  &I=K-L, 
  &&J=K-N.
\end{align}
Then 
\begin{align}
  &M=I+J,
  &&N=I-J+L,
  &&K=I+L.
\end{align}

Then the summation becomes
\begin{align}
  \sum_{I=0}^\infty\sum_{J=0}^L\Lambda_{I+J,\wt\nu}\Lambda_{I-J+L,\wt\nu}
  \Gamma\bgb
  I+L+d,I+\fr{d}{2},L-J+\fr{d}{2},J+\fr{d}{2}\\
  I+L+1+\fr{d}{2},I+1,L-J+1,J+1
  \edb
\end{align}
First, consider the $J$ sum:
\begin{align}
  \sum_{J=0}^L\Lambda_{I+J,\wt\nu}\Lambda_{I-J+L,\wt\nu}
  \Gamma\bgb
  L-J+\fr{d}{2},J+\fr{d}{2}\\
  L-J+1,J+1
  \edb.
\end{align}
Due to the factor $\Gamma[L-J+1,J+1]$ in the denominator, the summand automatically vanishes when $J$ takes integer values in the range $J<0$ and $J>L$. Therefore, we can expand the summation range to entire real integers:
\begin{align}
  \sum_{J=-\infty}^{\infty}\Lambda_{I+J,\wt\nu}\Lambda_{I-J+L,\wt\nu}
  \Gamma\bgb
  L-J+\fr{d}{2},J+\fr{d}{2}\\
  L-J+1,J+1
  \edb.
\end{align}
Then we can rewrite this summation as an integral:
\begin{align}
  &\oint\FR{\di J}{2\pi\ii}\pi\cot(\pi J)\Lambda_{I+J,\wt\nu}\Lambda_{I-J+L,\wt\nu}
  \Gamma\bgb
  L-J+\fr{d}{2},J+\fr{d}{2}\\
  L-J+1,J+1
  \edb\n\\
  =&\oint\FR{\di J}{2\pi\ii}(-\cos\pi J)\Lambda_{I+J,\wt\nu}\Lambda_{I-J+L,\wt\nu}  \Gamma\bgb
  L-J+\fr{d}{2},J+\fr{d}{2},-J\\
  L-J+1
  \edb.
\end{align}
At large $J$, the integrand goes as $|J|^{d-4}$. So if we choose the contour to be a circle around the complex infinity, the whole integral is zero. Now, let us evaluate the same integral with the residue theorem. The integrand has the following set of poles:
\begin{enumerate}
  \item $J=0,\cdots,L$, coming from $\Gamma(-J)/\Gamma(L-J+1)$, which simply gives the original sum.
  \item $J=d/2+L+n$, $n=0,1,2,\cdots$, coming from $\Gamma(L-J+d/2)$.
  \item $J=-d/2-n$, $n=0,1,2,\cdots$, coming from $\Gamma(J+d/2)$.
  \item $J=-I-d/2\pm\ii\wt\nu$, coming from $\Lambda_{I+J,\wt\nu}$.
  \item $J=I+L+d/2\pm\ii\wt\nu$, coming from $\Lambda_{I-J+L,\wt\nu}$.
\end{enumerate}
The second set and the third set of poles give identical results. Each of the two contributions reads: 
\begin{align} 
  -\cos (\pi d/2)\sum_{n=0}^\infty \Lambda_{I+L+d/2+n,\wt\nu}\Lambda_{I-d/2-n,\wt\nu}  
  \Gamma\bgb
  n+L+d,n+\fr{d}{2}\\
  n+1,n+L+\fr{d}{2}+1
  \edb. 
\end{align}
The fourth and fifth sets of poles again give identical contributions. Each of the two sets gives:
\begin{align}
 \FR{ \pi \cos[\pi (d/2-\ii\wt\nu)]}{\sin(-\ii\pi\wt\nu)} \Lambda_{2I+L+d/2-\ii\wt\nu,\wt\nu}  \Gamma\bgb
  L+I+d-\ii\wt\nu,I+d/2-\ii\wt\nu \\
  L+I+d/2-\ii\wt\nu+1,I+1-\ii\wt\nu
  \edb+(\wt\nu\to-\wt\nu).
\end{align}
Therefore, we get:
\begin{align}
  &\sum_{J=0}^L\Lambda_{I+J,\wt\nu}\Lambda_{I-J+L,\wt\nu}
  \Gamma\bgb
  L-J+\fr{d}{2},J+\fr{d}{2}\\
  L-J+1,J+1
  \edb\n\\
  =&~\bigg\{\FR{2\pi \cos[\pi (d/2-\ii\wt\nu)]}{\sin(\ii\pi\wt\nu)} \Lambda_{2I+L+d/2-\ii\wt\nu,\wt\nu}  \Gamma\bgb
  L+I+d-\ii\wt\nu,I+d/2-\ii\wt\nu \\
  L+I+d/2-\ii\wt\nu+1,I+1-\ii\wt\nu
  \edb+(\wt\nu\to-\wt\nu)\bigg\}\n\\
  &~+2\cos (\pi d/2)\sum_{n=0}^\infty \Lambda_{I+L+d/2+n,\wt\nu}\Lambda_{I-d/2-n,\wt\nu}  
  \Gamma\bgb
  n+L+d,n+\fr{d}{2}\\
  n+1,n+L+\fr{d}{2}+1
  \edb.
\end{align}
Next, consider the $I$ sum. There are two types of contributions from the $J$ sum. First, there is a $n$ sum which contributes the following terms:
\begin{align}
  \sum_{I=0}^\infty  
  \sum_{n=0}^\infty \Lambda_{I+L+d/2+n,\wt\nu}\Lambda_{I-d/2-n,\wt\nu}  
  \Gamma\bgb
  I+L+d,I+\fr{d}{2},n+L+d,n+\fr{d}{2} \\
  I+1 ,I+L+1+\fr{d}{2} ,n+1,n+L+\fr{d}{2}+1
  \edb.
\end{align}
As observed in \cite{Marolf:2010zp}, the $\Gamma$ products and $\Lambda_{I+L+d/2+n,\wt\nu}$ is invariant under the change of variables $I\leftrightarrow n$, but $\Lambda_{I-d/2-n,\wt\nu}$ changes sign. So this part of the $I$ sum vanishes. Next, consider the rest contributions from the $J$ sum: 
\begin{align}
\label{eq_Isum}
  &\sum_{I=0}^\infty  
  \Lambda_{2I+L+d/2-\ii\wt\nu,\wt\nu}
  \Gamma\bgb
  I+L+d,I+\fr{d}{2},L+I+d-\ii\wt\nu,I+\fr{d}{2}-\ii\wt\nu \\
  I+1 ,I+L+1+\fr{d}{2},L+I+\fr{d}{2}-\ii\wt\nu+1,I+1-\ii\wt\nu
  \edb\n\\
  =&\sum_{I=0}^\infty  
  \Gamma\bgb
  I+\fr{L+d-\ii\wt\nu}{2} +1,I+\fr{L+d}{2},I+\fr{L+d}{2}-\ii\wt\nu\\
  I+\fr{L+d-\ii\wt\nu}{2},I+\fr{L+d}{2}+1,I+\fr{L+d}{2}-\ii\wt\nu+1
  \edb
  \n\\
  &\times\Gamma\bgb
  I+L+d,I+\fr{d}{2},L+I+d-\ii\wt\nu,I+\fr{d}{2}-\ii\wt\nu \\
  I+1 ,I+L+1+\fr{d}{2},L+I+\fr{d}{2}-\ii\wt\nu+1,I+1-\ii\wt\nu
  \edb\n\\
  =&~{}_7\mathcal{F}_6\left[
  \bgm
  \fr{L+d-\ii\wt\nu}{2}+1,\fr{L+d}{2},\fr{L+d}{2}-\ii\wt\nu,L+d,\fr{d}{2},L+d-\ii\wt\nu,\fr{d}{2}-\ii\wt\nu\\
  \fr{L+d-\ii\wt\nu}{2},\fr{L+d}{2}+1,\fr{L+d}{2}-\ii\wt\nu+1,L+1+\fr{d}{2},L+\fr{d}{2}-\ii\wt\nu+1,1-\ii\wt\nu
  \edm
  \middle|1\right].
\end{align}
At this point, we make use of a connection formula between two generalized hypergeometric functions \cite{Slater:1966}: 
\begin{align}
\label{eq_7F6connect1}
    &_7\mathrm{F}_6\left[\begin{matrix}
        \mb{A} \\
        \mb{B}
    \end{matrix}\middle|1\right]=\Gamma\left[\begin{matrix}
        \mb{C} \\
        \mb{D}
    \end{matrix}\right]{}_7\mathrm{F}_6\left[\begin{matrix}
        \mb{E}\\
        \mb{F}
    \end{matrix}\middle|1\right],
\end{align}
where
\begin{align}
    \mb{A}=&~\{a, \fr a2+1, b, c, e, f, g\},\n \\
    \mb{B}=&~\{\fr a2, 1+a-b, 1+a-c, 1+a-e, 1+a-f, 1+a-g\},\n \\
    \mb{C}=&~\{1+a-f, 1+a-g, 2+2a-b-c-e, 2+2a-b-c-e-f-g\}, \n\\
    \mb{D}=&~\{1+a, 1+a-f-g, 2+2a-b-c-e-f, 2+2a-b-c-e-g\}, \n\\
    \mb{E}=&~\{1+2a-b-c-e, \tfrac32+a-\tfrac b2-\tfrac c2-\tfrac e2, \n\\
           &~~\,1+a-c-e, 1+a-b-e, 1+a-b-c, f, g\}, \n\\
    \mb{F}=&~\{\tfrac12+a-\tfrac b2-\tfrac c2-\tfrac e2, 1+a-b, 1+a-c, 1+a-e, \n\\
           &~~\,2+2a-b-c-e-f, 2+2a-b-c-e-g\}.
\end{align}
Now, we can make the following assignment of parameters: 
\begin{align}
    &a=L+d-\ii \wt{\nu}, 
    &&b=\frac d2-\ii \wt{\nu}, 
    &&c=L+d, e=\frac d2, \n\\
    &f=\frac{L+d}2-\ii \wt{\nu}, 
    &&g=\frac{L+d}2.
\end{align}
Then, we can rewrite the hypergeometric function in (\ref{eq_Isum}) as:
\begin{align}
\label{eq_Isum2}
    &\sum_{I=0}^\infty  
  \Lambda_{2I+L+d/2-\ii\wt\nu,\wt\nu}
  \Gamma\bgb
  I+L+d,I+\fr{d}{2},L+I+d-\ii\wt\nu,I+\fr{d}{2}-\ii\wt\nu \\
  I+1 ,I+L+1+\fr{d}{2},L+I+\fr{d}{2}-\ii\wt\nu+1,I+1-\ii\wt\nu
  \edb\n\\
  =&~\FR12\Gamma\left[\begin{matrix}
    \tfrac d2, 2-d,\tfrac d2-\ii \wt{\nu}, L+d, 2+L-\ii \wt{\nu}, \tfrac{L+d}2-\ii \wt{\nu}, \tfrac{L+d}2 \\
    1-\ii\wt{\nu}, L+\tfrac d2+1, L+1+\tfrac d2-\ii \wt{\nu}, 2+\tfrac{L-d}2-\ii \wt{\nu}, 2+\tfrac{L-d}2
  \end{matrix}\right]\n \\
  &\times{}_7F_6\left[\begin{matrix}
  1+L-\ii \wt{\nu}, \tfrac{3+L-\ii \wt{\nu}}2, 1-\tfrac d2, 1-\ii \wt{\nu}-\tfrac d2, 1+L, \tfrac{L+d}2-\ii \wt{\nu}, \tfrac{L+d}2 \\
  \tfrac{1+L-\ii \wt{\nu}}2, L+1+\tfrac d2-\ii \wt{\nu}, 1+L+\tfrac d2, 1-\ii\wt{\nu}, 2+\tfrac{L-d}2-\ii \wt{\nu}, 2+\tfrac{L-d}2
  \end{matrix}\middle|1\right].
\end{align}
It seems that we are making the expression more complicated. The reason for using the connection formula (\ref{eq_7F6connect1}) is the following. We anticipate that the spectral function is divergent when $d=3$, which is nothing but the familiar UV divergence of the 1-loop bubble diagram. In (\ref{eq_Isum}), this divergence is somewhat hidden in the complicated parameter dependence in the hypergeometric function, making it difficult to isolate and subtract the UV divergence. After using the connection formula (\ref{eq_7F6connect1}), we see that the divergence is manifested through the $\Gamma(2-d)$ factor, which is particularly easy to deal with. 

Now, combining (\ref{eq_BEdSsumMid}) and (\ref{eq_Isum2}), we get the final expression for the bubble function $B_{\wt{\nu}}\left(L\right)$ as:
\begin{align}
    &B_{\wt\nu}^\text{EdS}\left(L\right)=\frac1{8\pi^{d/2}}\frac{\cos\left[\pi\left(d/2-\ii \wt{\nu}\right)\right]}{\sin\left(i\pi \wt{\nu}\right)}\Gamma\left[\begin{matrix}
    2-d,\tfrac d2-\ii \wt{\nu}, L+1, 2+L-\ii \wt{\nu}, \tfrac{L+d}2-\ii \wt{\nu}, \tfrac{L+d}2 \\
    1-\ii\wt{\nu}, L+\tfrac d2+1, L+1+\tfrac d2-\ii \wt{\nu}, 2+\tfrac{L-d}2-\ii \wt{\nu}, 2+\tfrac{L-d}2
  \end{matrix}\right]\n \\
  &\times{}_7F_6\left[\begin{matrix}
  1+L-\ii \wt{\nu}, \tfrac{3+L-\ii \wt{\nu}}2, 1-\tfrac d2, 1-\ii \wt{\nu}-\tfrac d2, 1+L, \tfrac{L+d}2-\ii \wt{\nu}, \tfrac{L+d}2 \\
  \tfrac{1+L-\ii \wt{\nu}}2, L+1+\tfrac d2-\ii \wt{\nu}, 1+L+\tfrac d2, 1-\ii\wt{\nu}, 2+\tfrac{L-d}2-\ii \wt{\nu}, 2+\tfrac{L-d}2
  \end{matrix}\middle|1\right]+\left(\wt{\nu}\rightarrow-\wt{\nu}\right).
\end{align}

\paragraph{From the sphere to de Sitter.} Next, we derive the spectral function $\rho_{\wt\nu}^\text{dS}(\wt\nu')$ in dS from the bubble function in EdS as obtained above. Following \cite{Marolf:2010zp}, we use a Watson-Sommerfeld transformation to recast the summation in (\ref{eq_D2xySum}) into an integral:
\begin{equation} 
\label{eq_D2xyInt}
    \Big[D_{\wt{\nu}}(x,y)\Big]^2=\int_\gamma\frac{\di L}{2\pi\ii}\frac\pi{\sin(\pi L)}\frac{\Gamma(d/2)}{2\pi^{d/2+1}}(L+d/2)B_{\wt\nu}^\text{EdS}(L)\mathrm{C}_L^{d/2}(-Z_{xy}),
\end{equation}
where the contour $\gamma$ surrounds all the poles of $1/\sin\left(\pi L\right)$ at non-negative integer values of $L$. At large $|L|\gg1$ the integrand decays like $e^{-\pi|\text{Im}L|}$, so the contour $\gamma$ can be deformed to run along the imaginary axis with $\text{Re}\left[L\right]=-d/2$. Now, using the following formula for the Gegenbauer polynomial,
\begin{equation}
    \mathrm{C}_L^{d/2}(-Z_{xy})=\FR{(d)_L}{L!}{}_2\mathrm{F}_1\left[\begin{matrix}
        -L, d+L \\
        \tfrac{d+1}2
    \end{matrix}\middle|\frac{1+Z_{xy}}2\right],
\end{equation}
we find the following relation:
\begin{equation}
    \frac{(L+d/2)\Gamma(d/2)\mathrm{C}_L^{d/2}(-Z_{xy})}{2\pi^{d/2}\sin(\pi L)}\bigg|_{L=-d/2+\ii\wt{\nu}'}=-2\ii\wt{\nu}'D_{\wt{\nu}'}(x,y).
\end{equation}
So equation (\ref{eq_D2xyInt}) can be written as
\begin{align} 
    \Big[D_{\wt{\nu}}(x,y)\Big]^2=&-\int_{-\infty+\ii(-d/2+\ep)}^{\infty+\ii(-d/2+\ep)}\di\wt{\nu}'\FR{\wt{\nu}'}{\pi\ii}B_{\wt\nu}^\text{EdS}\Big(-\FR d2+\ii\wt{\nu}'\Big)D_{\wt{\nu}'}(x,y).
\end{align}
Now, the propagator in EdS $D_{\wt{\nu}}(x,y)$ can be analytically continued to the SK propagators $D_{\wt{\nu},\mathsf{ab}}(x,y)$ in dS:
\begin{align} 
\label{eq_D2AnaCont}
    \Big[D_{\wt{\nu},\aa\bb}(x,y)\Big]^2
   =&-\int_{-\infty-\ii\epsilon}^{\infty-\ii\epsilon}\di\wt{\nu}'\FR{\wt{\nu}'}{\pi\ii}B_{\wt\nu}^\text{EdS}\Big(-\frac d2+\ii\wt{\nu}'\Big)D_{\wt{\nu},\mathsf{ab}}(x,y)+ \text{poles from }\ii\wt{\nu}'\in (0,\fr d2 ).
\end{align}
Here we are moving the integration contour from $\text{Im}\,\wt\nu'=-d/2+\ep$ to $\text{Im}\,\wt\nu'=-\ep$. Therefore, we need to include any poles in the region $\ii\wt{\nu}'\in (0,\fr d2 )$, as indicated in the last term of (\ref{eq_D2AnaCont}). Below we shall show that there is actually no pole from this region for the parameters of our interest. 

Therefore, the spectral density function $\rho_{\wt{\nu}}\left(\wt{\nu}'\right)$ in $\mathrm{dS}^d$ is the opposite of which in the $d$-dimensional sphere $S^d$, $-B_{\wt\nu}^\text{EdS}\left(-d/2+\ii\wt{\nu}'\right)$.
\bge
  \rho_{\wt\nu}^\text{dS}(\wt\nu')=-B^\text{EdS}\left(-d/2+\ii\wt{\nu}'\right).
\ede
More explicitly,
\begin{align}
\label{eq_rhoOrigF}
    \rho_{\wt{\nu}}^\text{dS}(\wt{\nu}')
    =&~ \frac1{8\pi^{d/2}}\frac{\cos[\pi(-\frac d2+\ii \wt{\nu})]}{\sin(-\pi \ii \wt{\nu})} 
     \Gamma\left[\begin{matrix}
        2-d, \tfrac d2-\ii \wt{\nu}, 1-\tfrac d2+\ii \wt{\nu}', 2-\tfrac d2+\ii \wt{\nu}'-\ii \wt{\nu}, \tfrac{\ii \wt{\nu}'-2\ii \wt{\nu}+d/2}2, \tfrac{\ii \wt{\nu}'+d/2}2 \\
        1-\ii \wt{\nu}, 1+\ii \wt{\nu}', 1+\ii \wt{\nu}'-\ii \wt{\nu}, \tfrac{4+\ii \wt{\nu}'-2\ii \wt{\nu}-3d/2}2, \tfrac{4+\ii \wt{\nu}'-3d/2}2
    \end{matrix}\right]\n \\
    &\times {}_7\mathrm{F}_6\left[\left.\begin{matrix}
        1-\tfrac d2+\ii \wt{\nu}'-\ii \wt{\nu}, \tfrac{3-d/2+\ii \wt{\nu}'-\ii \wt{\nu}}2, 1-\tfrac d2, 1-\tfrac d2-\ii \wt{\nu}, 1-\tfrac d2+\ii \wt{\nu}', \tfrac{\ii \wt{\nu}'-2\ii \wt{\nu}+d/2}2, \tfrac{\ii \wt{\nu}'+d/2}2 \\
        \tfrac{1-d/2+\ii \wt{\nu}'-\ii \wt{\nu}}2, 1+\ii \wt{\nu}'-\ii \wt{\nu}, 1+\ii \wt{\nu}', 1-\ii \wt{\nu}, \tfrac{4+\ii \wt{\nu}'-3d/2}2, \tfrac{4+\ii \wt{\nu}'-2\ii \wt{\nu}-3d/2}2
    \end{matrix}\right|1\right]\n\\
    & +(\wt\nu\to-\wt\nu).
\end{align}
To analyze the analytical property of this spectral function, it is useful to rewrite it in terms of the regularized hypergeometric function ${}_7\wt{\mathrm{F}}_6$ as defined in (\ref{eq_RegF}):
\begin{align}
\label{eq_rhoRegF}
     \rho_{\wt{\nu}}^\text{dS}(\wt{\nu}') 
    =&~\frac1{8\pi^{d/2}}\frac{\cos[\pi(-\frac d2+\ii \wt{\nu})]}{\sin(-\pi \ii \wt{\nu})} \n\\
    &\times \Gamma\left[\begin{matrix}
        2-d, \tfrac d2-\ii \wt{\nu}, 1-\tfrac d2+\ii \wt{\nu}', 2-\tfrac d2+\ii \wt{\nu}'-\ii \wt{\nu}, \tfrac{\ii \wt{\nu}'-2\ii \wt{\nu}+d/2}2, \tfrac{\ii \wt{\nu}'+d/2}2 ,\tfrac{1-d/2+\ii \wt{\nu}'-\ii \wt{\nu}}2
    \end{matrix}\right]\n \\ 
    &\times {}_7\wt{\mathrm{F}}_6\left[\left.\begin{matrix}
        1-\tfrac d2+\ii \wt{\nu}'-\ii \wt{\nu}, \tfrac{3-d/2+\ii \wt{\nu}'-\ii \wt{\nu}}2, 1-\tfrac d2, 1-\tfrac d2-\ii \wt{\nu}, 1-\tfrac d2+\ii \wt{\nu}', \tfrac{\ii \wt{\nu}'-2\ii \wt{\nu}+d/2}2, \tfrac{\ii \wt{\nu}'+d/2}2 \\
        \tfrac{1-d/2+\ii \wt{\nu}'-\ii \wt{\nu}}2, 1+\ii \wt{\nu}'-\ii \wt{\nu}, 1+\ii \wt{\nu}', 1-\ii \wt{\nu}, \tfrac{4+\ii \wt{\nu}'-3d/2}2, \tfrac{4+\ii \wt{\nu}'-2\ii \wt{\nu}-3d/2}2
    \end{matrix}\right|1\right]\n\\
    & +(\wt\nu\to-\wt\nu).
\end{align}
It is possible to further simplify this expression by absorbing a lot of Euler $\Gamma$ factors into the generalized hypergeometric function, using the dressed hypergeometric function ${}_7\mathcal{F}_6$ defined in (\ref{eq_DressedF}). This result is shown in (\ref{eq_rhodS}).

\subsection{Pole structure of the spectral function}

In the main text, we use the contour integral and the residue theorem to compute the spectral integral. Therefore, the pole structure of the spectral function $\rho_{\wt\nu}^\text{dS}(\wt\nu')$ in the complex $\wt\nu'$ plane is needed. In this subsection, we present the poles and the corresponding residues.

In the expression (\ref{eq_rhoRegF}) for $\rho_{\wt\nu}^\text{dS}(\wt\nu')$, the regulated (generalized) hypergeometric function ${}_7\wt{\mathrm{F}}_6$ is an entire function of all its $7+6$ parameters, and therefore all possible poles come from the Euler $\Gamma$ factors and sine function in the denominator. Therefore, we see that there are superficially 5 sets of poles from the Euler Gamma factors, but 3 of them are canceled by the zeros in ${}_7\wt{\mathrm{F}}_6$. As a result, we have two sets of poles at the following positions: 
\begin{align}
    &(\text{Set A})&&\wt\nu'=\ii d/2\pm2\wt{\nu}+2\ii n,\\
    &(\text{Set B})&&\wt\nu'=\ii d/2+2\ii n.
\end{align}
The residues at these poles are: 
\begin{align}
\label{eq_rhoPoleA}
     \text{Res}\Big[\ii\rho_{\wt{\nu}}^\text{dS}(\wt{\nu}')\Big]_{\wt{\nu}'= \ii d/2\mp2\wt{\nu}+2\ii n}
     =&~\frac1{8\pi^{d/2}\Gamma(\tfrac d2)}\frac{\sin[\pi(-\fr{d}{2}\mp2\ii \wt{\nu})]}{\sin^2(\pi \ii \wt{\nu})}\n\\
     &\times\FR{(1+n)_{\frac{d}{2}-1}\big[(1\pm\ii\wt\nu+n)_{\frac{d}{2}-1}\big]^2(1\pm2\ii\wt\nu+n)_{\frac{d}{2}-1}}{(1\pm2\ii\wt\nu+2n)_{d-1}},
\end{align}
%
\begin{align} 
\label{eq_rhoPoleB}
  \text{Res}\Big[\ii\rho_{\wt{\nu}}^\text{dS}(\wt{\nu}')\Big]_{\wt{\nu}'=\ii d/2+2\ii n}
  =&~\frac1{4\pi^{d/2}\Gamma(\frac{d}{2})}\frac{\sin\fr{\pi d}{2}}{\sin^2 \pi \ii\wt{\nu} }
   \FR{[(1+n)_{\frac{d}{2}-1}]^2(1+\ii\wt\nu+n)_{\frac{d}{2}-1}(1-\ii\wt\nu+n)_{\frac{d}{2}-1}}{(1+2n)_{d-1}}.
\end{align}
Because there are no poles when $\text{Im}\,\wt{\nu}'<0$, the second term of equation (\ref{eq_D2AnaCont}) disappears. 

\subsection{Spectral function in 3+1 dimensions}
\label{app_rho3plus1}

The spectral function $\rho_{\wt\nu}^\text{dS}(\wt\nu')$ in 3 spatial dimensions is of special interest to us. As mentioned above, $\rho_{\wt\nu}^\text{dS}(\wt\nu')$ is divergent when $d=3$. The divergence appears as a simple pole, which comes from the factor $\Gamma(2-d)$. Here we want to compute the residue of the spectral function at this simple pole. Therefore, we need to compute $\rho_{\wt\nu}^\text{dS}(\wt\nu')/\Gamma(2-d)$, namely, the spectral function with the factor $\Gamma(2-d)$ removed. 

As shown in (\ref{eq_rhoOrigF}), the spectral function $\rho_{\wt\nu}^\text{dS}(\wt\nu')$ consists of two parts, one explicitly in (\ref{eq_rhoOrigF}), and the other represented by  ``$\wt\nu\to-\wt\nu$.'' After removing $\Gamma(2-d)$ and then taking $d=3$, the term explicitly displayed in (\ref{eq_rhoOrigF}) reads:
\begin{align}
\label{eq_rhoWOgamma1}
    &\frac1{8\pi^{3/2}}\Gamma\left[\begin{matrix}
        \tfrac32-\ii\wt{\nu},-\tfrac12+\ii\wt{\nu}', \tfrac12+\ii \wt{\nu}'-\ii \wt{\nu}, \tfrac{\ii \wt{\nu}'-2\ii \wt{\nu}+3/2}2, \tfrac{\ii \wt{\nu}'+3/2}2 \\
        1-\ii \wt{\nu}, 1+\ii \wt{\nu}', 1+\ii \wt{\nu}'-\ii \wt{\nu}, \tfrac{\ii \wt{\nu}'-2\ii \wt{\nu}-1/2}2, \tfrac{\ii \wt{\nu}'-1/2}2
    \end{matrix}\right]\n \\
    \times&{}_7\mathrm{F}_6\left[\left.\begin{matrix}
        -\tfrac12+\ii \wt{\nu}'-\ii \wt{\nu}, \tfrac{3/2+\ii \wt{\nu}'-\ii \wt{\nu}}2, -\tfrac12, -\tfrac12-\ii \wt{\nu}, -\tfrac12+\ii \wt{\nu}', \tfrac{\ii \wt{\nu}'-2\ii \wt{\nu}+3/2}2, \tfrac{\ii \wt{\nu}'+3/2}2 \\
        \tfrac{-1/2+\ii \wt{\nu}'-\ii \wt{\nu}}2, 1+\ii \wt{\nu}'-\ii \wt{\nu}, 1+\ii \wt{\nu}', 1-\ii \wt{\nu}, \tfrac{\ii \wt{\nu}'-1/2}2, \tfrac{\ii \wt{\nu}'-2\ii \wt{\nu}-1/2}2
    \end{matrix}\right|1\right].
\end{align}
To further simplify this expression, we use the following relation \cite{Slater:1966} \footnote{The corresponding formula in \cite{Slater:1966} may have a typographical error.}:
\begin{align}
    {}_3\mathrm{F}_2\left[\begin{matrix}
        x, y, z \\
        v, w
    \end{matrix}\middle|1\right]_{n-1}
    =&~\Gamma\left[\begin{matrix}
        v+w+n-1, x+n, y+n, z+n \\
        n, y+z+n, z+x+n, x+y+n
    \end{matrix}\right]\n \\
    &\times{}_7\mathrm{F}_6\left[\begin{matrix}
        a, \tfrac12a+1, w+n-1, v+n-1, x, y, z \\
        \tfrac12a, v, w, y+z+n, z+x+n, x+y+n
    \end{matrix}\middle|1\right],
\end{align}
where $a=x+y+z+n-1$, and
\begin{equation}
    {}_p\mathrm{F}_q\left[\begin{matrix}
        a_1, a_2, \cdots, a_p \\
        b_1, b_2, \cdots, b_q
    \end{matrix}\middle|z\right]_N
    \equiv\sum_{n=0}^N\frac{(a_1)_n(a_2)_n\cdots(a_q)_n}{(b_1)_n(b_2)_n\cdots(b_q)_n}\FR{z^n}{n!}.
\end{equation}
With the following parameter assignment:
\begin{align}
    &a=-\frac12+\ii\wt{\nu}'-\ii\wt{\nu}, 
    &&v=\frac{\ii\wt{\nu}'-1/2}2, 
    &&w=\frac{\ii\wt{\nu}'-2\ii\wt{\nu}-1/2}2,\n \\
    &x=-\frac12, 
    &&y=-\frac12-\ii\wt{\nu}, 
    &&z=-\frac12+\ii\wt{\nu}', 
    &&n=2,
\end{align}
we can simplify (\ref{eq_rhoWOgamma1}) as
\begin{align}
\label{eq_rhoTerm1d3}
    &\frac1{8\pi^{3/2}}\Gamma\left[\begin{matrix}
        -\tfrac12+\ii\wt{\nu}', \tfrac{\ii \wt{\nu}'-2\ii \wt{\nu}+3/2}2, \tfrac{\ii \wt{\nu}'+3/2}2 \\
        \tfrac32, \tfrac32+\ii\wt{\nu}', \tfrac{\ii \wt{\nu}'-2\ii \wt{\nu}-1/2}2, \tfrac{\ii \wt{\nu}'-1/2}2
    \end{matrix}\right]{}_3\mathrm{F}_2\left[\begin{matrix}
        -\tfrac12, -\tfrac12-\ii\wt{\nu}, -\frac12+\ii\wt{\nu}' \\
        \tfrac{\ii \wt{\nu}'-1/2}2, \tfrac{\ii \wt{\nu}'-2\ii \wt{\nu}-1/2}2
    \end{matrix}\middle|1\right]_1=\FR 1{(4\pi)^2}.
\end{align}
The ``$\wt\nu\to-\wt\nu$'' part of $\rho_{\wt\nu}^\text{dS}(\wt\nu')/\Gamma(2-d)$ in (\ref{eq_rhoOrigF}) can be simplified in the same way when $d=3$. The result is again $1/(4\pi)^2$. With these results, we see that the whole spectral function $\rho_{\wt\nu}^\text{dS}(\wt\nu')$ diverges at $d\to 3$ as\footnote{This result is slightly different from the one given in \cite{Marolf:2010zp}, which may have a typographical error.} 
\begin{equation}
    \lim_{d\to 3}\rho_{\wt{\nu}}^\text{dS}(\wt\nu')\sim -\frac1{8\pi^2(3-d)}+\mathcal{O}\Big((3-d)^0\Big).
\end{equation}
As is familiar in flat-space loop calculations, the dimensional regularization is often coupled with the modified minimal subtraction scheme ($\overline{\text{MS}}$ scheme). This is not a physical renormalization scheme but turns out particularly convenient. Therefore, we define a ``renormalized'' spectral function $\wh\rho_{\wt{\nu}}^\text{dS}(\wt\nu')$ under the $\overline{\text{MS}}$ scheme as in (\ref{eq_rhoMSbar}). It is possible to work out an explicit expression for $\wh\rho_{\wt{\nu}}^\text{dS}(\wt\nu')$ by using the two-argument hypergeometric function. The result is rather complicated and uninspiring, but it could be useful for numerical studies. (See the end of App.\ \ref{app_asympt1} for an example.) Here we present this result for completeness. To this end, it is useful to define the following shorthand:
\begin{align}
  \vartheta(a)
  \equiv \FR{1}{a}{}_7^{}\Theta_6^{(1)}
  \left[
  \bgm 1,1 \\ a_1+1 \edm \middle|
  \bgm a,a_1+1,\cdots, a_7+1 \\ 2, b_1+1,\cdots, b_6+1 \edm \middle|
  z,z\right],
\end{align}
where the function $\displaystyle {}_p^{}\Theta_q^{(1)}$ is defined in (\ref{eq_ThetaFunc}), and the parameters $(a_1,\cdots,a_7,b_1,\cdots, b_6)$ are the same as the ones appearing in the ${}_7\mathrm{F}_6$ function in (\ref{eq_rhoOrigF}). Then:
\begin{align}
\label{eq_rhodSMSbarExplicit}
    \wh\rho_{\wt{\nu}}^\text{dS} (\wt{\nu}')
    =&~\bigg[\frac1{16\pi^2}\Big\{-1-\cot(\pi \ii\wt{\nu})+\fr{1}{2}\log(16\pi)+\fr12\Big[\psi(\fr32-\ii\wt{\nu})-\psi(-\fr12+\ii\wt{\nu}') \n \\
    &-\psi(\fr12+\ii\wt{\nu}'-\ii\wt{\nu})+\fr12\psi(\fr{\ii\wt{\nu}'-2\ii\wt{\nu}+3/2}2)+\fr12\psi(\fr{\ii\wt{\nu}'+3/2}2)+\fr32(\fr{\ii\wt{\nu}'-1/2}2)+\fr32(\fr{\ii\wt{\nu}'-2\ii\wt{\nu}-1/2}2)\Big]\Big\}\n \\
    &-\FR{(\tfrac{3/2+\ii \wt{\nu}'-\ii \wt{\nu}}2)(-\tfrac12-\ii\wt{\nu})(\tfrac{\ii \wt{\nu}'-2\ii \wt{\nu}+3/2}2)(\tfrac{\ii \wt{\nu}'+3/2}2)}{8\pi^{3/2}}\Gamma\left[\begin{matrix}
        \tfrac32-\ii\wt{\nu},\tfrac12+\ii\wt{\nu}', \tfrac12+\ii \wt{\nu}'-\ii \wt{\nu}  \\
        2-\ii \wt{\nu}, 2+\ii \wt{\nu}', 2+\ii \wt{\nu}'-\ii \wt{\nu}
    \end{matrix}\right]
    \n \\
    &\times\Big\{-\fr12\vartheta(-1/2+\ii\wt{\nu}'-\ii\wt{\nu})-\fr14\vartheta(\fr{3/2+\ii\wt{\nu}'-\ii\wt{\nu}}2)-\fr12\vartheta(-\fr12)-\fr12\vartheta(-\fr12-\ii\wt{\nu})\n \\
    &-\fr12\vartheta(-\fr12+\ii\wt{\nu}')+\fr14\vartheta(\fr{\ii \wt{\nu}'-2\ii \wt{\nu}+3/2}2)+\fr14\vartheta(\fr{\ii \wt{\nu}'+3/2}2)+\fr14\vartheta(\fr{-1/2+\ii\wt{\nu}'-\ii\wt{\nu}}2)\n \\
    &+\fr34\vartheta(\fr{\ii\wt{\nu}'-1/2}2)+\fr34\vartheta(\fr{\ii\wt{\nu}'-2\ii\wt{\nu}-1/2}2)\Big\}\bigg]+(\wt{\nu}\rightarrow-\wt{\nu}).
\end{align}

\subsection[More on the $\Pi$ function]{More on the $\mb\Pi$ function}
\label{app_Pi}

At several places in the final result for the loop seed integral, the spectral function $\rho_{\wt\nu}^\text{dS}(\wt\nu')$ appears through the combination $\Pi_{\nu',d}(\wt\nu)\equiv \rho_{\wt\nu}^\text{dS}(-\ii\nu')-\rho_{\wt\nu}^\text{dS}(+\ii\nu')$ where $\nu'=\ii\wt\nu'$ is an integer or half-integer. In this appendix, we derive an alternative expression for $\Pi_{\nu',d}$ for general $d$ which is more convenient for some purposes. We also derive a closed form formula for $\Pi_{\nu',d}$ for $d=3$, and show that $\Pi_{n,d}=0$ when $n$ is an integer.  

Our starting point here is the original expression for the spectral function (\ref{eq_rhodS}). We apply the following connecting formula for the generalized hypergeometric function ${}_7\mathrm{F}_6$ \cite{nist:dlmf}:  
\begin{align}
\label{eq_7F6connect}
    {}_7\mathrm{F}_6\left[\begin{matrix}
        \mathbf{A} \\
        \mathbf{B}
    \end{matrix}\middle| 1\right]=\Gamma\bigg[\begin{matrix}
        \mathbf{C} \\
        \mathbf{D}
    \end{matrix}\bigg]-\Gamma\bigg[\begin{matrix}
        \mathbf{E} \\
        \mathbf{F}
    \end{matrix}\bigg]{}_7\mathrm{F}_6\left[\begin{matrix}
        \mathbf{G} \\
        \mathbf{H}
    \end{matrix}\middle| 1\right],
\end{align}
where {\allowdisplaybreaks
\begin{align}
    \mathbf{A}=&~\{a, 1+\tfrac a2, b, c, e, f, g\},\n \\
    \mathbf{B}=&~\{\tfrac a2, 1+a-b, 1+a-c, 1+a-e, 1+a-f, 1+a-g\},\n \\
    \mathbf{C}=&~\{1+a-c, 1+a-e, 1+a-f, 1+a-g, b+c-a, b+e-a, b+f-a, b+g-a\},\n \\
    \mathbf{D}=&~\{1+a, b-a, 1+a-e-f, 1+a-c-f, 1+a-c-e, \n\\
    &~~\,1+a-c-g, 1+a-e-g, 1+a-f-g\},\n \\
    \mathbf{E}=&~\{1+2b-a, b+c-a, b+e-a, b+f-a, b+g-a, a-b, \n\\
    &~~\,1+a-c, 1+a-e, 1+a-f, 1+a-g\},\n \\
    \mathbf{F}=&~\{1+b-c, 1+b-e, 1+b-f, 1+b-g, b-a, 1+a, c, e, f, g\},\n \\
    \mathbf{G}=&~\{2b-a, 1+b-\tfrac a2, b, b+c-a, b+e-a, b+f-a, b+g-a\},\n \\
    \mathbf{H}=&~\{b-\tfrac a2, 1+b-a, 1+b-c, 1+b-e, 1+b-f ,1+b-g\}.
\end{align} }
The formula (\ref{eq_7F6connect}) has the effect of transforming the ${}_7\mathrm{F}_6$ function in $\rho_{\wt\nu}^\text{dS}(-\ii\wt\nu')$ into the ${}_7\mathrm{F}_6$ function in $\rho_{\wt\nu}^\text{dS}(+\ii\wt\nu')$. Therefore, we apply it to the $\rho_{\wt\nu}^\text{dS}(-\ii\wt\nu')$ term in the $\Pi$ function, through the following assignment of parameters:
\begin{align}
   &a=1-\tfrac d2+\ii\wt{\nu}'\mp\ii\wt{\nu},
   &&b=1-\tfrac d2,
   &&c=1-\tfrac d2\mp\ii\wt{\nu},\n\\
   &e=1-\tfrac d2+\ii\wt{\nu}',
   &&f=\tfrac{\ii\wt{\nu}'\mp2\ii\wt{\nu}+d/2}2,
   &&g=\tfrac{\ii\wt{\nu}'+d/2}2.
\end{align}
Meanwhile, we keep the $\rho_{\wt\nu}^\text{dS}(+\ii\wt\nu')$ term in the $\Pi$ function intact. As a result, the various generalized hypergeometric functions in the $\Pi$ function can be combined. With some further simplifications, we get the following result:
{\allowdisplaybreaks
\begin{align}
\label{eq_PinupdNew}
     \Pi_{\nu',d}(\wt\nu)  
    =&-\frac1{8\pi^{d/2}}\bigg[\frac{\cot[\fr{\pi}{2}(d-2\ii \wt{\nu})]\sin[\pi(\ii\wt{\nu}-\nu')]\sin[\fr{\pi}{2}(-\nu'-2\ii\wt{\nu}+d/2)]}{\sin(\pi \ii \wt{\nu})\sin[\fr{\pi}{2}(\nu'-2\ii\wt{\nu}+d/2)]}\n \\
    &-\frac{\cot[\fr{\pi}{2}(d+2\ii \wt{\nu})]\sin[\pi(-\ii\wt{\nu}-\nu')]\sin[\fr{\pi}{2}(-\nu'+2\ii\wt{\nu}+d/2)]}{\sin(\pi \ii \wt{\nu})\sin[\fr{\pi}{2}(\nu'+2\ii\wt{\nu}+d/2)]}\bigg]\n \\
    &\times\Gamma\Bigg[\begin{matrix}
        1-\tfrac d2-\nu', \tfrac{-\nu'+d/2}2, \tfrac{-\nu'+2\ii\wt{\nu}+d/2}2, \tfrac{-\nu'-2\ii\wt{\nu}+d/2}2, 1-\tfrac d2+\nu', \tfrac{\nu'+d/2}2 \\
        \tfrac{-\nu'+2-d/2}2, \tfrac d2, \tfrac{\nu'+2-d/2}2, \tfrac{-\nu'-2\ii\wt{\nu}+2-d/2}2, \tfrac{-\nu'+2\ii\wt{\nu}+2-d/2}2
    \end{matrix}\Bigg]\n \\
    &+\bigg\{\frac1{8\pi^{d/2}}\bigg[\frac{\cot[\fr{\pi}{2}(-d+2\ii\wt{\nu})]\sin[\fr{\pi}{2}(d+2\ii\wt{\nu})]}{\sin(\pi \ii \wt{\nu})}-\frac{\cos[\fr{\pi}{2}(d+2\ii\wt{\nu})]}{\sin(\pi \ii \wt{\nu})}\bigg]\n \\
    &\times\Gamma\Big[ 
        2-d, \tfrac d2+\ii \wt{\nu}, 1-\tfrac d2-\nu', 2-\tfrac d2-\nu'+\ii \wt{\nu}, \tfrac{-\nu'+2\ii \wt{\nu}+d/2}2, \tfrac{-\nu'+d/2}2,\tfrac{1-d/2-\nu'+\ii \wt{\nu}}2 
     \Big]\n \\
    &\times{}_7\wt{\mathrm{F}}_6\left[\begin{matrix}
        1-\tfrac d2-\nu'+\ii \wt{\nu}, \tfrac{3-d/2-\nu'+\ii \wt{\nu}}2, 1-\tfrac d2, 1-\tfrac d2+\ii \wt{\nu}, 1-\tfrac d2-\nu', \tfrac{-\nu'+2\ii \wt{\nu}+d/2}2, \tfrac{-\nu'+d/2}2 \\
        1-\nu'+\ii \wt{\nu}, 1-\nu', 1+\ii \wt{\nu}, \tfrac{4-\nu'-3d/2}2, \tfrac{4-\nu'+2\ii \wt{\nu}-3d/2}2
    \end{matrix}\middle|1\right]\n \\
    &+(\wt\nu\to-\wt\nu)\bigg\}.
\end{align} }

We are particularly interested in the case of $d=3$. For the two terms in (\ref{eq_PinupdNew}), the $d\to 3$ limit of the first term (without ${}_7\mathrm{F}_6$) is straightforward, and the result is:
\begin{align}
  -\FR1{8\pi}\FR{\sin(\pi\nu')}{\sin(\pi\nu')-\cosh(2\pi\wt{\nu})}\Gamma\Bigg[\begin{matrix}
        -\tfrac14-\fr{\nu'}{2},-\tfrac14+\fr{\nu'}{2} , \tfrac{-\nu'+2\ii\wt{\nu}+3/2}2, \tfrac{-\nu'-2\ii\wt{\nu}+3/2}2 \\
        \tfrac14-\fr{\nu'}{2}, \tfrac14+\fr{\nu'}{2}, \tfrac{-\nu'-2\ii\wt{\nu}+1/2}2, \tfrac{-\nu'+2\ii\wt{\nu}+1/2}2  \\
    \end{matrix}\Bigg].
\end{align}
More care is needed to take the $d\to 3$ limit of the second term in (\ref{eq_PinupdNew}), namely the whole term within the curly bracket. Due to the factor $\Gamma(2-d)$, this term is seemingly divergent; However, the divergence is canceled by another divergent term in the ``$(\wt\nu\to-\wt\nu)$'' term. To see this, we note that we can isolate part of terms in the curly bracket of (\ref{eq_PinupdNew}) that is identical to (\ref{eq_rhoWOgamma1}) upon the replacement $\wt\nu\to -\wt\nu$ and $\ii\wt\nu'\to -\nu'$. Therefore, we can use the same treatment for (\ref{eq_rhoWOgamma1}) to compute the $d\to 3$ limit of this curly-bracket term. Comparing with (\ref{eq_rhoTerm1d3}), we see that
\begin{align}
    & \frac1{8\pi^{d/2}}\Gamma\Big[ 
         \tfrac d2+\ii \wt{\nu}, 1-\tfrac d2-\nu', 2-\tfrac d2-\nu'+\ii \wt{\nu}, \tfrac{-\nu'+2\ii \wt{\nu}+d/2}2, \tfrac{-\nu'+d/2}2,\tfrac{1-d/2-\nu'+\ii \wt{\nu}}2 
     \Big]\n \\
    &\times{}_7\wt{\mathrm{F}}_6\left[\begin{matrix}
        1-\tfrac d2-\nu'+\ii \wt{\nu}, \tfrac{3-d/2-\nu'+\ii \wt{\nu}}2, 1-\tfrac d2, 1-\tfrac d2+\ii \wt{\nu}, 1-\tfrac d2-\nu', \tfrac{-\nu'+2\ii \wt{\nu}+d/2}2, \tfrac{-\nu'+d/2}2 \\
        1-\nu'+\ii \wt{\nu}, 1-\nu', 1+\ii \wt{\nu}, \tfrac{4-\nu'-3d/2}2, \tfrac{4-\nu'+2\ii \wt{\nu}-3d/2}2
    \end{matrix}\middle|1\right]\n\\
    &=\FR{1}{(4\pi)^2}.
\end{align}
Therefore, the whole term inside the curly bracket in (\ref{eq_PinupdNew}) vanishes when $d=3$:
\begin{align}
  &~\FR{\Gamma(2-d)}{(4\pi)^2}\bigg[\frac{\cot[\fr{\pi}{2}(-d+2\ii\wt{\nu})]\sin[\fr{\pi}{2}(d+2\ii\wt{\nu})]}{\sin(\pi \ii \wt{\nu})}-\frac{\cos[\fr{\pi}{2}(d+2\ii\wt{\nu})]}{\sin(\pi \ii \wt{\nu})}\bigg]+(\wt\nu\to -\wt\nu) =0.
\end{align}
Therefore, we have got the following explicit expression for $\Pi_{n,3}$: 
\begin{keyeqn}
\begin{align}
\label{eq_Pin3}
  \Pi_{n,3}(\wt\nu)= \FR1{8\pi}\FR{\sin(\pi n)}{\cosh(2\pi\wt{\nu})-\sin(\pi n)}
  \FR{\big(\fr{1}{4}-\fr{n}{2}-\ii\wt\nu\big)_{\frac{1}{2}}\big(\fr{1}{4}-\fr{n}{2}+\ii\wt\nu\big)_{\frac{1}{2}}}{(-\fr{1}{4}-\fr{n}{2})_{\frac{1}{2}}(-\fr{1}{4}+\fr{n}{2})_{\frac{1}{2}}}.~~~~(n\in\mathbb{R})
\end{align}
\end{keyeqn}
Note that this result applies for general $n$, which is not required to be an integer. For our computation of loop seed integral in 3 spatial dimensions, it is worth noting that $\Pi_{n,3}(\wt\nu)=0$ when $n$ is an integer. The case of half-integer $n$ is also useful but is a bit complicated. First, when $n=1/2$, $\Pi_{1/2,3}(\wt\nu)$ is finite:
\begin{equation}
    \Pi_{1/2,3}(\wt\nu)=\FR{\wt{\nu}}{4\pi\sinh(2\pi\wt{\nu})}.
\end{equation}
Next, when $n=2k+1/2$ with $k=1,2,\cdots$, $\Pi_{n,3}(\wt\nu)=0$. Finally, when $n=2k+3/2$ with $k=0,1,2,\cdots$, $\Pi_{n,3}$ is divergent like a simple pole. 

It is also straightforward to find an expression for the $\Xi_{n,3}(\wt\nu)$ function defined in (\ref{eq_Xi}). This time we require $n$ to be an integer. The result is: 
\begin{align}
\label{eq_Xin3}
  \Xi_{n,3}(\wt\nu)=\FR{(-1)^n}{8}\FR{\big(\fr{1}{4}-\fr{n}{2}-\ii\wt\nu\big)_{\frac{1}{2}}\big(\fr{1}{4}-\fr{n}{2}+\ii\wt\nu\big)_{\frac{1}{2}}}{(-\fr{1}{4}-\fr{n}{2})_{\frac{1}{2}}(-\fr{1}{4}+\fr{n}{2})_{\frac{1}{2}}}\text{sech}(2\pi\wt\nu).~~~~(n\in\mathbb{Z})
\end{align}

\section{Asymptotic Behavior of the Spectral Function}
\label{app_asympt}

In several places of the main text, we need the asymptotic behavior of the spectral function $\rho_{\wt\nu}^\text{dS}(\wt\nu')$ when either $\wt\nu$ or $\wt\nu'$ is large. We examine these two limits in this appendix. 
 
\subsection[Large $\wt\nu$ limit]{Large $\bm{\wt\nu}$ limit}
\label{app_asympt1}

The spectral function $\rho_{\wt\nu}^\text{dS}(\wt\nu')$ as given in (\ref{eq_rhodS}) contains two terms, each of which is a product of a generalized hypergeometric function ${}_7\mathrm{F}_6$ and many Euler $\Gamma$ and trigonometric functions. To find the large $\wt\nu$ limit of this function, therefore, we should take the large $\wt\nu$ limit of all these factors. 

It turns out that the leading result of the large $\wt\nu$ limit is canceled between the explicitly displayed term and the $(\wt\nu\to-\wt\nu)$ term in (\ref{eq_rhodS}). To find the leading nonvanishing result, therefore, we need to know the first two leading terms of all these functions in the large $\wt\nu$ expansion. Below, we discuss them in turn.

\paragraph{Large argument limit of the $\mb\Gamma$ ratio.} First, we shall need the large $z$ limit of the ratio of two Euler $\Gamma$ functions \cite{nist:dlmf}:  
\begin{align} 
\label{eq_GammaRatioTaylor}
    &\frac{\Gamma(z+a)}{\Gamma(z+b)}\sim z^{a-b}\sum_{n=0}^\infty\FR{1}{z^n}{a-b \choose k}\mathrm{B}_k^{(a-b+1)}(a),
\end{align}
which holds so long as $z$ is away from the negative real axis. Here $\mathrm{B}_n^{(l)}$ are generalized Bernoulli polynomials \cite{nist:dlmf}. We shall need the first two of them:
\begin{align}
    &\mathrm{G}_0(a,b)=1, 
    &&\mathrm{G}_1(a,b)=\frac12(a-b)(a+b-1).
\end{align}
With (\ref{eq_GammaRatioTaylor}), we can easily find the following expansion for the products and ratios of Euler $\Gamma$ functions in (\ref{eq_rhodS}): 
\begin{align}
\label{eq_GammaProdExpand}
    &\FR1{8\pi^{d/2}}\FR{\cos[\pi(-\fr d2+\ii \wt{\nu})]}{\sin(-\pi\ii \wt{\nu})}\Gamma\Bigg[\begin{matrix}
        2-d, \tfrac d2-\ii \wt{\nu}, 1-\tfrac d2+\ii \wt{\nu}', 2-\tfrac d2+\ii \wt{\nu}'-\ii \wt{\nu}, \tfrac{\ii \wt{\nu}'-2\ii \wt{\nu}+d/2}2, \tfrac{\ii \wt{\nu}'+d/2}2 \\
        1-\ii \wt{\nu}, 1+\ii \wt{\nu}', 1+\ii \wt{\nu}'-\ii \wt{\nu}, \tfrac{4+\ii \wt{\nu}'-2\ii \wt{\nu}-3d/2}2, \tfrac{4+\ii \wt{\nu}'-3d/2}2
    \end{matrix}\Bigg]\n \\
    \sim&-\frac{\ii}{8\pi^{d/2}}\Gamma\Bigg[\begin{matrix}
        2-d, 1-\tfrac d2+\ii\wt{\nu}', \tfrac{\ii\wt{\nu}'+d/2}2 \\
        1+\ii\wt{\nu}', \tfrac{4+\ii\wt{\nu}'-3d/2}2
    \end{matrix}\Bigg]\wt{\nu}^{d-2}+\mathcal{O}\big(\wt{\nu}^{d-4}\big).
\end{align}
As mentioned above, we need to keep the first two orders in $\wt\nu$ in the expansion. However, it turns out that the next-to-leading order term, namely the $\order{\wt\nu^{d-3}}$ term, vanishes identically. Therefore, the omitted terms in (\ref{eq_GammaProdExpand}) start from $\order{\wt\nu^{d-4}}$. 

\paragraph{Large parameter limit of the generalized hypergeometric function. } It is not so trivial to get the large $\wt\nu$ expansion of the generalized hypergeometric function in (\ref{eq_rhodS}). Here we provide some details. To avoid unnecessarily lengthy expressions, we introduce the following four groups of parameters:  
\begin{align}
\label{eq_aParSet}
    \mathbf{a}=&~\Big\{1-\tfrac d2, 1-\tfrac d2+\ii \wt{\nu}', \tfrac{\ii \wt{\nu}'+d/2}2\Big\}, \\
\label{eq_bParSet}
    \mathbf{b}(\zeta)=&~\Big\{1-\tfrac d2+\ii \wt{\nu}'-\tfrac{\ii}{\zeta}, \tfrac{3-d/2+\ii \wt{\nu}'-\ii/\zeta}2, 1-\tfrac d2-\tfrac{\ii}{\zeta}, \tfrac{\ii \wt{\nu}'-2\ii/\zeta+d/2}2\Big\}, \\
\label{eq_cParSet}
    \mathbf{c}=&~\Big\{1+\ii \wt{\nu}', \tfrac{4+\ii \wt{\nu}'-3d/2}2\Big\}, \\
\label{eq_dParSet}
    \mathbf{d}(\zeta)=&~\Big\{\tfrac{1-d/2+\ii \wt{\nu}'-\ii/\zeta}2, 1+\ii \wt{\nu}'-\tfrac{\ii}{\zeta}, 1-\tfrac{\ii}{\zeta}, \tfrac{4+\ii \wt{\nu}'-2\ii/\zeta-3d/2}2\Big\}.
\end{align}
Here and below, for a group of parameters $\mb a=\{a_1,\cdots, a_p\}$, we use the shorthand notation $(\mb a)_\ell \equiv (a_1)_\ell\cdots(a_p)_\ell$, and $(\mb a+n)_{\ell}\equiv (a_1+n)_\ell\cdots(a_p+n)_\ell$. The group labels such as $\mb a$ and $\mb b(\zeta)$ can also freely appear in the hypergeometric functions or the products of Euler $\Gamma$ functions, which should be self-explanatory.

Now, if we take $\zeta=1/\wt\nu$, then the hypergeometric function in (\ref{eq_rhodS}) can be written and expanded in the following way around $\zeta=0$:
\begin{align}
  {}_7\mathrm{F}_6\left[\bgm \mb a,\mb b(\zeta)\\ \mb c,\mb d(\zeta)\edm\middle|1\right]
  =\lim_{\zeta\to 0}{}_7\mathrm{F}_6\left[\bgm \mb a,\mb b(\zeta)\\ \mb c,\mb d(\zeta)\edm\middle|1\right]
  +\zeta\lim_{\zeta\to 0}\FR{\pd}{\pd\zeta}{}_7\mathrm{F}_6\left[\bgm \mb a,\mb b(\zeta)\\ \mb c,\mb d(\zeta)\edm\middle|1\right]+\order{\zeta^2}.
\end{align}

First, consider the $\zeta^0$-term: 
\begin{align}
    \lim_{\zeta\rightarrow0}{}_7\mathrm{F}_6\left[\begin{matrix}
        \mathbf{a}, \mathbf{b}(\zeta) \\
        \mathbf{c}, \mathbf{d}(\zeta)
    \end{matrix}\middle|1\right]={}_3\mathrm{F}_2\left[\begin{matrix}
        \mathbf{a} \\
        \mathbf{c}
    \end{matrix}\middle|1\right]=\pi^{1/2}\Gamma\Bigg[\begin{matrix}
        \tfrac{4+\ii\wt{\nu}'-3d/2}2, 1+\ii\wt{\nu}', \tfrac{3+\ii\wt{\nu}'-d/2}2 \\
        \tfrac{3-d}2, \tfrac{2+\ii\wt{\nu}'-d/2}2, 2-\tfrac d2+\ii\wt{\nu}', \tfrac{1+\ii\wt{\nu}'+d/2}2
    \end{matrix}\Bigg].
\end{align}

Then, consider the $\zeta^1$-term. Here we need to take derivatives of the hypergeometric function with respect to its parameters, as summarized in (\ref{eq_dFda}) and (\ref{eq_dFdb}). To save some space, we again use the following shorthand:
\begin{align}
  \vartheta(a)
  \equiv \FR{1}{a}{}_7^{}\Theta_6^{(1)}
  \left[
  \bgm 1,1 \\ a_1+1 \edm \middle|
  \bgm a,a_1+1,\cdots, a_7+1 \\ 2, b_1+1,\cdots, b_6+1 \edm \middle|
  z,z\right],
\end{align}
where the parameters $(a_1,\cdots,a_7,b_1,\cdots,b_6)$ are given by $(\mb a,\mb b(\zeta),\mb c,\mb d(\zeta))$, respectively.  Then,
\begin{align}
    &\pdd{}{\zeta}{}_7\mathrm{F}_6\left[\begin{matrix}
        \mathbf{a}, \mathbf{b}(\zeta) \\
        \mathbf{c}, \mathbf{d}(\zeta)
    \end{matrix}\middle|1\right]\n \\
    =&~\FR{\ii}{\zeta^2}
    \FR{(\mathbf{a})_1(\mathbf{b}(\zeta))_1}{(\mathbf{c})_1(\mathbf{d}(\zeta))_1}.
    \Big[\vartheta\Big(1-\tfrac d2+\ii \wt{\nu}'-\tfrac{\ii}{\zeta}\Big)+\fr12\vartheta\Big(\tfrac{3-d/2+\ii \wt{\nu}'-\ii/\zeta}2\Big)+\vartheta\Big(1-\tfrac d2-\tfrac{\ii}{\zeta}\Big)+\vartheta\Big(\tfrac{\ii \wt{\nu}'-2\ii/\zeta+d/2}2\Big)\n \\
    &-\fr12\vartheta\Big(\tfrac{1-d/2+\ii \wt{\nu}'-\ii/\zeta}2\Big)-\vartheta\Big(1+\ii \wt{\nu}'-\tfrac{\ii}{\zeta}\Big)-\vartheta\Big(1-\tfrac{\ii}{\zeta}\Big)-\Big(\tfrac{4+\ii \wt{\nu}'-2\ii/\zeta-3d/2}2\Big)\Big].
\end{align}
Now, we use the definition of the two-argument hypergeometric function ${}_p\Theta_q^{(1)}$ in (\ref{eq_ThetaFunc}). Then we get 
\begin{align}
    &\pdd{}{\zeta}{}_7\mathrm{F}_6\left[\begin{matrix}
        \mathbf{a}, \mathbf{b}(\zeta) \\
        \mathbf{c}, \mathbf{d}(\zeta)
    \end{matrix}\middle|1\right]\n \\
    =&~\frac{\ii}{\zeta^2}\frac{(\mathbf{a})(\mathbf{b}(\zeta))}{(\mathbf{c})(\mathbf{d}(\zeta))}\sum_{\ell_1,\ell_2=0}^\infty \frac{(1)_{\ell_1}(1)_{\ell_2}(\mathbf{a}+1)_{\ell_1+\ell_2}(\mathbf{b}(\zeta)+1)_{\ell_1+\ell_2}}{\ell_1!\ell_2!(2)_{\ell_1+\ell_2}(\mathbf{c}+1)_{\ell_1+\ell_2}(\mathbf{d}(\zeta)+1)_{\ell_1+\ell_2}}
    \bigg[\frac{(2-d/2+\ii \wt{\nu}'-\ii/\zeta)_{\ell_1-1}}{(2-d/2+\ii \wt{\nu}'-\ii/\zeta)_{\ell_1}}\n \\
    &+\frac{((5-d/2+\ii \wt{\nu}'-\ii/\zeta)/2)_{\ell_1-1}}{2((5-d/2+\ii \wt{\nu}'-\ii/\zeta)/2)_{\ell_1}}+\frac{(2-d/2-\ii/\zeta)_{\ell_1-1}}{(2-d/2-\ii/\zeta)_{\ell_1}}+\frac{((2+\ii \wt{\nu}'-2\ii/\zeta+d/2)/2)_{\ell_1-1}}{((2+\ii \wt{\nu}'-2\ii/\zeta+d/2)/2)_{\ell_1}}\n \\
    &-\frac{((3-d/2+\ii \wt{\nu}'-\ii/\zeta)/2)_{\ell_1-1}}{2((3-d/2+\ii \wt{\nu}'-\ii/\zeta)/2)_{\ell_1}}-\frac{(2+\ii \wt{\nu}'-\ii/\zeta)_{\ell_1-1}}{(2+\ii \wt{\nu}'-\ii/\zeta)_{\ell_1}}-\frac{(2-\ii/\zeta)_{\ell_1-1}}{(2-\ii/\zeta)_{\ell_1}}\n \\
    &-\frac{((6+\ii \wt{\nu}'-2\ii/\zeta-3d/2)/2)_{\ell_1-1}}{((6+\ii \wt{\nu}'-2\ii/\zeta-3d/2)/2)_{\ell_1}}\bigg].
\end{align}
With a bit of simplification, the above expression can be rewritten in the following way:
\begin{align}
    &\pdd{}{\zeta}{}_7\mathrm{F}_6\left[\begin{matrix}
        \mathbf{a}, \mathbf{b}(\zeta) \\
        \mathbf{c}, \mathbf{d}(\zeta)
    \end{matrix}\middle|1\right]     
    = \frac{\ii}{\zeta^2}\frac{(\mathbf{a})(\mathbf{b}(\zeta))}{(\mathbf{c})(\mathbf{d}(\zeta))}\sum_{\ell_1,\ell_2=0}^\infty \frac{(\mathbf{a}+1)_{\ell_1+\ell_2}(\mathbf{b}(\zeta)+1)_{\ell_1+\ell_2}}{(2)_{\ell_1+\ell_2}(\mathbf{c}+1)_{\ell_1+\ell_2}(\mathbf{d}(\zeta)+1)_{\ell_1+\ell_2}}\n \\
    &\times\bigg[\frac{d/2}{(1-d/2+\ii \wt{\nu}'-\ii/\zeta+\ell_1)(1+\ii \wt{\nu}'-\ii/\zeta+\ell_1)}+\frac{d/2}{(1-d/2-\ii/\zeta+\ell_1)(1-\ii/\zeta+\ell_1)}\n \\
    &+\frac{2-d}{((\ii \wt{\nu}'-2\ii/\zeta+d/2)/2+\ell_1)((4+\ii \wt{\nu}'-2\ii/\zeta-3d/2)/2+\ell_1)}\n \\
    &-\frac1{2((3-d/2+\ii \wt{\nu}'-\ii/\zeta)/2+\ell_1)((1-d/2+\ii \wt{\nu}'-\ii/\zeta)/2+\ell_1)}\bigg].
\end{align}
Now, if we directly take the limit $\zeta\to 0$, the above result will reduce to a ``$0\times\infty$''-type expression. To regulate this singular behavior, we introduce a positive infinitesimal parameter $\ep$ through the following redefinition of the parameter sets:
\begin{align}
\label{eq_bEpParSet}
    \mathbf{b}(\zeta)\rightarrow\mathbf{b}_\epsilon(\zeta)=&\Big\{1-\tfrac d2+\ii \wt{\nu}'-\tfrac{\ii}{\zeta}-\epsilon, \tfrac{3-d/2+\ii \wt{\nu}'-\ii/\zeta}2, 1-\tfrac d2-\tfrac{\ii}{\zeta}, \tfrac{\ii \wt{\nu}'-2\ii/\zeta+d/2}2\Big\}, \\
\label{eq_cEpParSet}
    \mathbf{c}\rightarrow\mathbf{c}_\epsilon=&\Big\{1+\ii \wt{\nu}'+\epsilon, \tfrac{4+\ii \wt{\nu}'-3d/2}2\Big\}.
\end{align}
With this $\ep$-regulator, the $\zeta\to 0$ limit can be properly taken:
\begin{align}
\label{eq_d7F6dzetaMid}
    &\lim_{\zeta\rightarrow0}\pdd{}{\zeta}{}_7\mathrm{F}_6\left[\begin{matrix}
        \mathbf{a}, \mathbf{b}(\zeta) \\
        \mathbf{c}, \mathbf{d}(\zeta)
    \end{matrix}\middle|1\right] 
    = \lim_{\epsilon\rightarrow0^+}\lim_{\zeta\rightarrow0}\pdd{}{\zeta}{}_7\mathrm{F}_6\left[\begin{matrix}
        \mathbf{a}, \mathbf{b}_\epsilon(\zeta) \\
        \mathbf{c}_\epsilon, \mathbf{d}(\zeta)
    \end{matrix}\middle|1\right] \n\\
    = &-\lim_{\epsilon\rightarrow0^+}\ii\epsilon\frac{(\mathbf{a})}{(\mathbf{c}_\epsilon)}\sum_{\ell_1,\ell_2=0}^\infty \frac{(\mathbf{a}+1)_{\ell_1+\ell_2}}{(2)_{\ell_1+\ell_2}(\mathbf{c}_\epsilon+1)_{\ell_1+\ell_2}} 
    = -\lim_{\epsilon\rightarrow0^+}\ii\epsilon\frac{(\mathbf{a})}{(\mathbf{c}_\epsilon)}\sum_{\ell=0}^\infty\frac{(\ell+1)(\mathbf{a}+1)_\ell}{(2)_\ell(\mathbf{c}_\epsilon+1)_\ell} \n\\
    = &-\lim_{\epsilon\rightarrow0^+}\ii\epsilon\frac{(\mathbf{a})}{(\mathbf{c}_\epsilon)}\sum_{\ell=0}^\infty\frac{(\mathbf{a}+1)_\ell}{\ell!(\mathbf{c}_\epsilon+1)_\ell} 
    = -\lim_{\epsilon\rightarrow0^+}\ii\epsilon\frac{(\mathbf{a})}{(\mathbf{c}_\epsilon)}{}_3\mathrm{F}_2\left[\begin{matrix}
        \mathbf{a}+1 \\
        \mathbf{c}_\epsilon+1
    \end{matrix}\middle|1\right].
\end{align}
It is possible to further simplify this result by applying a generalization of Dixon's theorem \cite{Slater:1966}:
\begin{equation}
\label{eq_DixonThm}
    {}_3\mathrm{F}_2\left[\begin{matrix}
        a, b, c \\
        e, f
    \end{matrix}\middle|1\right]=\Gamma\bigg[\begin{matrix}
        e, f, s \\
        a, b+s, c+s
    \end{matrix}\bigg]{}_3\mathrm{F}_2\left[\begin{matrix}
        e-a, f-a, s \\
        s+b, s+c
    \end{matrix}\middle|1\right],
\end{equation}
where $s=e+f-a-b-c$. Comparing the ${}_3\mathrm{F}_2$ function in the last expression of (\ref{eq_d7F6dzetaMid}) with (\ref{eq_DixonThm}), together with (\ref{eq_aParSet}) and (\ref{eq_cEpParSet}), we see that we can make the following assignment:
\begin{align}
    &a=2-\FR d2,
    &&b=2-\FR d2+\ii \wt{\nu}',
    &&c=\FR{2+\ii \wt{\nu}'+d/2}2,\n\\
    &e=2+\ii \wt{\nu}'+\epsilon,
    &&f=\FR{6+\ii \wt{\nu}'-3d/2}2,
\end{align}
then we have $s=\ep$. Using (\ref{eq_DixonThm}), we finally get the coefficient of the $\order{\wt\nu}$ term in the large $\wt\nu$ expansion: 
\begin{equation}
    \lim_{\zeta\rightarrow0}\pdd{}{\zeta}{}_7\mathrm{F}_6\left[\begin{matrix}
        \mathbf{a}, \mathbf{b}(\zeta) \\
        \mathbf{c}, \mathbf{d}(\zeta)
    \end{matrix}\middle|1\right]=-\ii\Gamma\bigg[\begin{matrix}
        \mathbf{c} \\
        \mathbf{a}
    \end{matrix}\bigg].
\end{equation}
Combining the above results, we see that the hypergeometric function has the following asymptotic behavior in the large $\wt\nu$ limit:
\begin{align}
    {}_7\mathrm{F}_6\left[\begin{matrix}
        \mathbf{a}, \mathbf{b}(\zeta) \\
        \mathbf{c}, \mathbf{d}(\zeta)
    \end{matrix}\middle|1\right]={}_3\mathrm{F}_2\left[\begin{matrix}
        \mathbf{a} \\
        \mathbf{c}
    \end{matrix}\middle|1\right]-\frac{\ii}{\wt{\nu}}\Gamma\bigg[\begin{matrix}
        \mathbf{c} \\
        \mathbf{a}
    \end{matrix}\bigg]+\mathcal{O}\Big(\wt{\nu}^{-2}\Big).
\end{align}

\paragraph{Final result.}
Therefore, the asymptotic expression of spectral density is
\begin{align}
    \lim_{\wt\nu\to \infty}\rho_{\wt{\nu}}^\text{dS}(\wt{\nu}')\sim&-\frac1{4\pi^{d/2}}\Gamma\Bigg[\begin{matrix}
        2-d, 1-\tfrac d2+\ii\wt{\nu}', \tfrac{\ii\wt{\nu}'+d/2}2, \mathbf{c} \\
        1+\ii\wt{\nu}', \tfrac{4+\ii\wt{\nu}'-3d/2}2, \mathbf{a}
    \end{matrix}\Bigg]\wt{\nu}^{d-3}+\mathcal{O}\Big(\wt{\nu}^{d-4}\Big).
\end{align}
After further simplifications, we finally get:
\begin{keyeqn}
\begin{align} 
\label{eq_rhoLargeNu}
   \lim_{\wt\nu\to \infty}\rho_{\wt{\nu}}^\text{dS}(\wt{\nu}')\sim&-\frac1{(4\pi)^{(d+1)/2}}\Gamma\bigg(\frac{3-d}2\bigg)\wt{\nu}^{d-3}+\mathcal{O}\Big(\wt{\nu}^{d-4}\Big).
\end{align}
\end{keyeqn}
In particular, when $d\to 3$, we find that the renormalized spectral function with $\overline{\text{MS}}$, as defined in (\ref{eq_rhoMSbar}), has the following asymptotic behavior in the large-mass limit:
\begin{equation}
\label{eq_rhodS3largeNu}
    \lim_{\wt\nu\to \infty}\wh{\rho}_{\wt{\nu}}^\text{dS}(\wt{\nu}')\sim  \frac1{16\pi^2} \log(\wt{\nu}^2) + \mathcal{O}(\wt\nu^{-1}).
\end{equation}

\begin{figure}[t]
 \centering
  \includegraphics[width=0.32\textwidth]{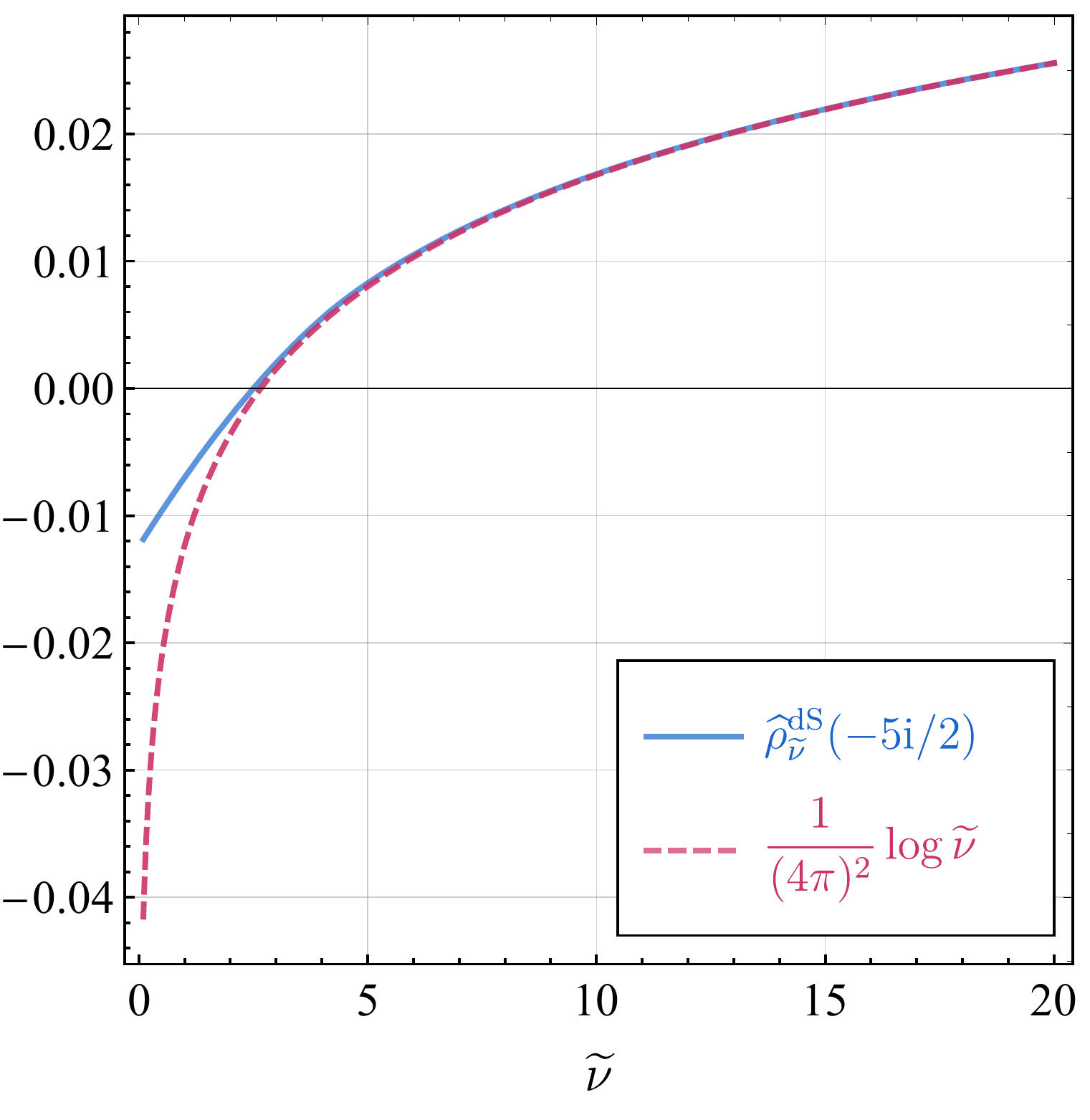} 
  \includegraphics[width=0.32\textwidth]{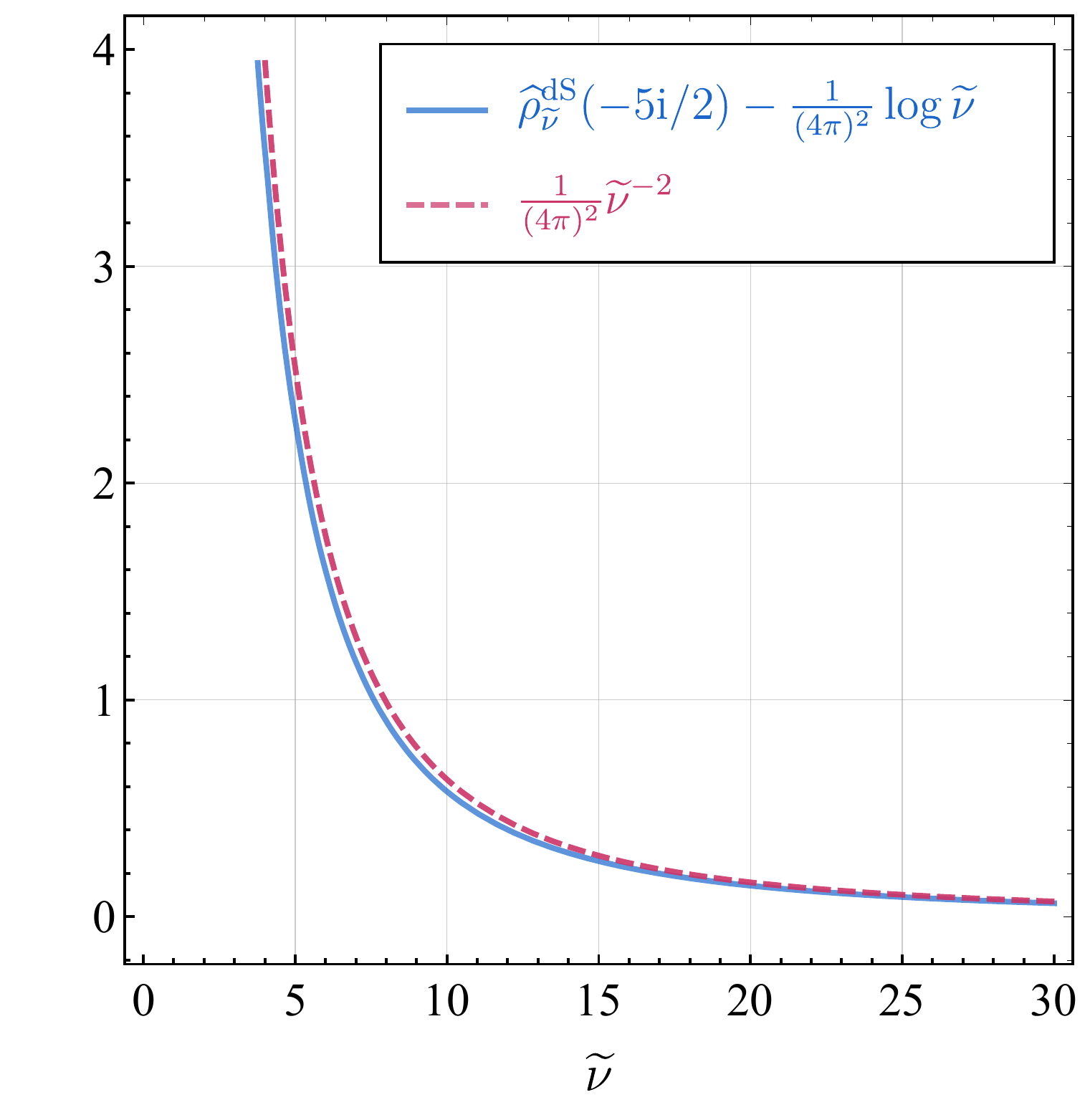} 
  \includegraphics[width=0.32\textwidth]{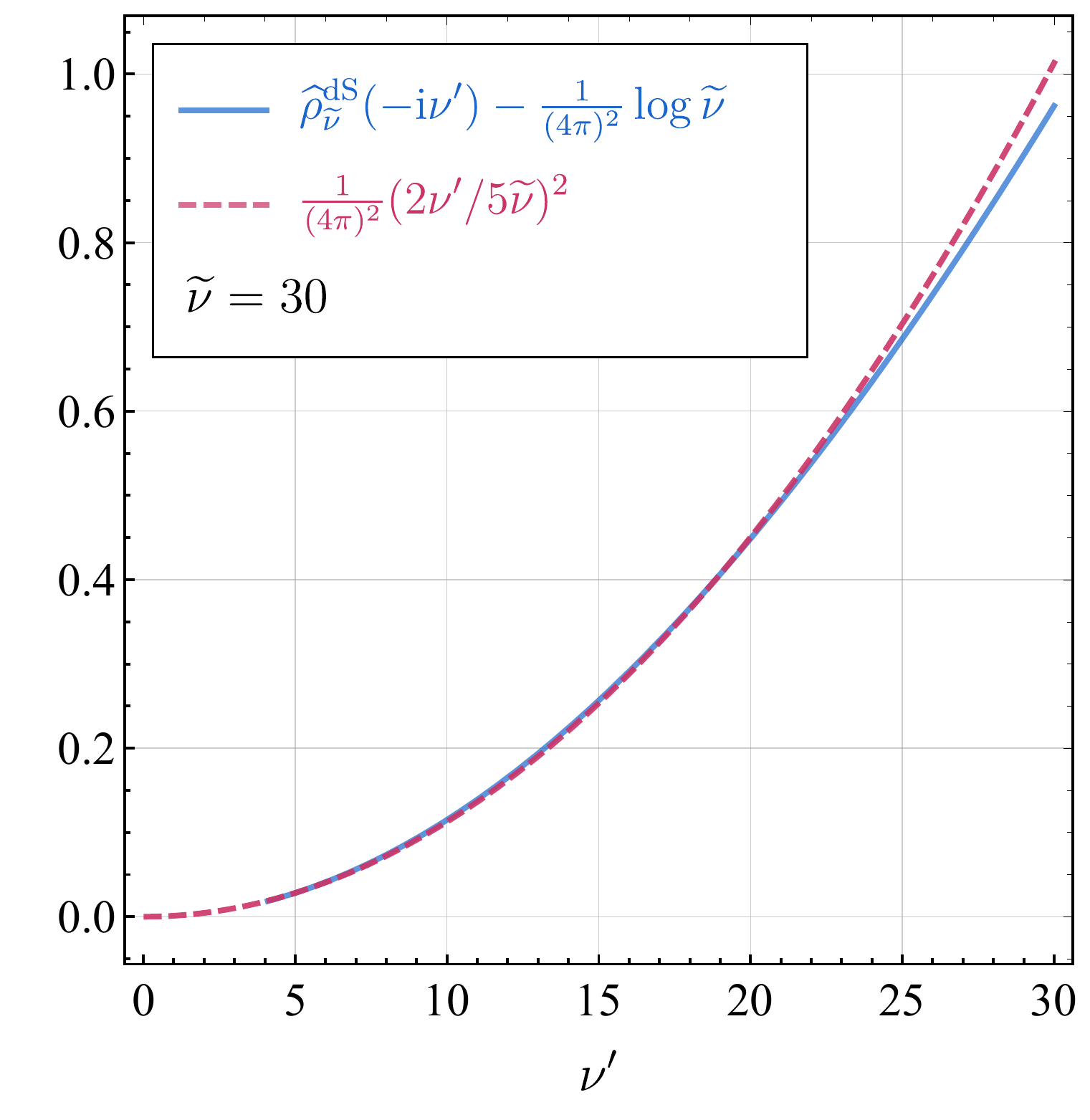} 
\caption{The renormalized spectral function $\wh{\rho}_{\wt{\nu}}^\text{dS}(-\ii {\nu'})$ in (\ref{eq_rhodSMSbarExplicit}) and its approximations in the large $\wt\nu$ limit. The left panel compares the leading (logarithmic) order behavior for a range of $\wt\nu$, the middle panel compares the subleading (inverse quadratic) order behavior, and the right panel shows the dependence on $\nu'$ for a fixed value of $\wt\nu=30$. }
\label{fig_rhoFit} 
\end{figure}

\paragraph{Numerical fit of the spectral function at $\bm{\mathcal{O}(\wt\nu^{-2})}$.} It is satisfactory to see that the leading order result (\ref{eq_rhodS3largeNu}) of the spectral function in the large $\wt\nu$ limit agrees with what we would expect from the flat space. (See App.\ \ref{app_mink}.) However, we know that this logarithmic dependence on $\wt\nu$ is from the 1-loop renormalization, and can be subtracted by choosing the renormalization scale properly. Therefore, it would useful to have the behavior of the spectral function at the next order in $\wt\nu$. The flat-space result suggests that the $\order{\wt\nu^{-1}}$ contribution should be zero, and the first nonvanishing subleading contribution should come from $\order{\wt\nu^{-2}}$. To get the coefficient of this order requires us to take the derivative of parameters of ${}_7\mathrm{F}_6$ function thrice, which is analytically nontrivial. Fortunately, it is easy to do a numerical fit from the explicit expression of $\wh\rho_{\wt\nu}^\text{dS}(\wt\nu')$ in (\ref{eq_rhodSMSbarExplicit}). For large $\wt\nu$ (and not too large $\nu'$), we find the following expression a perfect fit:
\begin{equation}
\label{eq_rhodS3largeNu2}
    \lim_{\wt\nu\to \infty}\wh{\rho}_{\wt{\nu}}^\text{dS}(-\ii {\nu'})\sim  \frac1{16\pi^2} \bigg[\log(\wt{\nu}^2) + \Big(\FR{2\nu'}{5\wt\nu}\Big)^2\bigg].
\end{equation}
In Fig.\ \ref{fig_rhoFit}, we show the full spectral function with the $\overline{\text{MS}}$ scheme against several analytical approximations. We see good asymptotic agreements between the full spectral function $\wh{\rho}_{\wt{\nu}}^\text{dS}(-\ii {\nu'})$, evaluated with the explicit formula (\ref{eq_rhodSMSbarExplicit}), and the analytical approximations given in (\ref{eq_rhodS3largeNu2}).

\subsection[Large $\wt\nu'$ limit]{Large $\bm{\wt\nu'}$ limit}
\label{app_largenup}

The large $\wt\nu'$ limit of the spectral function $\rho_{\wt{\nu}}^\text{dS}(\wt{\nu}')$ can be worked out similarly. We shall be brief. First, the large $\wt\nu'$ limit of the ${}_7\mathrm{F}_6$ function can be worked out as:
\begin{align}
    &~{}_7\mathrm{F}_6\left[\begin{matrix}
        1-\tfrac d2+\ii \wt{\nu}'-\ii \wt{\nu}, \tfrac{3-d/2+\ii \wt{\nu}'-\ii \wt{\nu}}2, 1-\tfrac d2, 1-\tfrac d2-\ii \wt{\nu}, 1-\tfrac d2+\ii \wt{\nu}', \tfrac{\ii \wt{\nu}'-2\ii \wt{\nu}+d/2}2, \tfrac{\ii \wt{\nu}'+d/2}2 \\
        \tfrac{1-d/2+\ii \wt{\nu}'-\ii \wt{\nu}}2, 1+\ii \wt{\nu}'-\ii \wt{\nu}, 1+\ii \wt{\nu}', 1-\ii \wt{\nu}, \tfrac{4+\ii \wt{\nu}'-3d/2}2, \tfrac{4+\ii \wt{\nu}'-2\ii \wt{\nu}-3d/2}2
    \end{matrix}\middle|1\right]\n \\
    \sim&~{}_2\mathrm{F}_1\left[\begin{matrix}
        1-\tfrac d2, 1-\tfrac d2-\ii \wt{\nu} \\
        1-\ii \wt{\nu}
    \end{matrix}\middle|1\right]+\mathcal{O}(\wt{\nu}'^{-1}).
\end{align}
Next, the pre-factors of the ${}_7\mathrm{F}_6$ function in (\ref{eq_rhodS}) have the following asymptotic behavior for large $\wt\nu'$:
\begin{align}
    &~\FR1{8\pi^{d/2}}\FR{\cos[\frac{\pi}{2}(-d+2\ii \wt{\nu})]}{\sin(\pi \ii \wt{\nu})}\Gamma\Bigg[\begin{matrix}
        2-d, \tfrac d2-\ii \wt{\nu}, 1-\tfrac d2+\ii \wt{\nu}', 2-\tfrac d2+\ii \wt{\nu}'-\ii \wt{\nu}, \tfrac{\ii \wt{\nu}'-2\ii \wt{\nu}+d/2}2, \tfrac{\ii \wt{\nu}'+d/2}2 \\
        1-\ii \wt{\nu}, 1+\ii \wt{\nu}', 1+\ii \wt{\nu}'-\ii \wt{\nu}, \tfrac{4+\ii \wt{\nu}'-2\ii \wt{\nu}-3d/2}2, \tfrac{4+\ii \wt{\nu}'-3d/2}2
    \end{matrix}\Bigg]\n \\
    \sim&~\FR1{2^{2d-1}\pi^{d/2}}\FR{\cos[\frac{\pi}{2}(-d+2\ii \wt{\nu})]}{\sin(\pi \ii \wt{\nu})}\Gamma\Bigg[\begin{matrix}
        2-d, \tfrac d2-\ii \wt{\nu} \\
        1-\ii \wt{\nu}
    \end{matrix}\Bigg](\ii\wt{\nu}')^{d-3}+\mathcal{O}(\wt{\nu}'^{d-4}).
\end{align}
Combining the above two results, we conclude that the spectral density $\rho_{\wt{\nu}}^\text{dS}(\wt{\nu}')\sim(\wt{\nu}')^{d-3}$ when $|\wt{\nu}'|\gg1$.

\section{Tree Seed Integral in Arbitrary Dimensions}
\label{App_TreeSeed}

The 4-point function with tree-level massive exchange is one of the ingredients for our bootstrap program of 1-loop correlators. To implement the dimensional regularization, we need the result for the tree-level correlator in general $d$ spatial dimensions. In this appendix, we compute the tree seed integral defined in (\ref{eq_TreeSeed}) for arbitrary $(p_1,p_2,d)$.

We shall use the partial Mellin-Barnes representation to compute the integral and we will compute the four terms in the summation in (\ref{eq_TreeSeed}) one by one. The computation is largely in parallel with the one given in \cite{Qin:2022fbv}, in which the readers can find more details about the method. Here we shall be brief. 

The four terms in the summation in (\ref{eq_TreeSeed}) are given by: (We omit the label for the mass $\wt\nu$.)
\begin{equation}
\label{eq_Iab}
    \mathcal{I}_{\mathsf{ab}}^{p_1p_2}(r_1,r_2)
    \equiv-\mathsf{ab}\, k_s^{d+2+p_1+p_2}\int_{-\infty}^0\di\tau_1\di\tau_2\left(-\tau_1\right)^{p_1}\left(-\tau_2\right)^{p_2}\text{e}^{\ii\mathsf{a} k_{12}\tau_1+\ii\mathsf{b} k_{34}\tau_2}D_{\mathsf{a}\mathsf{b}}\left(k_s;\tau_1,\tau_2\right).
\end{equation}
In $d$ spatial dimensions, the Mellin-Barnes representation for the two homogeneous propagators in (\ref{eq_Dgreater}) are:
\begin{align}
    D_{\gtrless}\left(k_s;\tau_1,\tau_2\right)
    =&~\frac{1}{4\pi}\int_{-\ii\infty}^{+\ii\infty}\frac{\di s_1}{2\pi\ii}\frac{\di s_2}{2\pi\ii}\text{e}^{\pm \ii\pi\left(s_1-s_2\right)}\Big(\frac{k_s}2\Big)^{-2s_{12}}\left(-\tau_1\right)^{-2s_1+d/2}\left(-\tau_2\right)^{-2s_2+d/2}\n \\
    &~\times\Gamma\bigg[s_1-\frac{\ii\wt{\nu}}2,s_1+\frac{\ii\wt{\nu}}2,s_2-\frac{\ii\wt{\nu}}2,s_2+\frac{\ii\wt{\nu}}2\bigg].
\end{align}  
Together with (\ref{eq_Dpmpm}) and (\ref{eq_Dpmmp}), we can finish the time integrals in the seed integral. For the two opposite-sign integrals $\mathcal{I}_{\mp\pm}^{p_1p_2}$, we have 
\begin{align}
    &\mathcal{I}_{\mp\pm}^{p_1p_2}(r_1,r_2)
    =\frac{1}{4\pi}\text{e}^{\pm\ii\pi\left(p_1-p_2\right)/2}r_1^{d/2+1+p_1}r_2^{d/2+1+p_2}
    \int_{-\ii\infty}^{+\ii\infty}\frac{\di s_1}{2\pi\ii}\frac{\di s_2}{2\pi\ii}\left(\frac{r_1}2\right)^{-2s_1}\left(\frac{r_2}2\right)^{-2s_2}\n \\
    &\times\Gamma\Big[s_1-\fr{\ii\wt{\nu}}{2},s_1+\fr{\ii\wt{\nu}}{2},s_2-\fr{\ii\wt{\nu}}{2},s_2+\fr{\ii\wt{\nu}}{2}, 
    -2s_1+\fr d2+1+p_1,-2s_2+\fr d2+1+p_2\Big].
\end{align}
The two same-sign integrals $\mathcal{I}_{\pm\pm}^{p_1p_2}$ can be separated into two parts, one called the ``factorized'' (F) and the other called the ``time-ordered'' (TO): (See \cite{Qin:2022fbv} for the details.)
\bge
  \mathcal{I}_{\pm\pm}^{p_1p_2}(r_1,r_2)
  =\mathcal{I}_{\pm\pm,\text{F},>}^{p_1p_2}(r_1,r_2)+\mathcal{I}_{\pm\pm,\text{TO},>}^{p_1p_2}(r_1,r_2).~~~~(r_1<r_2)
\ede
More explicitly, 
\begin{align}
   &\mathcal{I}_{\pm\pm,\text{F},>}^{p_1p_2}(r_1,r_2)
   = \frac{\text{e}^{-\ii\pi (p_{12}+d )/2}}{4\pi} r_1^{d/2+1+p_1}r_2^{d/2+1+p_2}\int_{-\ii\infty}^{\ii\infty}\frac{\di s_1}{2\pi\ii}\frac{\di s_2}{2\pi\ii}
   \text{e}^{2\ii\pi s_1 }\Big(\frac{r_1}2\Big)^{-2s_1}\Big(\frac{r_2}2\Big)^{-2s_2} \n \\
    &\times  
    \Gamma\Big[s_1-\fr{\ii\wt{\nu}}{2},s_1+\fr{\ii\wt{\nu}}{2},s_2-\fr{\ii\wt{\nu}}{2},s_2+\fr{\ii\wt{\nu}}{2},-2s_1+\fr d2+1+p_1,-2s_2+\fr d2+1+p_2\Big],
\end{align}
and,
\begin{align}
   &\mathcal{I}_{\pm\pm,\text{TO},>}^{p_1p_2}(r_1,r_2)
   =-\frac{\text{e}^{\mp\ii\pi (p_{12}+d )/2}}{4\pi}r_1^{d+2+p_{12}} \int_{-\ii\infty}^{\ii\infty}\frac{\di s_1}{2\pi\ii}\frac{\di s_2}{2\pi\ii}
    \Gamma\Big[s_1-\fr{\ii\wt{\nu}}{2},s_1+\fr{\ii\wt{\nu}}{2},s_2-\fr{\ii\wt{\nu}}{2},s_2+\fr{\ii\wt{\nu}}{2}\Big]\n \\
    &\times 
    (\text{e}^{\pm2\ii\pi s_1}-\text{e}^{\pm2\ii\pi s_2}) \Big(\frac{r_1}2\Big)^{-2s_{12}}  
       {}_2\mathcal{F}_1\left[\bgm
         -2s_2+\fr d2+p_2+1, -2s_{12}+d+p_{12}+2 \\
         -2s_2+\fr d2+p_2+2
    \edm\middle|-\FR{r_1}{r_2}\right] .
\end{align}
 
To finish the integration over the Mellin variables $s_1$ and $s_2$, we close the contours from the left, since we are assuming $r_1<r_2<1$. Then, the enclosed poles can be classified into two groups: 
\begin{align}
        & s_1=-n_1\mp\frac{\ii\wt{\nu}}2,\ &&s_2=-n_2\pm\frac{\ii\wt{\nu}}2, \\
        & s_1=-n_1\mp\frac{\ii\wt{\nu}}2,\ &&s_2=-n_2\mp\frac{\ii\wt{\nu}}2.  
\end{align}
Using the residue theorem for all these poles, and also summing over the four integrals in (\ref{eq_Iab}), we can find the final result for the tree seed integral. As shown in  (\ref{eq_Itotal}), the tree seed integral can be broken into three parts:
\begin{equation}
    \sum_{\mathsf{a},\mathsf{b}=\pm}\mathcal{I}_{\mathsf{ab}}^{p_1p_2}(r_1,r_2)=\mathcal{I}_{\text{NL},\wt\nu}^{p_1p_2}(r_1,r_2)+\mathcal{I}_{\text{L},\wt\nu}^{p_1p_2}(r_1,r_2)+\mathcal{I}_{\text{BG},\wt\nu}^{p_1p_2}(r_1,r_2).
\end{equation}
The three pieces are respectively given by:
\begin{align}
    \mathcal{I}_{\text{NL},\wt\nu}^{p_1p_2}
    =&~\frac1{4\pi}r_1^{d/2+1+p_1}r_2^{d/2+1+p_2}\sum_{n_1,n_2=0}^\infty \frac{ (-1 )^{n_{12}}}{n_1!n_2!}\Big[\text{e}^{\ii\pi\overline p_{12}/2}+\text{e}^{ \pi   \wt{\nu}-\ii\pi(p_{12}+d )/2 }\Big]\Big(\frac{r_1}2\Big)^{2n_1+\ii \wt{\nu}}\Big(\frac{r_2}2\Big)^{2n_2+\ii \wt{\nu}}\n \\ 
   &  \times\Gamma\Big[2n_1+\fr d2+1+p_1+\ii \wt{\nu},2n_2+\fr d2+1+p_2+\ii \wt{\nu},-n_1-\ii \wt{\nu},-n_2-\ii \wt{\nu}\Big] +\text{c.c.}\n \\
    =&~\frac1{2\pi}\Big[\cos \fr{\pi\overline p_{12}}2 +\cos \fr{\pi(2\ii \wt{\nu}+p_{12}+d)}{2}\Big]r_1^{d/2+1+p_1}r_2^{d/2+1+p_2}\Big(\frac{r_1r_2}4\Big)^{\ii \wt{\nu}}\n \\
    &\times\Gamma\Big[\fr d2+1+p_1+\ii \wt{\nu},\fr d2+1+p_2+\ii \wt{\nu},-\ii \wt{\nu},-\ii \wt{\nu}\Big]\n \\
    &\times{}_2F_1\left[ \begin{matrix}
        \fr {d+2p_1+2}{4}+\tfrac{\ii \wt{\nu}}2, \fr{d+2p_1+4}{4}+\tfrac{\ii \wt{\nu}}2 \\
        1+\ii \wt{\nu} 
    \end{matrix}\middle|r_1^2\right]
    {}_2F_1\left[ \begin{matrix}
        \fr {d+2p_2+2}{4}+\tfrac{\ii \wt{\nu}}2, \fr{d+2p_2+4}{4}+\tfrac{\ii \wt{\nu}}2 \\
        1+\ii \wt{\nu}
    \end{matrix}\middle|r_2^2\right]+\text{c.c.};
\end{align}
\begin{align}
    \mathcal{I}_{\text{L},\wt\nu}^{p_1p_2}
    =&~\frac1{4\pi}r_1^{d/2+1+p_1}r_2^{d/2+1+p_2}\sum_{n_1,n_2=0}^\infty \frac{ (-1 )^{n_{12}}}{n_1!n_2!}\Big[\text{e}^{\ii\pi\overline p_{12}/2}+\text{e}^{ \pi   \wt{\nu}-\ii\pi(p_{12}+d )/2 }\Big]\Big(\frac{r_1}2\Big)^{2n_1+\ii \wt{\nu}}\Big(\frac{r_2}2\Big)^{2n_2-\ii \wt{\nu}}\n \\ 
     &\times\Gamma\Big[2n_1+\fr d2+1+p_1+\ii \wt{\nu},2n_2+\fr d2+1+p_2-\ii \wt{\nu},-n_1-\ii \wt{\nu},-n_2+\ii \wt{\nu}\Big] +\text{c.c.}\n \\
    =&~\frac1{2\pi}\Big[\cos \fr{\pi\overline p_{12}}2 +\cos \fr{\pi(2\ii \wt{\nu}+p_{12}+d)}{2}\Big]r_1^{d/2+1+p_1}r_2^{d/2+1+p_2} \Big(\frac{r_1}{r_2}\Big)^{\ii \wt{\nu}} \n \\
    \times&\Gamma\Big[\fr d2+1+p_1+\ii \wt{\nu},\fr d2+1+p_2-\ii \wt{\nu},-\ii \wt{\nu},+\ii \wt{\nu}\Big]\n \\
    &\times{}_2F_1\left[ \begin{matrix}
        \fr {d+2p_1+2}{4}+\tfrac{\ii \wt{\nu}}2, \fr{d+2p_1+4}{4}+\tfrac{\ii \wt{\nu}}2 \\
        1+\ii \wt{\nu} 
    \end{matrix}\middle|r_1^2\right]
    {}_2F_1\left[ \begin{matrix}
        \fr {d+2p_2+2}{4}-\tfrac{\ii \wt{\nu}}2, \fr{d+2p_2+4}{4}-\tfrac{\ii \wt{\nu}}2 \\
        1-\ii \wt{\nu}
    \end{matrix}\middle|r_2^2\right]+\text{c.c.};
\end{align}
\begin{align}
  \label{eq_IBGpMB}
    \mathcal{I}_{\text{BG},\wt\nu}^{p_1p_2}
    =&~\frac{-1}{ \ii\wt\nu}\sin\big[\fr{\pi}{2}(p_{12}+d)\big]r_1^{d+p_{12}+2}\sum_{n_1,n_2=0}^\infty\bigg\{\frac{(-1)^{n_{12}}}{n_1!n_2!}\Big(\frac{r_1}2\Big)^{2n_1+2n_2}(-\ii\wt\nu)_{-n_1}(\ii\wt\nu)_{-n_2} \n \\
   & \times {}_2\mathcal{F}_1\left[ \begin{matrix}
        2n_2-\ii \wt{\nu}+\tfrac d2+p_2+1, 2n_{12}+p_{12}+d+2 \\
        2n_2-\ii \wt{\nu}+\tfrac d2+p_2+2
    \end{matrix}\middle|-\frac{r_1}{r_2}\right]\bigg\}+\text{c.c.}.
\end{align}
The nonlocal and local signal pieces, namely $\mathcal{I}_{\text{NL},\wt\nu}^{p_1p_2}$ and $\mathcal{I}_{\text{L},\wt\nu}^{p_1p_2}$, can be written in more compact forms, as shown in (\ref{eq_INL}) and (\ref{eq_IL}). On the other hand, the result for the background piece $\mathcal{I}_{\text{BG},\wt\nu}^{p_1p_2}$ in (\ref{eq_IBGpMB}) has resummed all $r_1/r_2$ dependence into a hypergeometric function, which is typically what we would get from a partial Mellin-Barnes representation \cite{Qin:2022fbv}. On the other hand, the result shown in (\ref{eq_IBG}) was expressed as a double series in $r_1$ and $r_1/r_2$, which is typically what we would find from recursively solving the bootstrap equation \cite{Arkani-Hamed:2018kmz,Qin:2022fbv}. In \cite{Qin:2022fbv}, it was checked numerically that the partially resummed result (\ref{eq_IBGpMB}) agrees with the bootstrapped double series (\ref{eq_IBG}). We shall prove analytically the equivalence of these two results in the next appendix. 

\section{From Partial Mellin-Barnes to Bootstrapped Series}
\label{app_pMBtoBoot}

As commented at the end of the previous appendix, the background piece of the tree seed integral can be computed in two different ways, namely the partial Mellin-Barnes representation and solving the bootstrap equations. They yield two results of different looks, given respectively by (\ref{eq_IBGpMB}) and (\ref{eq_IBG}). In this appendix, we shall show the equivalence between the two by deriving (\ref{eq_IBG}) from (\ref{eq_IBGpMB}). 

First, we use the definition of the (dressed) hypergeometric function (\ref{eq_HGF}) and (\ref{eq_DressedF}) to rewrite ${}_2\mathcal{F}_1$ in (\ref{eq_IBGpMB}) as a series:
\begin{align}
    \mathcal{I}_{\text{BG},\wt\nu}^{p_1p_2}
    =&~\frac{-1}{ \ii\wt\nu}\sin\big[\fr{\pi}{2}(p_{12}+d)\big]r_1^{d+p_{12}+2}\sum_{n_1,n_2,\ell=0}^\infty \frac{(-1)^{n_{12}+\ell}}{n_1!n_2!\ell!}\Big(\frac{r_1}2\Big)^{2n_{12}}\Big(\FR{r_1}{r_2}\Big)^{\ell}(-\ii\wt\nu)_{-n_1}(\ii\wt\nu)_{-n_2} \n \\
   &\times\FR{\Gamma(\ell+2n_{12}+p_{12}+d+2)}{\ell+2n_2-\ii \wt{\nu}+\tfrac d2+p_2+1} 
   +\text{c.c.}.
\end{align}
Using the new summation variable $m=n_1+n_2$, we can rewrite the summation as 
\begin{align}
\label{eq_IBGexpandedMid1}
    \mathcal{I}_{\text{BG},\wt\nu}^{p_1p_2}
    =&~\frac{-1}{ \ii\wt\nu}\sin\big[\fr{\pi}{2}(p_{12}+d)\big]r_1^{d+p_{12}+2}\sum_{m,\ell=0}^\infty\sum_{n_2=0}^m\binom{m}{n_2} \frac{(-1)^{m+\ell}}{m!\ell!}\Big(\frac{r_1}2\Big)^{2m}\Big(\FR{r_1}{r_2}\Big)^{\ell}(-\ii\wt\nu)_{n_2-m}(\ii\wt\nu)_{-n_2} \n \\
   &\times\FR{\Gamma(\ell+2m+p_{12}+d+2)}{\ell+2n_2-\ii \wt{\nu}+\tfrac d2+p_2+1} \n\\
   &~+\frac{1}{ \ii\wt\nu}\sin\big[\fr{\pi}{2}(p_{12}+d)\big]r_1^{d+p_{12}+2}\sum_{m,\ell=0}^\infty\sum_{n_2=0}^m\binom{m}{n_2} \frac{(-1)^{m+\ell}}{m!\ell!}\Big(\frac{r_1}2\Big)^{2m}\Big(\FR{r_1}{r_2}\Big)^{\ell}(+\ii\wt\nu)_{n_2-m}(-\ii\wt\nu)_{-n_2} \n \\
   &\times\FR{\Gamma(\ell+2m+p_{12}+d+2)}{\ell+2n_2+\ii \wt{\nu}+\tfrac d2+p_2+1} 
\end{align}
There are two summations in this expression. For the second summation, we change the summation variable $n_2\to m-n_2$:
\begin{align}
 & \sum_{m,\ell=0}^\infty\sum_{n_2=0}^m\binom{m}{n_2} \frac{(-1)^{m+\ell}}{m!\ell!}\Big(\frac{r_1}2\Big)^{2m}\Big(\FR{r_1}{r_2}\Big)^{\ell}(+\ii\wt\nu)_{n_2-m}(-\ii\wt\nu)_{-n_2}  
    \FR{\Gamma(\ell+2m+p_{12}+d+2)}{\ell+2n_2+\ii \wt{\nu}+\tfrac d2+p_2+1} \n\\
  =&\sum_{m,\ell=0}^\infty\sum_{n_2=0}^m\binom{m}{n_2} \frac{(-1)^{m+\ell}}{m!\ell!}\Big(\frac{r_1}2\Big)^{2m}\Big(\FR{r_1}{r_2}\Big)^{\ell}(+\ii\wt\nu)_{-n_2}(-\ii\wt\nu)_{n_2-m}  
    \FR{\Gamma(\ell+2m+p_{12}+d+2)}{\ell+2(m-n_2)+\ii \wt{\nu}+\tfrac d2+p_2+1}.
\end{align}
Then, combining this result with the first summation in (\ref{eq_IBGexpandedMid1}), we get:
\begin{align} 
   \mathcal{I}_{\text{BG},\wt\nu}^{p_1p_2}
    =&~\frac{-1}{ \ii\wt\nu}\sin\big[\fr{\pi}{2}(p_{12}+d)\big]r_1^{d+p_{12}+2} 
    \sum_{m,\ell=0}^\infty\sum_{n_2=0}^m\binom{m}{n_2} \frac{(-1)^{m+\ell}}{m!\ell!}\Big(\frac{r_1}2\Big)^{2m}\Big(\FR{r_1}{r_2}\Big)^{\ell}(-\ii\wt\nu)_{n_2-m}(\ii\wt\nu)_{-n_2}\n\\
    &\times 
    \FR{2(m-2n_2+\ii\wt\nu)\Gamma(\ell+2m+p_{12}+d+2) }{(\ell+2n_2-\ii \wt{\nu}+\tfrac d2+p_2+1)[\ell+2(m-n_2)+\ii \wt{\nu}+\tfrac d2+p_2+1]}.
\end{align}
Next, we finish the summation over $n_2$. Isolating the terms dependent on $n_2$, we have:
\begin{align} 
    &\sum_{n_2=0}^m\binom{m}{n_2}
    \FR{(n_2-\fr{m+\ii\wt\nu}{2})(-\ii\wt\nu)_{n_2-m}(\ii\wt\nu)_{-n_2} }{(n_2+\fr{\ell-\ii \wt{\nu}+d/2+p_2+1}{2})(n_2-m-\fr{\ell+\ii \wt{\nu}+d/2+p_2+1}{2})}\n\\
    =&\sum_{n_2=0}^m\FR{1}{n_2!} \FR{\Gamma(-m+n_2)}{\Gamma(-m)}
    \FR{(-\ii\wt\nu) (n_2-\fr{m+\ii\wt\nu}{2}) }{(n_2+\fr{\ell-\ii \wt{\nu}+d/2+p_2+1}{2})(n_2-m-\fr{\ell+\ii \wt{\nu}+d/2+p_2+1}{2})}\Gamma\bgb-m+n_2-\ii\wt\nu\\ 1+n_2-\ii\wt\nu\edb\n\\
    =&~\FR{-\ii\wt\nu}{\Gamma(-m)}{}_5\mathcal{F}_4\left[\bgm
   -m,-\fr{m+\ii\wt\nu}{2}+1,\fr{\ell-\ii \wt{\nu}+d/2+p_2+1}{2},-m-\fr{\ell+\ii \wt{\nu}+d/2+p_2+1}{2},-m-\ii\wt\nu \\
   -\fr{m+\ii\wt\nu}{2},\fr{\ell-\ii \wt{\nu}+d/2+p_2+3}{2},-m-\fr{\ell+\ii \wt{\nu}+d/2+p_2-1}{2}, 1-\ii\wt\nu
    \edm\middle|1\right].
\end{align}
The dressed hypergeometric function ${}_5\mathcal{F}_4$ in the last line can be simplified by using the Rogers–Dougall very well-poised sum for ${}_5\mathrm{F}_4$ \cite{Slater:1966,nist:dlmf}:
\begin{align}
    &{}_5F_4\left[\begin{matrix}
        a, \tfrac a2+1, b, c, e \\
        \tfrac a2, a-b+1, a-c+1, a-e+1
    \end{matrix}\middle|1\right]\n\\
    &=\Gamma\left[\begin{matrix}
        a-b+1, a-c+1, a-e+1, a-b-c-e+1 \\
        a+1, a-b-c+1, a-b-e+1, a-c-e+1
    \end{matrix}\right],
\end{align}
when $\text{Re}(b+c+d-a)<1$, or when the series terminates with $e=-n$($n=0,1,2,\cdots$). Here we can choose
\begin{align}
    &a=-m-\ii \wt{\nu}, &&b=\frac{\ell-\ii \wt{\nu}+p_2+1}2+\frac d4, &&c=-m-\frac{\ell+\ii \wt{\nu}+p_2+1}2-\frac d4, &&e=-m.
\end{align}
Then, we get
\begin{align}
    \mathcal{I}_{\text{BG},\wt\nu}^{p_1p_2}
    =&~\FR12\sum_{m,\ell=0}^\infty \frac{\left(-1\right)^{m+\ell}\sin[\fr{\pi}{2}(p_{12}+d)]}{\ell!}\Big(\frac{r_1}2\Big)^{2m}\Big(\frac{r_1}{r_2}\Big)^{\ell}r_1^{d+p_{12}+2}\n \\
    &\times\Gamma\left[\begin{matrix}
        \tfrac{\ell-\ii \wt{\nu}'+p_2+1}2+\tfrac d4, -m-\tfrac{\ell+\ii \wt{\nu}'+p_2+1}2-\tfrac d4, 2m+\ell+d+p_{12}+2 \\
        -\tfrac{\ell+\ii \wt{\nu}'+p_2-1}2-\tfrac d4, m+\tfrac{\ell-\ii \wt{\nu}'+p_2+3}2+\tfrac d4
    \end{matrix}\right]\n \\
    =&\sum_{m,\ell=0}^\infty\frac{(-1)^{\ell+1}\sin[\fr{\pi}{2}(p_{12}+d)](\ell+1)_{2m+d+p_{12}+1}}{2^{2m+1}\big(\tfrac{\ell-\ii \wt{\nu}'+p_2+1}2+\tfrac d4\big)_{m+1}\big(\tfrac{\ell+\ii \wt{\nu}'+p_2+1}2+\tfrac d4\big)_{m+1}}r_1^{2m+d+p_{12}+2}\Big(\frac{r_1}{r_2}\Big)^{\ell},
\end{align}
which is nothing but the double series (\ref{eq_IBG}).

\section{Bootstrapping Loop Correlators in Minkowski Spacetime}
\label{app_mink}

In this section, we compute the 4-point correlation with 1-loop massive scalar exchange in the $s$-channel in Minkowski spacetime. The diagram is again given by the left-hand side of Fig.\ \ref{fig_LoopSD}. It turns out a direct computation of SK time integrals is already complicated enough in $(d+1)$-dimensional Minkowski space. Therefore, we also take the spectral decomposition approach here. Below, we first compute the corresponding tree-level correlator, namely the right-hand side of Fig.\ \ref{fig_LoopSD}, and then derive a spectral function in flat space. Finally, we compute the 1-loop correlator by finishing the spectral integral over the tree-level correlator. 

\subsection{Tree-level correlator}

We again apply the SK diagrammatic method \cite{Chen:2017ryl} to compute the correlator. In flat space, algebra is much simplified. For instance, one of the homogeneous propagators for a scalar of mass $m$ is given by 
\bge
  D_{m,>}(\mb k;t_1,t_2)=\FR{e^{-\ii \sqrt{\mb k^2+m^2}(t_1-t_2)}}{2\sqrt{\mb k^2+m^2}},
\ede
and $D_{m,<}$ is the complex conjugation of $D_{m,>}$. The four SK propagators $D_{\aa\bb}$ can be built as in (\ref{eq_Dpmpm}) and (\ref{eq_Dpmmp}). 
Then, the $s$-channel diagram of the 4-point function can be computed as
\begin{align}
\label{eq_MinkTreeSKInt}
  \mathcal{T}^\text{Mink}_{m,\aa\bb} 
  =&- \aa\bb\int_{-\infty}^0\di t_1\di t_2\,G_{\aa}(\mb k_1;\tau_1)G_{\aa}(\mb k_2;\tau_1)G_{\bb}(\mb k_3;\tau_2)G_{\bb}(\mb k_4;\tau_2)D_{m,\aa\bb}(\mb k_s;\tau_1,\tau_2).
\end{align}
Here $G_{\aa}(\mb k_i,\tau)$ are boundary propagators for the four external legs, and $D_{m,\aa\bb}$ is the bulk propagator for the intermediate states. We take all these lines to be massive scalars. The simple algebra of Minkowski space allows us to introduce (different) masses to all these legs without any pain. So, we associate $m_i$ $(i=1,2,3,4)$ to the four external legs and $m$ to the intermediate state. Then, we have $E_i\equiv\sqrt{\mb k_i^2+m_i^2}$ and $E_s\equiv\sqrt{\mb k_s^2+m^2}$. With these quantities, the time integrals can be done directly. First, the two same-sign integrals read:
\begin{align}
  \mathcal{T}^\text{Mink}_{m,\pm\pm}
  =&-\FR{1}{32E_1E_2E_3E_4E_s}\bigg[\int_{-\infty}^0\di t_1\int_{-\infty}^{t_1}\di t_2\,e^{\pm\ii (E_{12}-E_s)t_1\pm\ii (E_{34}+E_s)t_2} \n\\
  &+\int_{-\infty}^0\di t_1\int_{t_1}^{0}\di t_2\,e^{\pm\ii (E_{12}+E_s)t_1\pm\ii (E_{34}-E_s)t_2}\bigg]\n\\
  =&~\FR{1}{32E_1E_2E_3E_4E_s(E_{12}+E_s)(E_{34}+E_s)}\bigg(1+\FR{2E_s}{E_{1234}}\bigg).
\end{align}
Then, the two opposite-sign integrals read:
\begin{align}
  \mathcal{T}^\text{Mink}_{m,s,\pm\mp}
  =&~\FR{1}{32E_1E_2E_3E_4E_s} \int_{-\infty}^0\di t_1 \di t_2\,e^{\pm\ii (E_{12}+E_s)t_1\mp\ii (E_{34}+E_s)t_2}\n\\
  =&~\FR{1}{32E_1E_2E_3E_4E_s(E_{12}+E_s)(E_{34}+E_s)} .
\end{align}
Summing up the four integrals, we get the final answer for the tree-level 4-point diagram with $s$-channel massive scalar exchange: 
\begin{align}
\label{eq_MinkTree}
  \mathcal{T}_{m}^\text{Mink}\equiv \sum_{\aa,\bb=\pm}\mathcal{T}_{m,\aa\bb}^\text{Mink}
  =\FR{1}{8E_1E_2E_3E_4E_s(E_{12}+E_s)(E_{34}+E_s)}\bigg(1+\FR{E_s}{E_{1234}}\bigg).
\end{align}
This will be the resource for us to bootstrap the 1-loop correlator.

\subsection{Spectral function}

Now we turn to the spectral decomposition in flat space. The goal is to find a spectral function $\rho^\text{Mink}_{m^2}(m'^{\,2})$ in Minkowski spacetime such that
\bge
\label{eq_MinkRho}
  \Big[D_m(x,y)\Big]^2=\int_0^\infty\di m'\FR{m'}{\pi\ii}\rho_{m^2}^\text{Mink}(m'^{\,2})D_{m'}(x,y),
\ede
or, in spatial-momentum space,
\bge
  \int\FR{\di^d\mb q}{(2\pi)^d}D_{m,\aa\bb}\Big(|\mb q|;t_1,t_2\Big)D_{m,\aa\bb}\Big(|\mb q-\mb k|;t_1,t_2\Big)=\int_0^\infty\di m'\FR{m'}{\pi\ii}\rho_{m^2}^\text{Mink}(m'^{\,2})D_{m',\aa\bb}\Big(|\mb k|;t_1,t_2\Big).
\ede

It turns out convenient to start from the Euclidean version. Let us consider the product of a pair of scalar propagators $[D_m^\text{Euc}(x,y)]^2$ in Euclidean space, and take the Fourier transform:
\bge
\label{eq_EucBubble}
  \Big[D_m^\text{Eucl}(x,y)\Big]^2=\int\FR{\di^{d+1}k}{(2\pi)^{d+1}}\,B_{m^2}^\text{Eucl}(k^2)\text{e}^{\ii k_\mu(x_1-x_2)^\mu},
\ede
where $B_{m^2}^\text{Eucl}(k^2)$ is the 1-loop bubble function in Euclidean space. By the symmetry of the theory, the bubble function $B_{m^2}^\text{Eucl}(k^2)$ depends only on the magnitude of the (Euclidean) momentum $k^2=k_\mu k^\mu$. 

Now, suppose that we know the bubble function $B_{m^2}^\text{Eucl}(k^2)$. Then, it is straightforward to express the spectral function $\rho^\text{Mink}_{m^2}(m'^{\,2})$ in terms of the bubble function. More explicitly, we have:
\begin{align} 
\label{eq_EucBubbleInt}
    &\int\frac{\di^{d+1}k}{(2\pi)^{d+1}}B_{m^2}^{\text{Eucl}}(k^2)\text{e}^{\ii k_\mu(x-y)^\mu}\n \\
    =&~\FR{1}{2^d\pi^{d/2+1}\Gamma(\frac{d}{2})}\int_0^\infty\di k\,k^d \int_0^\pi\di\theta\,\sin^{d-1}\theta\,\text{e}^{\ii kr\cos\theta}B_{m^2}^{\text{Eucl}}(k^2)\n \\
    =&\int_0^\infty\frac{\di k}{(2\pi)^{(d+1)/2}}\frac{k^{(d+1)/2}}{r^{(d-1)/2}}\mathrm{J}_{(d-1)/2}(kr)B_{m^2}^{\text{Eucl}}(k^2)\n \\
    =&\int_0^\infty\frac{\di m'}{(2\pi)^{(d+1)/2}}\frac{\ii}\pi\frac{m'^{\,(d+1)/2}}{r^{(d-1)/2}}\mathrm{K}_{(d-1)/2}(m'r)\Big[B_{m^2}^{\text{Eucl}}(m'^{\,2}\text{e}^{\ii\pi})-B_{m^2}^{\text{Eucl}} (m'^{\,2}\text{e}^{-\ii\pi} ) \Big]\n \\
    =&\int_0^\infty\di m'\,\frac{m'}{\pi\ii} D_{m'}^\text{Eucl}(x,y)\Big[B_{m^2}^{\text{Eucl}}(m'^{\,2}\text{e}^{-\ii\pi})-B_{m^2}^{\text{Eucl}} (m'^{\,2}\text{e}^{\ii\pi} ) \Big],
\end{align}
where $\mathrm{J}_\nu(z)$ is the Bessel function and $\mathrm{K}_{\nu}(z)$ is a modified Bessel function. Here we have used the fact that the Euclidean propagator for a scalar of mass $m$ is given by 
\begin{align}
  D_m^\text{Eucl}(x,y)=\FR{1}{(2\pi)^{(d+1)/2}}\Big(\FR{m}{r}\Big)^{(d-1)/2}\mathrm{K}_{(d-1)/2}(mr).
\end{align}

Now we can Wick-rotate back to the Minkowski spacetime. Then, comparing (\ref{eq_EucBubble}), (\ref{eq_EucBubbleInt}), and (\ref{eq_MinkRho}), we find
\bge
\label{eq_rhoMinkInB}
  \rho_{m^2}^\text{Mink}(m'^{\,2})=B_{m^2}^{\text{Eucl}}(m'^{\,2}\text{e}^{-\ii\pi})-B_{m^2}^{\text{Eucl}} (m'^{\,2}\text{e}^{\ii\pi} ).
\ede

It remains to find the bubble function $B_{m^2}^\text{Eucl}(k^2)$. To this end, we Fourier transform each of the two propagators $D_{m}^\text{Eucl}(x,y)$ separately:
\begin{align}
\label{eq_EucLoop}
  \Big[D_m^\text{Eucl}(x,y)\Big]^2
  =&\int\FR{\di q_1^{d+1}}{(2\pi)^{d+1}}\FR{\di q_2^{d+1}}{(2\pi)^{d+1}}\FR{e^{\ii(q_1+q_2)_\mu(x-y)^\mu}}{(q_1^2+m^2)(q_2^2+m^2)}\n\\
  =&\int\FR{\di p^{d+1}}{(2\pi)^{d+1}}\FR{e^{\ii p_\mu(x-y)^\mu}}{[(p-q)^2+m^2](q^2+m^2)},
\end{align}
where $p=q_1+q_2$ and $q=q_2$. Comparing (\ref{eq_EucLoop}) with (\ref{eq_EucBubble}), we find:
\begin{align}
  B_{m^2}^\text{Eucl}(k^2)
  =&\int\FR{\di^{d+1}q}{(2\pi)^{d+1}}\FR{1}{((k-q)^2+m^2)(q^2+m^2)} \n\\
  =&~ \FR{\Gamma(\frac{3-d}2)}{(4\pi)^{(d+1)/2}}\int_0^1 \FR{\di\xi}{[m^2+\xi(1-\xi)k^2]^{(3-d)/2}},
\end{align} 
where we use the standard Feynman parametrization in the last step. Finally, using (\ref{eq_rhoMinkInB}), we get an expression for the 1-loop spectral function in Minkowski spacetime:
\begin{align}
\label{eq_MinkRhoResult}
  \rho_{m^2}^\text{Mink}(m'^{\,2})
  =&~\FR{\Gamma(\frac{3-d}2)}{(4\pi)^{(d+1)/2}}\int_0^1\di\xi \bigg\{\FR{1}{[m^2+\xi(1-\xi)m'^{\,2}e^{-\ii\pi}]^{(3-d)/2}}-\FR{1}{[m^2+\xi(1-\xi)m'^{\,2}e^{\ii\pi}]^{(3-d)/2}}\bigg\}\n\\
  =&~\FR{\Gamma(\frac{3-d}2)}{(4\pi)^{(d+1)/2}}\int_0^1\di\xi\,2\ii\,\text{Im}\, \FR{1}{[m^2-\xi(1-\xi)m'^{\,2}-\ii\ep]^{(3-d)/2}}.
\end{align}

\subsection{1-loop correlator}

Now we are ready to bootstrap the 1-loop correlator. Using the Feynman rule in SK formalism, the 1-loop correlator has the following form:
\begin{align}
  \mathcal{L}^\text{Mink}_{m}\equiv\sum_{\aa,\bb=\pm} \mathcal{L}^\text{Mink}_{m,\aa\bb}
  =&-\FR{1}{2}\sum_{\aa,\bb=\pm}\aa\bb\int_{-\infty}^0\di t_1\di t_2\,G_{\aa}(\mb k_1;\tau_1)G_{\aa}(\mb k_2;\tau_1)G_{\bb}(\mb k_3;\tau_2)G_{\bb}(\mb k_4;\tau_2)\n\\
   &\times\int\FR{\di^d\mb q}{(2\pi)^d}D_{m,\aa\bb}\Big(|\mb q|;t_1,t_2\Big)D_{m,\aa\bb}\Big(|\mb q-\mb k_s|;t_1,t_2\Big).
\end{align}
Comparing this expression with the tree amplitude (\ref{eq_MinkTreeSKInt}) and using the spectral function (\ref{eq_MinkRho}), we have
\begin{align}
  \mathcal{L}^\text{Mink}_{m}=\int_0^\infty\di m'\FR{m'}{2\pi\ii}\rho_{m^2}^\text{Mink}(m'^{\,2})\mathcal{T}^\text{Mink}_{m'}.
\end{align}

Now, with the explicit expressions for the tree-level correlator $\mathcal{T}^\text{Mink}_{m}$ in (\ref{eq_MinkTree}) and the spectral function $\rho_{m^2}^\text{Mink}(m'^{\,2})$ in (\ref{eq_MinkRhoResult}), we can compute the 1-loop correlator as
\begin{align}
  \mathcal{L}_{m}^\text{Mink}=\int_0^\infty\di m'\FR{m'}{2\pi\ii}\FR{(1+E_s/E_{1234})}{8E_1E_2E_3E_4E_s(E_{12}+E_s)(E_{34}+E_s)} \rho_{m^2}^\text{Mink}(m'^{\,2}).
\end{align}
Change the integration variable from $m'$ to $E_s=\sqrt{\mb k_s^2+m'^{\,2}}$, we have $\di m'(m'/E_s)=\di E_s$. Then,
\begin{align}
  \mathcal{L}_{m}^\text{Mink} 
 =&\int_0^1\di\xi\int_{E_\text{min}}^\infty\FR{\di E_s}{\pi}\FR{(1+E_s/E_{1234})}{8E_1E_2E_3E_4(E_{12}+E_s)(E_{34}+E_s)}\n\\
  &~\times\FR{\Gamma(\frac{3-d}2)}{(4\pi)^{(d+1)/2}}\FR{-\cos(\pi d/2)}{[\xi(1-\xi)(E_s^2-E_\text{min}^2)]^{(3-d)/2}}.
\end{align}
Here $E_\text{min}\equiv \sqrt{\mb k_s^2+m^2/[\xi(1-\xi)]}$. The $E_s$ integral can be finished directly. In the $d\to 3$ limit, the result is
\begin{align}
  \mathcal{L}_{m}^\text{Mink} 
  =&~\FR{1}{256\pi^2E_1E_2E_3E_4E_{1234}}\bigg[\FR{2}{3-d}-\ga_E+\log4\pi+2\n\\
  &~+\FR{2}{E_{12}-E_{34}}\int_0^1\di\xi\,\bigg(E_{34}\log\FR{E_{12}+E_\text{min}}{\mu_R}-E_{12}\log\FR{E_{34}+E_\text{min}}{\mu_R}\bigg) \bigg].
\end{align}
This is the final result for the 1-loop correlator in Minkowski spacetime. We can now directly use $\overline{\text{MS}}$ scheme to subtract the divergent piece $\propto 2/(3-d)-\ga_E+\log4\pi$. It is also of interest to have an expression for the renormalized loop correlator in the large mass limit $m\to \infty$: 
\begin{align}
\label{eq_MinkLoopLargeM}
  \Big[\mathcal{L}_{m}^\text{Mink}\Big]_{\overline{\text{MS}}}
  =&~\FR{1}{256\pi^2E_1E_2E_3E_4E_{1234}}\bigg[\log\FR{\mu_R^2}{m^2}-\FR{E_{12}E_{34}+\mb k_s^2}{6m^2}+\mathcal{O}\Big(\FR{1}{m^{3}}\Big)\bigg].
\end{align}

\end{appendix}

\newpage
\bibliography{CosmoCollider} 
\bibliographystyle{utphys}

\end{document}